\documentclass[aps,prd,showpacs,tightenlines,twocolumn,floats,superscriptaddress,nofootinbib]{revtex4-1}
\usepackage{amssymb}
\usepackage{amsmath}
\usepackage{amsfonts}
\usepackage{bm}
\usepackage{graphicx}
\usepackage{color}
\usepackage{here}
\usepackage{subfigure}
\usepackage{natbib}
\usepackage{dsfont}
\usepackage{enumitem}
\usepackage{hyperref}
\usepackage{braket}

\usepackage{epsf,latexsym,bbm,euscript}
\usepackage{times,graphics}
\usepackage[normalem]{ulem}

\newcommand{\ii}{\mathrm{i}}

\begin{document}

\title{Superadditivity of channel capacity through quantum fields}

\author{Koji Yamaguchi}
\affiliation{Graduate School of Science, Tohoku University, Sendai, 980-8578, Japan}

\author{Aida Ahmadzadegan}
\affiliation{Department of Applied Mathematics, University of Waterloo, Waterloo, Ontario, N2L 3G1, Canada}
\affiliation{Perimeter Institute for Theoretical Physics, Waterloo, Ontario N2L 2Y5, Canada}

\author{Petar Simidzija}
\affiliation{Department of Physics and Astronomy, University of British Columbia, Vancouver, British Columbia, V6T 1Z4, Canada}

\author{Achim Kempf}
\affiliation{Department of Applied Mathematics, University of Waterloo, Waterloo, Ontario, N2L 3G1, Canada}
\affiliation{Institute for Quantum Computing, University of Waterloo, Waterloo, Ontario, N2L 3G1, Canada}
\affiliation{Perimeter Institute for Theoretical Physics, Waterloo, Ontario N2L 2Y5, Canada}
\affiliation{Department of Physics and Astronomy, University of Waterloo, Waterloo, Ontario, N2L 3G1, Canada}

\author{Eduardo Mart\'in-Mart\'inez}
\affiliation{Department of Applied Mathematics, University of Waterloo, Waterloo, Ontario, N2L 3G1, Canada}
\affiliation{Institute for Quantum Computing, University of Waterloo, Waterloo, Ontario, N2L 3G1, Canada}
\affiliation{Perimeter Institute for Theoretical Physics, Waterloo, Ontario N2L 2Y5, Canada}

\begin{abstract}
Given that any communication is communication through quantum fields, we here study the scenario where a sender, Alice, causes information-carrying disturbances in a quantum field. We track the exact spread of these disturbances in space and time by using the technique of quantum information capsules (QIC).  
We find that the channel capacity between Alice and a receiver, Bob, is enhanced by Bob placing detectors not only inside but in addition also outside the causal future of Alice's encoding operation. Intuitively, this type of superadditivity arises because the field outside the causal future of Alice is entangled with the field inside Alice's causal future. 
Hence, the quantum noise picked up by Bob's detectors outside Alice's causal future is correlated with the noise of Bob's detectors inside Alice's causal future. In effect, this correlation allows Bob to improve the signal-to-noise ratio of those of his detectors which are in the causal future of Alice. Further, we develop the multimode generalization of the QIC technique. This allows us to extend the analysis to the case where Alice operates multiple localized and optionally entangled emitters.
We apply the new techniques to the case where Alice enhances the channel capacity by operating multiple emitters that are suitably lined up and pre-timed to generate a quantum shockwave in the field. 
\end{abstract}
\maketitle

\section{Introduction}
Given the progress in wireless communication technologies, it is becoming increasingly important to fully develop the underlying theory, namely to fully take into account that the emitters, the field, and the receivers are quantum systems. 
In addition to the prospect of new technological applications, e.g., for quantum communication and quantum cryptography, these studies also reveal fundamental new insights into the relationship between the flow of information, quantum phenomena and relativistic effects. 

For example, it has been shown to be possible to send information  from a sender to a receiver  without transmitting energy \cite{PhysRevLett.114.110505}. Since the receiver needs to provide energy to detect the signal, the protocol may be referred to as quantum collect calling. Another novel protocol \cite{ahmadzadegan2018quantum} shows that in a setup of multiple emitters it is possible to shape the beam that they emit not only through the modulation of amplitudes and phases of the emitters but also through the modulation of the initial entanglement of the emitters. It was shown, in particular, that a suitable array of pre-timed emitters can emit a 
quantum shockwave that is modulated by the entanglement of the emitters.  
The results of Ref. \cite{ahmadzadegan2018quantum}  demonstrate, therefore, that the presently ubiquitously used multiple input multiple output (MIMO) systems (i.e., systems with multiple senders and multiple receivers) can be improved, in principle, by making use of the quantum nature of the systems involved. 

An aspect of wireless communication that does not change when taking into account the quantum nature of the emitters, receivers and the field is the role of the strong Huygens principle. Indeed, also when fully quantized, \cite{AIHPA_1974__20_2_153_0}, communication via a massless field is still restricted to lightlike separated senders and receivers in flat spacetimes in $(3+1)$-dimensions, (while communication is possible on \it and in \rm the future light cone of the emitter in $(1+1), (2+1)$ and general $(2n+1)$ dimensions, as well as in cases of nonvanishing generic curvature in spacetimes of any dimension.) 

In addition, there exist features of fully quantized wireless communication that possess no analog in classical systems, i.e., that arise only 
when taking into account the quantum nature of the emitters, receivers and the field. In particular, quantum emitters, receivers and fields can establish a communication channel that possesses quantum channel capacity, i.e., that can transmit entanglement. 

Quantum channel capacity has delicate properties without classical analogs. For example, quantum channel capacity is subject to  the no-cloning theorem \cite{wootters1982single}, which translates here into the constraint that it is generally impossible to broadcast quantum information to multiple disjoint receivers. This was originally shown to be the case for communication protocols via quantum fields in (1+1)-dimensions~\cite{Jonsson_2018}, while a further understanding of the phenomenon in general dimensions was reached in Ref. \cite{simidzija2019transmission}. It has also been established, for example, that in order for an emitter or receiver system to even transmit quantum channel capacity into or out of a quantum field, the emitter or receiver system should not interact too briefly with the field, as very short interactions with a quantum field tend to be entanglement breaking \cite{No-go}. For a strategy for maximizing the quantum channel capacity, see Ref. \cite{simidzija2019transmission}. 

At the heart of the new phenomena that appear when taking into account the quantum nature of emitter, receiver and fields is the fact that the local degrees of freedom of any quantum field are generally entangled at timelike, null and also spacelike distances \cite{Alegbra1,Alegbra2}, even if the field is in the vacuum state. This means that when quantized emitters and receivers couple to a quantum field then they nontrivially couple to an extended system which possesses pre-existing entanglement. For example, two localized quantum systems that briefly couple to the field while at spacelike separations can become entangled (e.g., among others, Refs. \cite{Valentini1991,Reznik2003,Reznik1,reznik2,Retzker2005,Nick,Olson2011,Olson2012,Salton:2014jaa,Pozas-Kerstjens:2015,Pozas2016,Nambu2013,Sabinprl,Farming,BeiLok1,Kukita2017,Kukita20172,Ng2,Henderson2018,Henderson2019,PetarHarv,Cong2019,Brown1,Brown2}) because they generically swap entanglement from the field. 

For our study of communication through quantum fields here, we will make use of techniques developed in Refs. \cite{YAMAGUCHI20191255,yamaguchi2019quantum}. There, it was shown how, when a system couples to a large entangled system (such as a quantum field), one can identify the exact degrees of freedom that pick up information from that coupling. These degrees of freedom have been named quantum information capsules (QICs).
Concretely, Refs. \cite{YAMAGUCHI20191255,yamaguchi2019quantum} investigated encoding processes in the form of an interaction Hamiltonian consisting of a single Hermitian operator. It was shown that there always exists a subsystem characterized by a subalgebra such that the subsystem is in a pure state and the encoding operation is generated by the subalgebra. This subsystem is called a QIC. The purity of the QIC implies that no information is shared with its complement subsystem. Thus, a QIC can be used as a unit of memory of encoded information. The existence of a QIC has been shown for multiple-qubit systems \cite{YAMAGUCHI20191255} and multiple-qudit systems \cite{yamaguchi2019quantum} in a general entangled state. Furthermore, for continuous-valued systems, i.e., multiple harmonic oscillators and quantum fields, in a Gaussian state, a formula to identify a QIC mode has been proven \cite{yamaguchi2019quantum}. We will refer to this formula as the single-mode QIC formula. 

In the present paper, we  use the formalism of QICs to identify new phenomena that arise when taking into account the quantum nature of emitters, receivers and fields. Concretely, we first investigate the communication setup where a sender (Alice) encodes information by using a single Unruh-Dewitt (UDW) particle detector \cite{PhysRevD.14.870, DeWitt} (i.e. a first quantized system, such as a qubit, or an atom) which instantaneously couples to a scalar field. In this case, the information carrier is uniquely identified by the single-mode QIC formula since the encoding operation is generated by a single Hermitian operator. We illustrate the utility of the new method by calculating the Huygens-principle-related difference in the time evolution of the QIC in $(3+1)$- and $(2+1)$-dimensional Minkowski spacetimes.

We then investigate the classical channel capacity for setups in which Alice uses one emitter to message Bob who uses multiple detector systems at various  locations. It is clear that Bob can increase the channel capacity from Alice to him by placing more detectors on or in the future light cone of Alice's emission. However, as we here show, Bob can increase the channel capacity from Alice to him also by placing detectors outside the future lightcone - where Alice's signal cannot reach. The reason for the occurrence of this new type of superadditivity of the channel capacity is that those of Bob's detectors that are outside Alice's future light cone can record quantum noise of the field. Due to the entanglement in the quantum field, this noise is correlated with the quantum noise in the field that Bob's detectors in Alice's causal future are picking up. Bob can use this fact to better separate the signal from the noise in those of his detectors that are inside the future light cone of Alice. We therefore arrive at a novel way to enhance the channel capacity between Alice and Bob, namely by using entanglement-induced non-local correlations in the noise at the receivers. 

Technically, we will show here that the QIC mode that Alice creates in the quantum field generally has a tail through all of space, even if Alice encodes her information by a local operation. This is because quantum fields possess entanglement and correspondingly correlated quantum field fluctuations even across spacelike distances. 

We then go beyond this setup and consider the case where Alice possesses multiple emitters. To this end, we generalize the single-mode QIC formula of Ref. \cite{yamaguchi2019quantum}. We then show that when Alice makes use of $k$ emitters, then (at most) $k$ modes in a pure state are the information carriers, which we call a $k$-mode QIC. Finally, we demonstrate the utility of the new $k$-mode QIC technique by applying it to the scenario of Ref. \cite{ahmadzadegan2018quantum}, where Alice uses her emitters to communicate by creating quantum shockwaves.


\noindent Throughout this paper, we adopt natural units, $\hbar=c=1$.

\section{Information propagation through quantum fields}\label{sec_te}
In this section, we investigate the propagation of information encoded by an UDW detector by using the single-mode QIC formula. An UDW detector is a first quantized system which is linearly coupled to the quantum field~\cite{DeWitt}. In particular, we will take the UDW detector to be a qubit which couples to a free scalar field. Despite its simplicity,  this model provides an accurate description of the light-matter interaction between atoms and the electromagnetic field (i.e. a vector field) in cases where the exchange of angular momentum can be ignored~\cite{Pozas2016,Pablo}.

\subsection{Setup} 
Consider a scalar field $\hat{\phi}(t,\bm{x})$ and its conjugate momentum $\hat{\Pi}(t,\bm{x})$ in a $(d+1)$-dimensional Minkowski spacetime. They are expanded by using plane wave solutions of the equation of motion and given by
\begin{align}
 \hat{\phi}(t,\bm{x})&=\int \frac{\text{d}^d\bm{k}}{\sqrt{(2\pi)^d 2|\bm{k}}|}\nonumber\\
 &\quad \times\left(\hat{a}_{\bm{k}}e^{-\ii(|\bm{k}|t-\bm{k}\cdot\bm{x})}+\hat{a}_{\bm{k}}^\dag e^{\ii(|\bm{k}|t-\bm{k}\cdot\bm{x})}\right),\\
\hat{\Pi}(t,\bm{x}) & =\int \frac{\text{d}^d\bm{k}}{\sqrt{(2\pi)^d 2|\bm{k}}|}(-\ii|\bm{k}|)\nonumber\\
&\quad \times\left(\hat{a}_{\bm{k}}e^{-\ii(|\bm{k}|t-\bm{k}\cdot\bm{x})}-\hat{a}_{\bm{k}}^\dag e^{\ii(|\bm{k}|t-\bm{k}\cdot\bm{x})}\right),
\end{align}
where the creation and annihilation operators satisfy
\begin{align}
\left[\hat{a}_{\bm{k}},\hat{a}_{\bm{k}'}\right]= \left[\hat{a}_{\bm{k}}^\dag,\hat{a}_{\bm{k}'}^\dag\right]=0,\quad \left[\hat{a}_{\bm{k}},\hat{a}_{\bm{k}'}^\dag\right]=\delta^{(d)}\left(\bm{k}-\bm{k}'\right).
\end{align} 

Suppose that Alice wants to encode information of a qubit in the scalar field by a UDW-type interaction between the qubit and field. For an inertial qubit, the interaction Hamiltonian is given by
\begin{align}
 \hat{H}_{\mathrm{int}}(t)=\lambda\chi(t)\hat{\mu}(t)\otimes \hat{O}(t)
\end{align}
in the interaction picture. Here $\lambda$ is the coupling constant, $\chi(t)$ is the switching function, and $\hat{\mu}(t)$ and $\hat{O}(t)$ are observables of the qubit and the field, respectively. The field operator $\hat{O}(t)$ is assumed to be given by
\begin{align}
 \hat{O}(t)=\int \text{d}^d\bm{x}\left(v^{(1)}(\bm{x})\hat{\phi}(t,\bm{x})+v^{(2)}(\bm{x})\hat{\Pi}(t,\bm{x})\right),
\end{align}
where $v^{(1)}(\bm{x})$ and $v^{(2)}(\bm{x})$ are called the smearing functions, which characterize the spatial extent of the detector.

We further assume that the switching function is given by a delta function: $\chi(t)=\delta(t-t_0)$, which enables a non-perturbative analysis \cite{PhysRevD.96.065008}. In the interaction picture of time evolution, the encoding process is now expressed by the unitary operator
\begin{align}
 \hat{U}=e^{-i\lambda\hat{\mu}(t_0)\otimes \hat{O}(t_0)}.\label{eq_enc_uni}
\end{align}
Since the encoding process is expressed by a single Hermitian operator $\hat{O}(t_0)$, we can uniquely identify the carrier of information by using the QIC formula \cite{yamaguchi2019quantum}. Hereafter, for notational simplicity, $ \hat{O}(t_0)$ is denoted by $\hat{O}$. In addition, we assume that the initial state of the field is in a Gaussian state $\ket{\Psi}$ with vanishing first moments: $\braket{\Psi|\hat{\phi}(t,\bm{x})|\Psi}=\braket{\Psi|\hat{\Pi}(t,\bm{x})|\Psi}=0$. 

Now, let us introduce a linear map $f_\Psi$, mapping local field operators to local field operators, defined by
\begin{align}
&f_{\Psi}\left(\hat{O}\right) \nonumber\\
&\equiv 2 \int \text{d}^d\bm{x}\left(-\mathrm{Re}\left(\Braket{\Psi|\hat{O}\hat{\Pi}(t_0,\bm{x})|\Psi}\right)\hat{\phi}(t_0,\bm{x})\right.\nonumber\\
&\quad\left.+\mathrm{Re}\left(\Braket{\Psi|\hat{O}\hat{\phi}(t_0,\bm{x})|\Psi}\right)\hat{\Pi}(t_0,\bm{x})\right).\label{eq_lin_map}
\end{align}
It can be shown \cite{yamaguchi2019quantum} that
\begin{align}
\left[\hat{O},f_{\Psi}\left(\hat{O}\right)\right]&= 2\ii\Braket{\Psi|\hat{O}^2|\Psi}\label{eq_ccr},\\
\Braket{\Psi|\left(f_{\Psi}\left(\hat{O}\right)\right)^2|\Psi}&=\Braket{\Psi|\hat{O}^2|\Psi},\label{eq_sqic1}\\
\mathrm{Re}\left(\Braket{\Psi|\hat{O}f_{\Psi}\left(\hat{O}\right)|\Psi}\right)&=0,\label{eq_sqic2}
\end{align}
hold for pure Gaussian states $\ket{\Psi}$. Equation \eqref{eq_ccr} implies that the set of field operators
\begin{equation}
    \Bigg\{\hat{O},\frac{1}{2\Braket{\Psi|\hat{O}^2|\Psi}}f_\Psi\left(\hat{O}\right)\Bigg\},
\end{equation}
satisfies the canonical commutation relationship, meaning that it characterizes a mode as a subsystem of the scalar field. Since the operators are given by linear combinations of canonical variables, the mode is also in a Gaussian state. Equations \eqref{eq_sqic1} and \eqref{eq_sqic2} show the determinant of covariance matrix for this mode is $\frac{1}{4}$. This condition holds if and only if the mode is in a pure state (see, e.g., Ref. \cite{serafini2017quantum}). 
Since the encoding unitary operation in Eq. \eqref{eq_enc_uni} is a unitary operation on this mode, the composite system of qubit and the mode remains in a pure state after the encoding process. Therefore, no information is leaked outside the mode, which is called a quantum information capsule (QIC). 
The QIC mode is uniquely determined under the assumption that the operators characterizing the mode are given by linear combinations of canonical variables \cite{yamaguchi2019quantum}. 

For future convenience, we adopt another convention for operators characterizing the QIC mode. Introducing a normalization factor
\begin{align}
 \alpha&\equiv \sqrt{2\Braket{\Psi|\hat{O}^2|\Psi}},
\end{align}
we define
\begin{align}
 \hat{Q}&\equiv \frac{1}{\alpha}\hat{O},\quad \hat{P} \equiv \frac{1}{\alpha}f_\Psi\left(\hat{O}\right).
\end{align}
The QIC mode is characterized by $(\hat{Q},\hat{P})$ satisfying $\left[\hat{Q},\hat{P}\right]=\ii$. In this convention, the mode is initially in a pure Gaussian state in the standard form, i.e., 
\begin{align}
\begin{split}
     &\Braket{\Psi|\hat{Q}^2|\Psi}=\Braket{\Psi|\hat{P}^2|\Psi}=\frac{1}{2},\\
& \mathrm{Re}\left(\Braket{\Psi|\hat{Q}\hat{P}|\Psi}\right)=0.
\end{split}
\label{eq_std_form}
\end{align}
This implies that the initial Gaussian state is decomposed into the following form:
\begin{align}
 \ket{\Psi}=\ket{0}\otimes \ket{\Psi'},
\end{align}
where $\ket{0}$ is the ``vacuum'' state annihilated by $\frac{1}{\sqrt{2}}(\hat{Q}+\ii\hat{P})$ and $\ket{\Psi'}$ is a Gaussian state for modes orthogonal to the QIC modes. The encoding unitary operator \eqref{eq_enc_uni} is now regarded as an interaction between a qubit and a harmonic oscillator characterized by $(\hat{Q},\hat{P})$ which is non-locally embedded in the scalar field.

It should be noted that
\begin{align}
 f_{\Psi}\left(f_{\Psi}\left(\hat{O}\right)\right)=-\hat{O} \label{eq_f^2}
\end{align}
holds for any operator $\hat{O}$ given by linear combination of canonical variables. For a simple proof, let us consider an operator $f_\Psi(\hat{P})$. From the uniqueness of QIC operators and the normalization condition Eq.\eqref{eq_std_form}, we get
\begin{align}
 f_\Psi(\hat{P})=-\hat{Q},\label{eq_fqmp}
\end{align}
where the minus sign appears from the fact that $[\hat{P},-\hat{Q}]=\ii$ holds. Since the map $f_\Psi$ is linear, Eq.\eqref{eq_f^2} is proven. Equation \eqref{eq_fqmp} will be used to extend the QIC formula for multiple modes in Sec. \ref{sec_mqic}.

The propagation of information can be visualized by investigating the time evolution of the QIC mode.
The functions $v^{(1)}(t,\bm{x}),v^{(2)}(t,\bm{x}),u^{(1)}(t,\bm{x}),u^{(2)}(t,\bm{x})$ satisfying
\begin{align}
 \hat{O}&=\int \text{d}^d\bm{x}\left(v^{(1)}(t,\bm{x})\hat{\phi}(t,\bm{x})+v^{(2)}(t,\bm{x})\hat{\Pi}(t,\bm{x})\right),\\
 f_\Psi\left(\hat{O}\right)&=\int \text{d}^d\bm{x}\left(u^{(1)}(t,\bm{x})\hat{\phi}(t,\bm{x})+u^{(2)}(t,\bm{x})\hat{\Pi}(t,\bm{x})\right)
\end{align}
can be calculated by
\begin{align}
\begin{split}
  v^{(1)}(t,\bm{x})&\equiv\frac{1}{\ii}\Braket{\Psi|\left[\hat{O},\hat{\Pi}(t,\bm{x})\right]|\Psi}\\
  &=-\partial_t v^{(2)}(t,\bm{x}),\\
 v^{(2)}(t,\bm{x})&\equiv -\frac{1}{\ii}\Braket{\Psi|\left[\hat{O},\hat{\phi}(t,\bm{x})\right]|\Psi}\\
&=-2 \mathrm{Im}\left(\int \text{d}^d\bm{y}\left(v^{(1)}(\bm{y})W(t_0,\bm{y},t,\bm{x})\right.\right.\\
&\quad\quad\quad\left.\left.+v^{(2)}(\bm{y})\partial_{t_0}W(t_0,\bm{y},t,\bm{x})\right)\right),\\
u^{(1)}(t,\bm{x})&\equiv \frac{1}{\ii}\Braket{\Psi|\left[f_\Psi\left(\hat{O}\right),\hat{\Pi}(t,\bm{x})\right]|\Psi}\\
&=-\partial_t u^{(2)}(t,\bm{x}),\\
 u^{(2)}(t,\bm{x})&\equiv-\frac{1}{\ii}\Braket{\Psi|\left[f_\Psi\left(\hat{O}\right),\hat{\phi}(t,\bm{x})\right]|\Psi}\\
&=-2 \mathrm{Im}\left(\int\text{d}^d\bm{y}\left(u^{(1)}(\bm{y})W(t_0,\bm{y},t,\bm{x})\right.\right.\\
&\quad\quad\quad\left.\left.+u^{(2)}(\bm{y})\partial_{t_0}W(t_0,\bm{y},t,\bm{x})\right)\right),
\end{split}
\label{eq_te_qic}
\end{align}
where $W(t,\bm{x},t',\bm{x})\equiv \Braket{\Psi|\hat{\phi}(t,\bm{x})\hat{\phi}(t',\bm{x}')|\Psi}$ is the Wightman function and
\begin{align}
\begin{split}
     u^{(1)}(\bm{x})&\equiv -2\mathrm{Re}\left(\Braket{\Psi|\hat{O}\hat{\Pi}(t_0,\bm{x})|\Psi)}\right) ,\\
     u^{(2)}(\bm{x})&\equiv 2\mathrm{Re}\left(\Braket{\Psi|\hat{O}\hat{\phi}(t_0,\bm{x})|\Psi)}\right) .
\end{split}
\end{align}
The mode carrying information at $t>t_0$ is visualized by four functions $F^{(1)},F^{(2)},G^{(1)},G^{(2)}$
\begin{align}
 \hat{Q}&=\int \text{d}^d\bm{x}\left(F^{(1)}(t,\bm{x})\hat{\phi}(t,\bm{x})+F^{(2)}(t,\bm{x})\hat{\Pi}(t,\bm{x})\right),\\
\hat{P} &=\int \text{d}^d\bm{x}\left(G^{(1)}(t,\bm{x})\hat{\phi}(t,\bm{x})+G^{(2)}(t,\bm{x})\hat{\Pi}(t,\bm{x})\right) ,
\end{align}
where
\begin{align}
 F^{(l)}(t,\bm{x})=\frac{1}{\alpha} v^{(l)}(t,\bm{x}),\quad G^{(l)}(t,\bm{x})=\frac{1}{\alpha}u^{(l)}(t,\bm{x}) 
\end{align}
for $l=1,2$.
We call these four functions weighting functions of the mode. 
It should be noted that the mass dimensions of $(F^{(1)},G^{(1)})$ and $(F^{(2)},G^{(2)})$ defined here are given by $\frac{d+1}{2}$ and $\frac{d-1}{2}$ respectively, since $\hat{Q}$ and $\hat{P}$ are dimensionless.

A common and important example is the cases where the field starts with its vacuum state $\ket{0}$. The Wightman function for $\ket{0}$ is given by
\begin{align}
 W(t,\bm{x},t',\bm{x}')=\int \frac{\text{d}^d\bm{k}}{(2\pi)^d2|\bm{k}|} e^{-\ii\left(|\bm{k}|(t-t') -\bm{k}\cdot(\bm{x}-\bm{x}')\right)}.
\end{align}
Let us further assume that the detector only couples to the field $\hat\phi$ (and not the conjugate momentum field $\hat\Pi$), i.e. we set $v_2(\bm{x})=0$. In this case, the operator $f_\Psi(\hat{O})$ is simplified and characterized by
\begin{align}
 u^{(1)}(\bm{x})&=-2\int \text{d}^d\bm{x} v^{(1)}(\bm{y})\mathrm{Re}\left(\Braket{0|\hat{\phi}(t_0,\bm{y})\hat{\Pi}(t_0,\bm{x})|0}\right)\nonumber\\
 &=0 \label{eq_offdiagonal},\\
 u^{(2)}(\bm{x})&= 2\mathrm{Re}\left(\int \text{d}^d\bm{y}v^{(1)}(\bm{y}) \int\frac{ \text{d}^d\bm{k}}{(2\pi)^d|\bm{k}|} e^{\ii \bm{k}\cdot (\bm{y}-\bm{x})}\right)\nonumber\\
 &=2\mathrm{Re}\left(\int \frac{\text{d}^d\bm{k}}{(2\pi)^d 2|\bm{k}|}e^{-\ii\bm{k}\cdot \bm{x}}\tilde v^{(1)}(\bm{k})\right),
\end{align}
where we have defined the Fourier transformation $\tilde{f}$ of a function $f$ by
\begin{align}
 \tilde{f}(\bm{k})\equiv \int \text{d}^d\bm{x}f(\bm{x})e^{\ii\bm{k}\cdot \bm{x}}.
\end{align}
From Eq.\eqref{eq_te_qic}, the QIC mode at $t>t_0$ is characterized by the functions
\begin{align}
\begin{split}
  v^{(2)}(t,\bm{x})
 &=-2\mathrm{Im}\left(\int \frac{\text{d}^d\bm{k}}{(2\pi)^d 2|\bm{k}|}e^{-\ii|\bm{k}|(t_0-t)}e^{-\ii\bm{k}\cdot \bm{x}}\tilde{v}^{(1)}(\bm{k})\right)\\
 u^{(2)}(t,\bm{x})&=2\mathrm{Re}\left(\int \frac{\text{d}^d\bm{k}}{(2\pi)^d 2|\bm{k}|}e^{-\ii|\bm{k}|(t_0-t)}e^{-\ii\bm{k}\cdot\bm{x}}\tilde{v}^{(1)}(\bm{k})\right)
\end{split}\label{eq_int}
\end{align}
and their derivatives with respect to $t$. 
On the other hand, the normalization factor is calculated from
\begin{align}
 \Braket{0|\hat{O}^2|0}=\int \frac{\text{d}^d\bm{k}}{(2\pi)^d 2|\bm{k}|}\left|\tilde{v}^{(1)}(\bm{k})\right|^2.
\end{align}

\subsection{Propagation of information in $(3+1)$-dimensional Minkowski spacetime}
Let us investigate the propagation of information in $(3+1)$-dimensional Minkowski spacetime. We adopt a Gaussian smearing
\begin{align}
 v^{(1)}(\bm{x})=e^{-\frac{|\bm{x}-\bm{x}_0|^2}{2\sigma^2}},
\end{align}
and $v^{(2)}(\bm{x})=0$ for the UDW detector which encodes the information in the field (i.e. the UDW detector of the sender). Its Fourier transformation is given by
\begin{align}
 \tilde{v}^{(1)}(\bm{k})=\sqrt{(2\pi\sigma^2)^3}e^{-\frac{\sigma^2}{2}|\bm{k}|^2}e^{\ii\bm{k}\cdot \bm{x}_0}.
\end{align}

The integral in Eq. \eqref{eq_int} is calculated as
\begin{align}
& \int \frac{d^3\bm{k}}{(2\pi)^3 2|\bm{k}|}e^{-\ii|\bm{k}|(t_0-t)}e^{-\ii\bm{k}\cdot\bm{x}}\tilde{v}^{(1)}(\bm{k})\nonumber\\
 &= \frac{2\pi\sqrt{(2\pi\sigma^2)^3}}{(2\pi)^3 2\ii|\bm{x_0-\bm{x}|}}\nonumber\\
 &\times\int_0^\infty dke^{-\ii k(t_0-t)} e^{-\frac{\sigma^2}{2}k^2}\left(e^{\ii k|\bm{x}_0-\bm{x}|}-e^{-\ii k|\bm{x}_0-\bm{x}|}\right)\nonumber\\
 &= \frac{\sigma^2}{4\ii|\bm{x}_0-\bm{x}|}\nonumber\\
& \times\left(e^{-\frac{((t_0-t)-|\bm{x}_0-\bm{x}|)^2}{2\sigma^2}}\left(1-\mathrm{Erf}\left(\ii\frac{((t_0-t)-|\bm{x}_0-\bm{x}|)}{\sqrt{2\sigma^2}}\right)\right)\right.\nonumber\\
&\left. \quad -e^{-\frac{((t_0-t)+|\bm{x}_0-\bm{x}|)^2}{2\sigma^2}}\left(1-\mathrm{Erf}\left(\ii\frac{((t_0-t)+|\bm{x}_0-\bm{x}|)}{\sqrt{2\sigma^2}}\right)\right)\right)
\end{align}
where we have used
\begin{align}
 \int_0^\infty dk e^{-ak^2}e^{\ii b k}=\frac{\sqrt{\pi}}{2\sqrt{a}}e^{-\frac{b^2}{4a}}\left(1+\mathrm{Erf}\left(\ii\frac{b}{2\sqrt{a}}\right)\right)
\end{align}
and the error function defined by
\begin{align}
 \mathrm{Erf}(\xi)\equiv \frac{2}{\sqrt{\pi}}\int_0^\xi dt e^{-t^2}.
\end{align}

On the other hand, the expectation value of the generator is evaluated as
\begin{align}
& \Braket{0|\hat{O}^2|0}
 =\pi \sigma^4 .
\end{align}
Therefore, the normalization factor is determined as
\begin{align}
 \alpha=\sqrt{2\pi}\sigma^2.
\end{align}

\begin{figure}[H]
    \centering
     \includegraphics[keepaspectratio, scale=0.55]{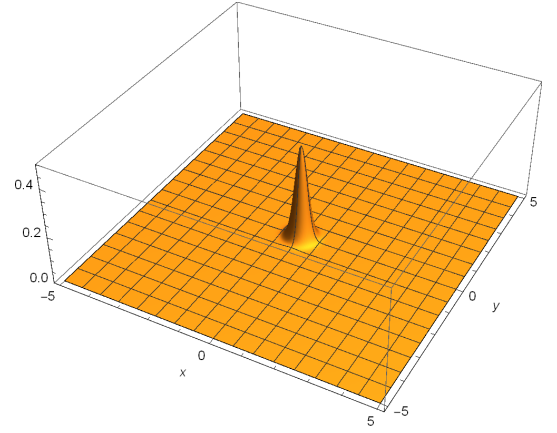}
        \caption{$\sigma^2 F^{(1)}(t,x,y,0)$ at $t=0$.}
        \label{fig_t0f11_3d}
\end{figure}

\begin{figure}[H]
    \centering
     \includegraphics[keepaspectratio, scale=0.60]{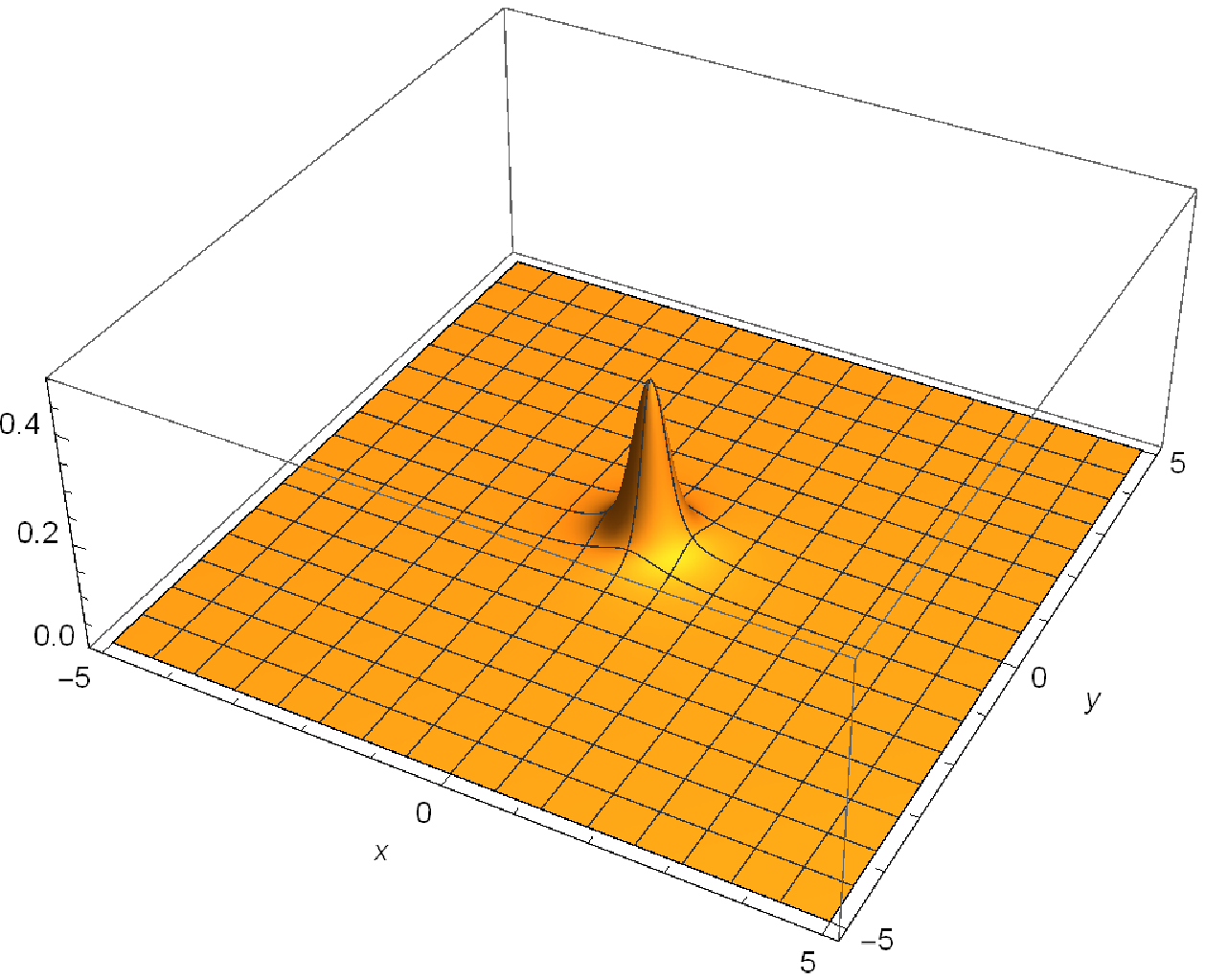}
    \caption{$\sigma G^{(2)}(t,x,y,0)$ at $t=0$.}
    \label{fig_t0g12_3d}
\end{figure}

\begin{figure}[H]
    \centering
     \includegraphics[keepaspectratio, scale=0.60]{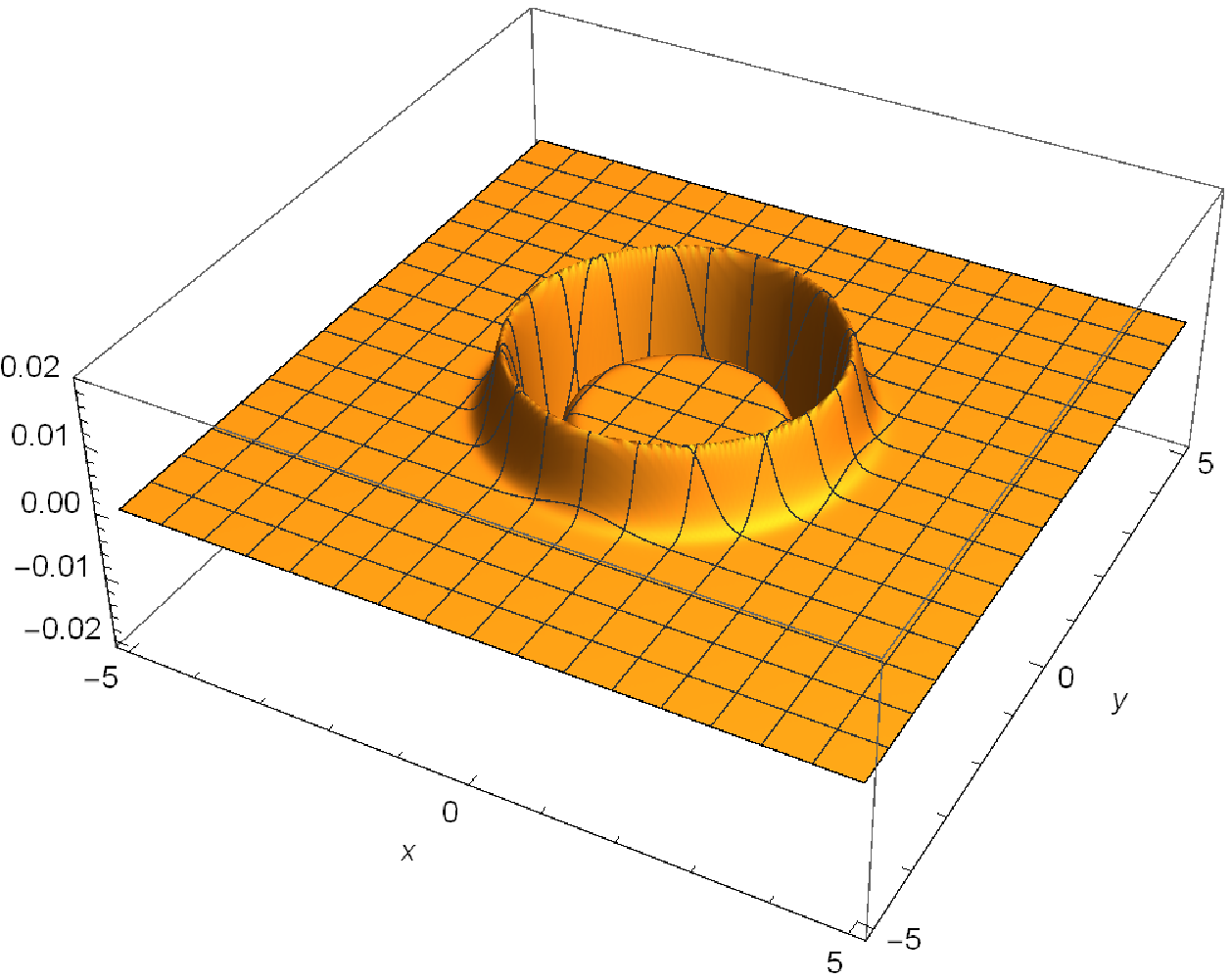}
        \caption{$\sigma^2 F^{(1)}(t,x,y,0)$ at $t=2$.}
        \label{fig_t2f11_3d}
\end{figure}

\begin{figure}[H]
    \centering
    \includegraphics[keepaspectratio, scale=0.60]{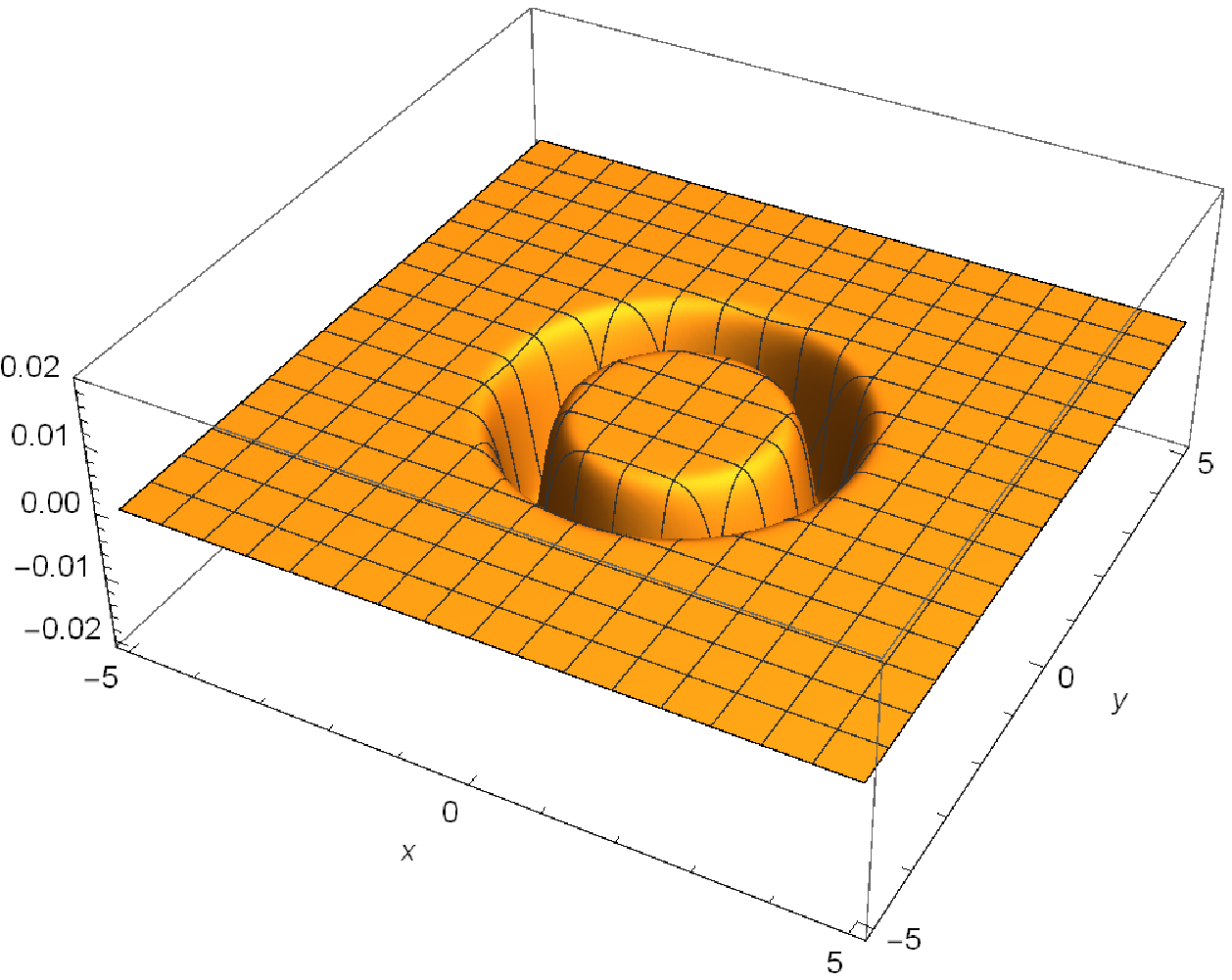}
      \caption{$\sigma F^{(2)}(t,x,y,0)$ at $t=2$.}
       \label{fig_t2f12_3d}
\end{figure}

\begin{figure}[H]
    \centering
    \includegraphics[keepaspectratio, scale=0.60]{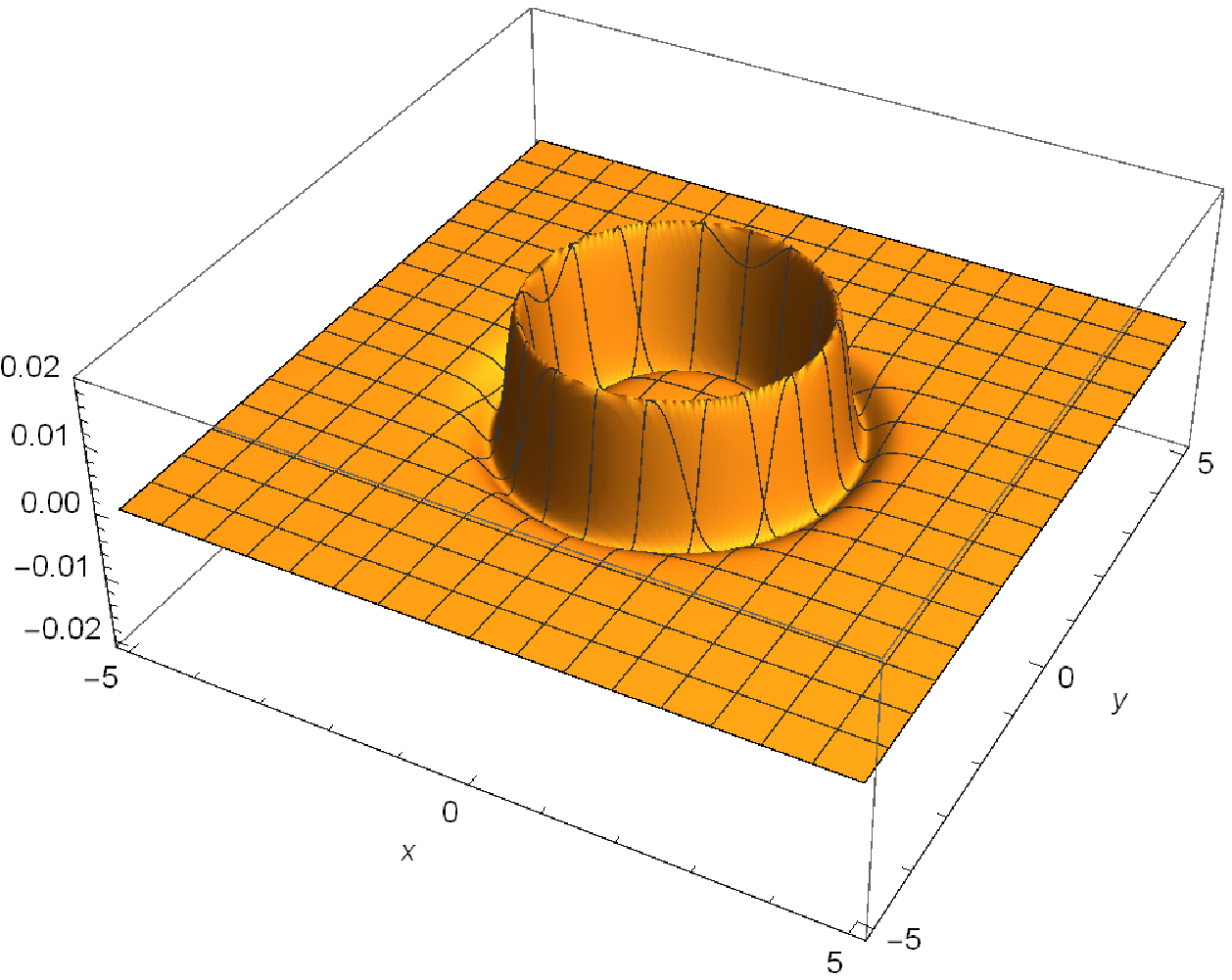}
    \caption{$\sigma^2 G^{(1)}(t,x,y,0)$ at $t=2$.}
    \label{fig_t2g11_3d}
\end{figure}

\begin{figure}[H]
    \centering
     \includegraphics[keepaspectratio, scale=0.60]{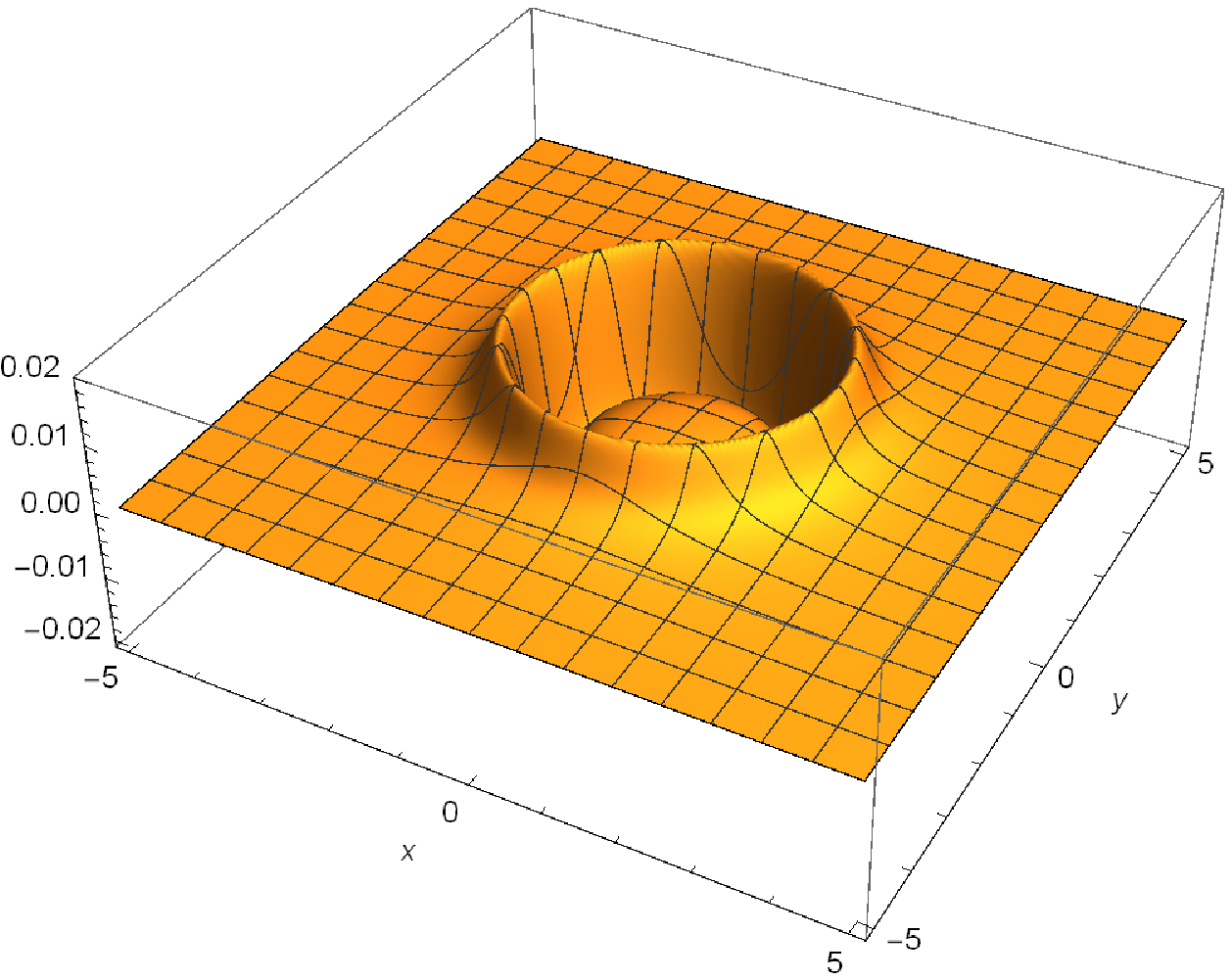}
    \caption{$\sigma G^{(2)}(t,x,y,0)$ at $t=2$.}
    \label{fig_t2g12_3d}
\end{figure}

\begin{figure}[H]
    \centering
     \includegraphics[keepaspectratio, scale=0.60]{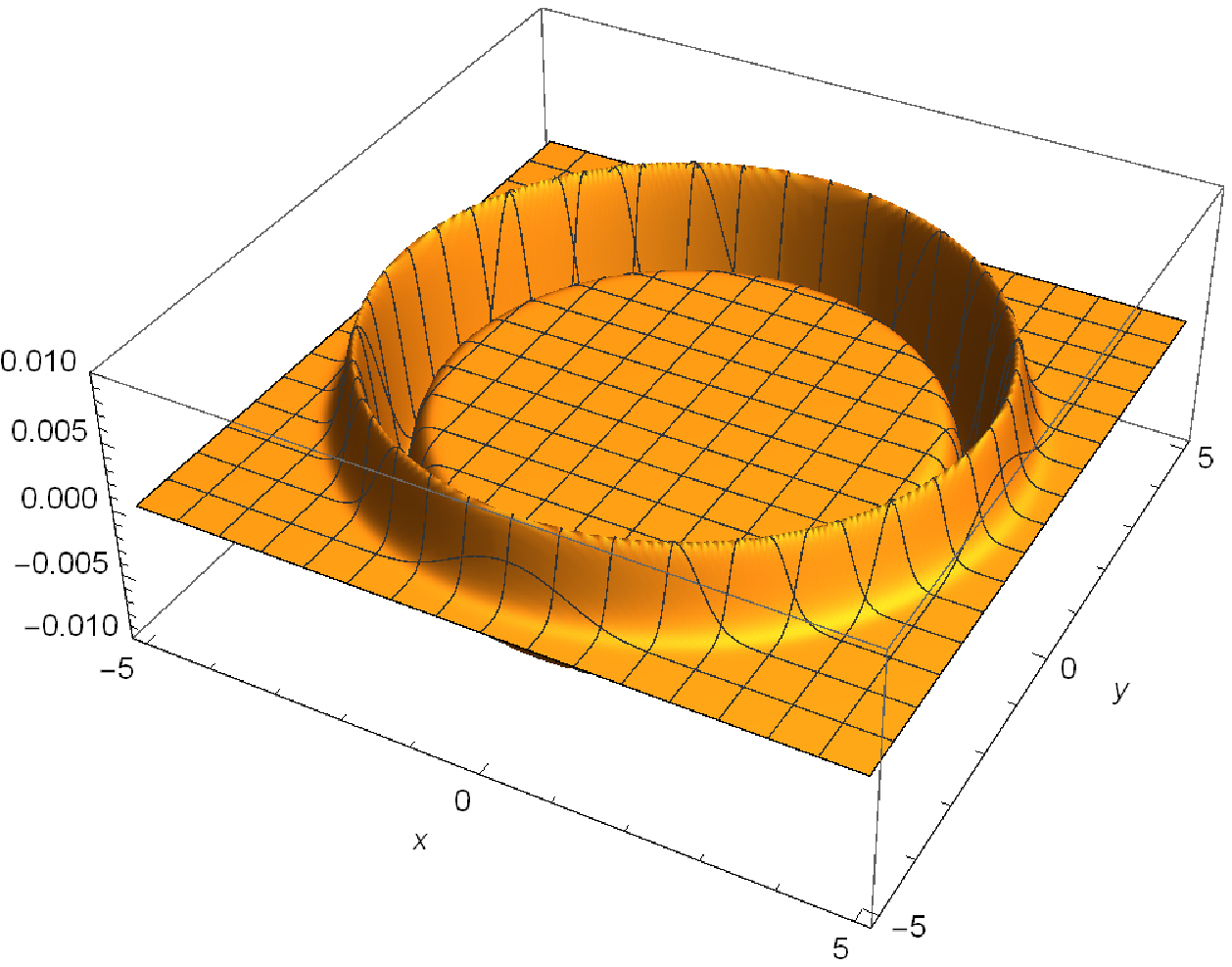}
        \caption{$\sigma^2 F^{(1)}(t,x,y,0)$ at $t=4$.}
        \label{fig_t4f11_3d}
\end{figure}

\begin{figure}[H]
    \centering
    \includegraphics[keepaspectratio, scale=0.60]{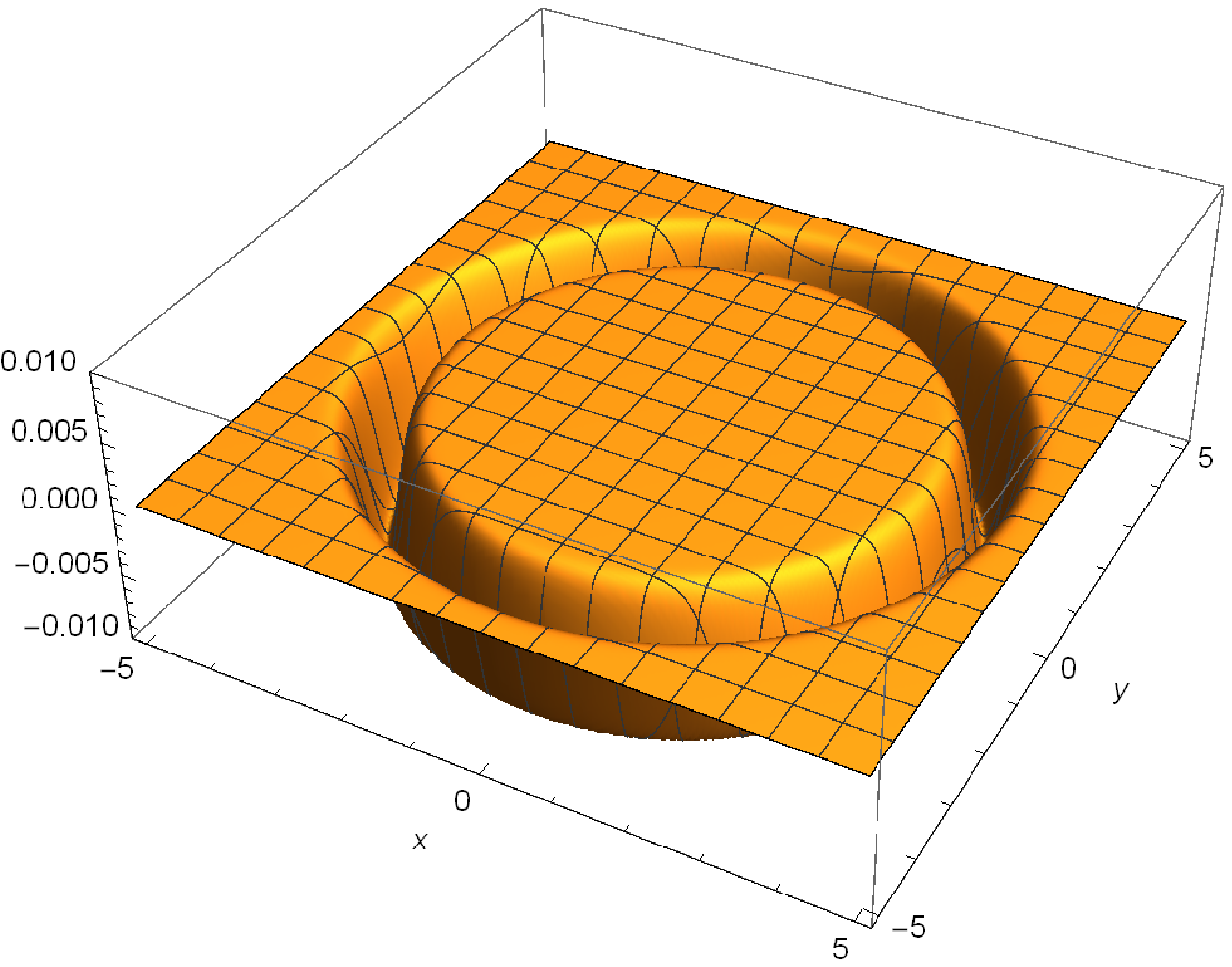}
      \caption{$\sigma F^{(2)}(t,x,y,0)$ at $t=4$.}
       \label{fig_t4f12_3d}
\end{figure}

\begin{figure}[H]
    \centering
\includegraphics[keepaspectratio, scale=0.60]{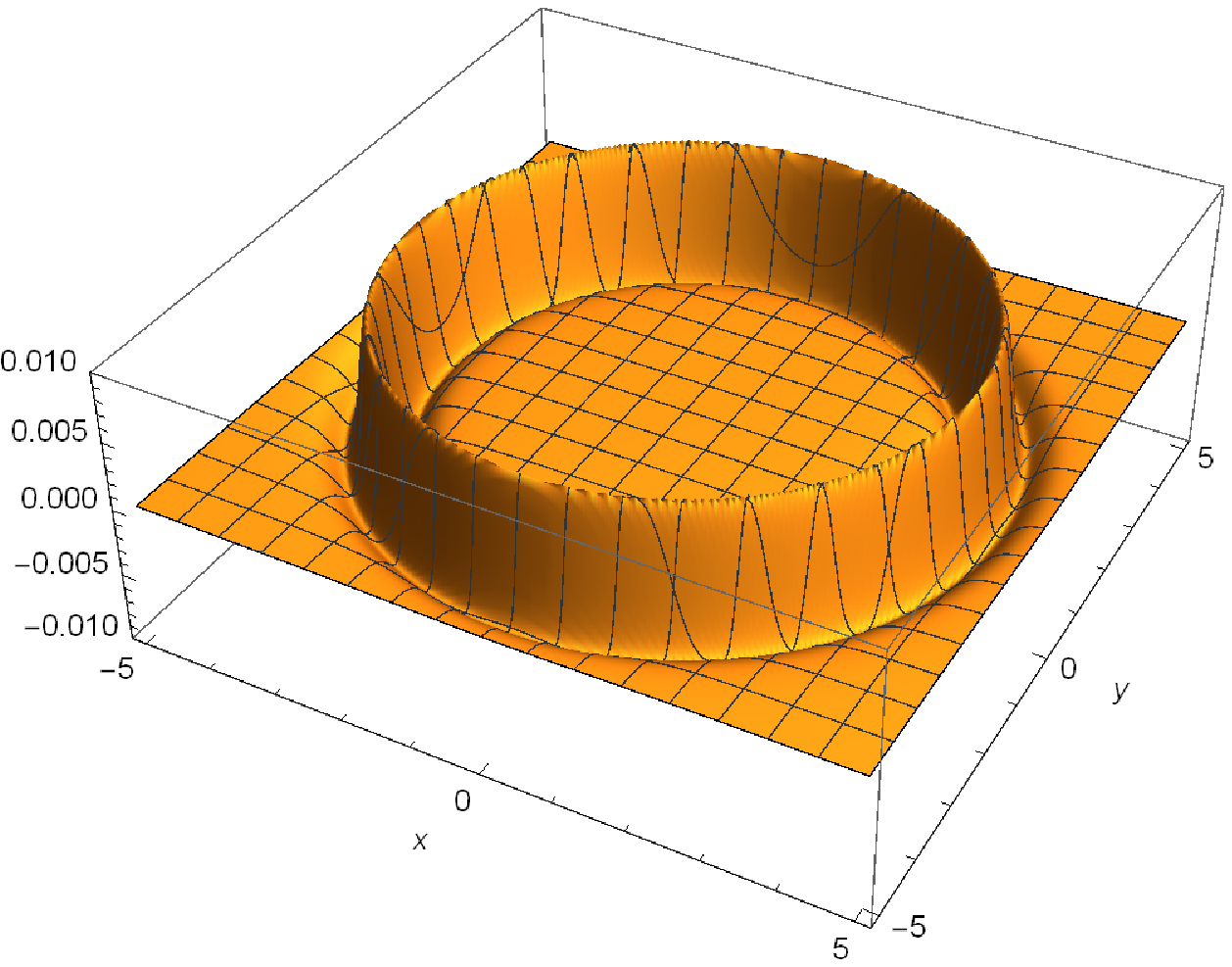}
    \caption{$\sigma^2 G^{(1)}(t,x,y,0)$ at $t=4$.}
    \label{fig_t4g11_3d}
\end{figure}

\begin{figure}[H]
    \centering
     \includegraphics[keepaspectratio, scale=0.60]{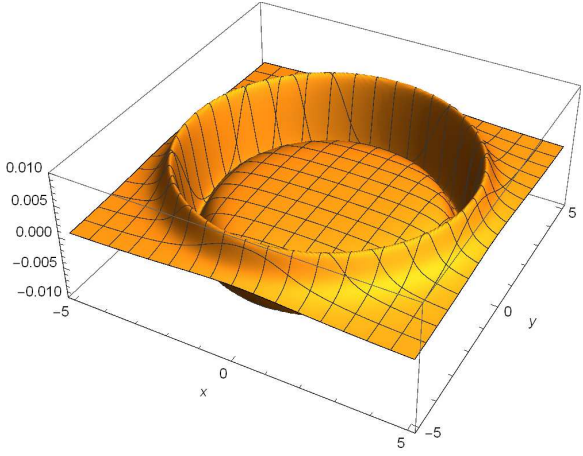}
    \caption{$\sigma G^{(2)}(t,x,y,0)$ at $t=4$.}
    \label{fig_t4g12_3d}
\end{figure}

Figures \ref{fig_t0f11_3d}-\ref{fig_t4g12_3d} show the time evolution of QIC mode. In these figures, the weighting functions are made to be dimensionless by using $\sigma$ and plotted at $z=0$. The parameters characterizing the detector are fixed as $\sigma=0.2$ and $(t_0,\bm{x}_0)=0$. At $t=0$, $F^{(2)}(0,\bm{x})=G^{(1)}(0,\bm{x})=0$ as is seen from Eq.\eqref{eq_offdiagonal}. The tail of $G^{(2)}(0,\bm{x})$ is broader than that of $F^{(1)}(0,\bm{x})$, which shows that an encoding operation by $\hat{O}$ affects non-local correlations. At $t=2$, four weighting functions are non-vanishing and localized around the circle with radius $2$, reflecting the fact that the massless scalar field propagates at the speed of light $c=1$. At $t=4$, four weighting functions are localized around the circle with radius $4$. 
\vspace{-0.3cm}
\vspace{-0.3cm}

\subsection{Propagation of information in $(2+1)$-dimensional Minkowski spacetime}
We now investigate the time evolution of the QIC in $(2+1)$-dimensional Minkowski spacetime, where the strong Huygens principle is violated.
We again adopt a Gaussian smearing:
\begin{align}
 v^{(1)}(\bm{x})=e^{-\frac{|\bm{x}-\bm{x}_0|^2}{2\sigma^2}}
\end{align}
and $v^{(2)}(\bm{x})=0$.
Its Fourier transformation is given by
\begin{align}
 \tilde{v}^{(1)}(\bm{k})=(2\pi\sigma^2)e^{-\frac{\sigma^2}{2}|\bm{k}|^2}e^{\ii\bm{k}\cdot \bm{x}_0}.
\end{align}

The integral in Eq. \eqref{eq_int} is calculated as
\begin{align}
& \int \frac{d^2\bm{k}}{(2\pi)^2 2|\bm{k}|}e^{-\ii|\bm{k}|(t_0-t)}e^{-\ii\bm{k}\cdot\bm{x}}\tilde{v}^{(1)}(\bm{k})\nonumber\\
 &=\frac{ 2\pi\sigma^2}{(2\pi)^22}\int \frac{d^2\bm{k}}{|\bm{k}|} e^{-\frac{\sigma^2}{2}|\bm{k}|^2}e^{-\ii|\bm{k}|(t_0-t)}e^{\ii\bm{k}\cdot(\bm{x}_0-\bm{x})}\nonumber\\
 &=  \frac{ 2\pi\sigma^2}{(2\pi)^22}\int_0^\infty dk e^{-\frac{\sigma^2}{2}k^2}e^{-\ii k(t_0-t)}\int_0^{2\pi} d\theta e^{\ii k|\bm{x}_0-\bm{x}|\cos{\theta}}\nonumber\\
 &=\frac{\sigma^2}{2}\int_0^\infty dk e^{-\frac{\sigma^2}{2}k^2} e^{-\ii k(t_0-t)}J_0(k|\bm{x}_0-\bm{x}|),\label{eq_2d_conjfunc}
\end{align}
where we have used the integral representation of Bessel function:
\begin{align}
 J_0(\xi)=\frac{1}{2\pi}\int_0^{2\pi}d\theta e^{\ii\xi \cos{\theta}}.
\end{align}
Equation\eqref{eq_2d_conjfunc} can be numerically evaluated. 

On the other hand, the normalization constant $\alpha$ is given by $\alpha=\pi^{\frac{3}{4}}\sigma^{\frac{3}{2}}$ since
\begin{align}
\Braket{0|\hat{O}^2|0}&= \int \frac{d^2\bm{k}}{(2\pi)^2 2|\bm{k}|}|\tilde{v}^{(1)}(\bm{k})|^2
= \frac{\pi^{\frac{3}{2}}\sigma^3}{2}.
\end{align}
 
\begin{figure}[H]
    \centering
    \includegraphics[keepaspectratio, scale=0.60]{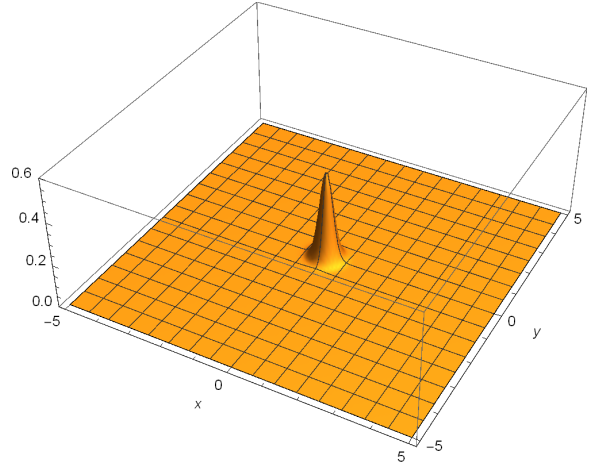}
    \caption{$\sigma^{3/2} F^{(1)}(t,x,y)$ at $t=0$.}
    \label{fig_t0f11_2d}
\end{figure}

\begin{figure}[H]
    \centering
\includegraphics[keepaspectratio, scale=0.60]{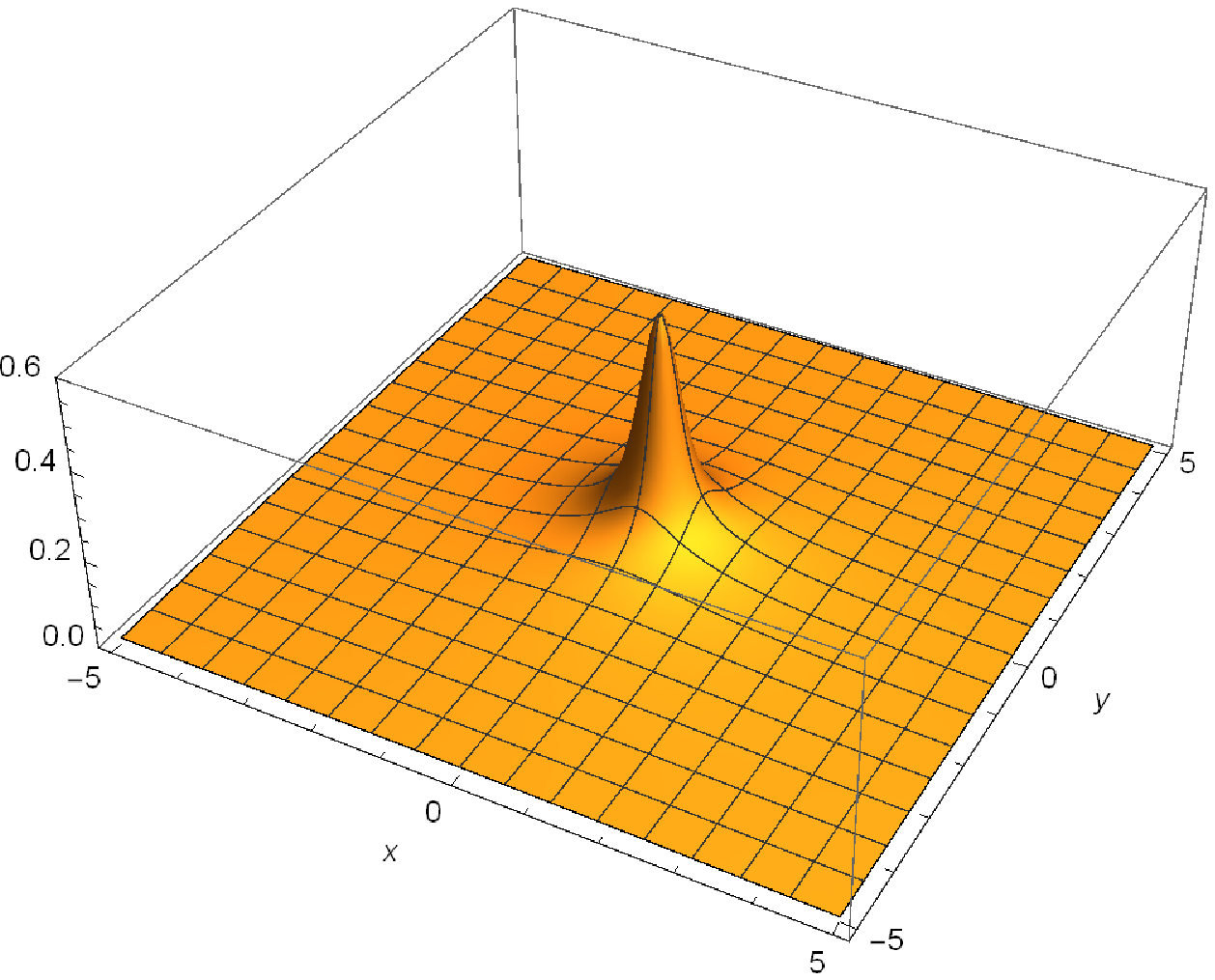}
    \caption{$\sigma^{1/2} G^{(2)}(t,x,y)$ at $t=0$.}
    \label{fig_t0g12_2d}
\end{figure}

\begin{figure}[H]
    \centering
    \includegraphics[keepaspectratio, scale=0.60]{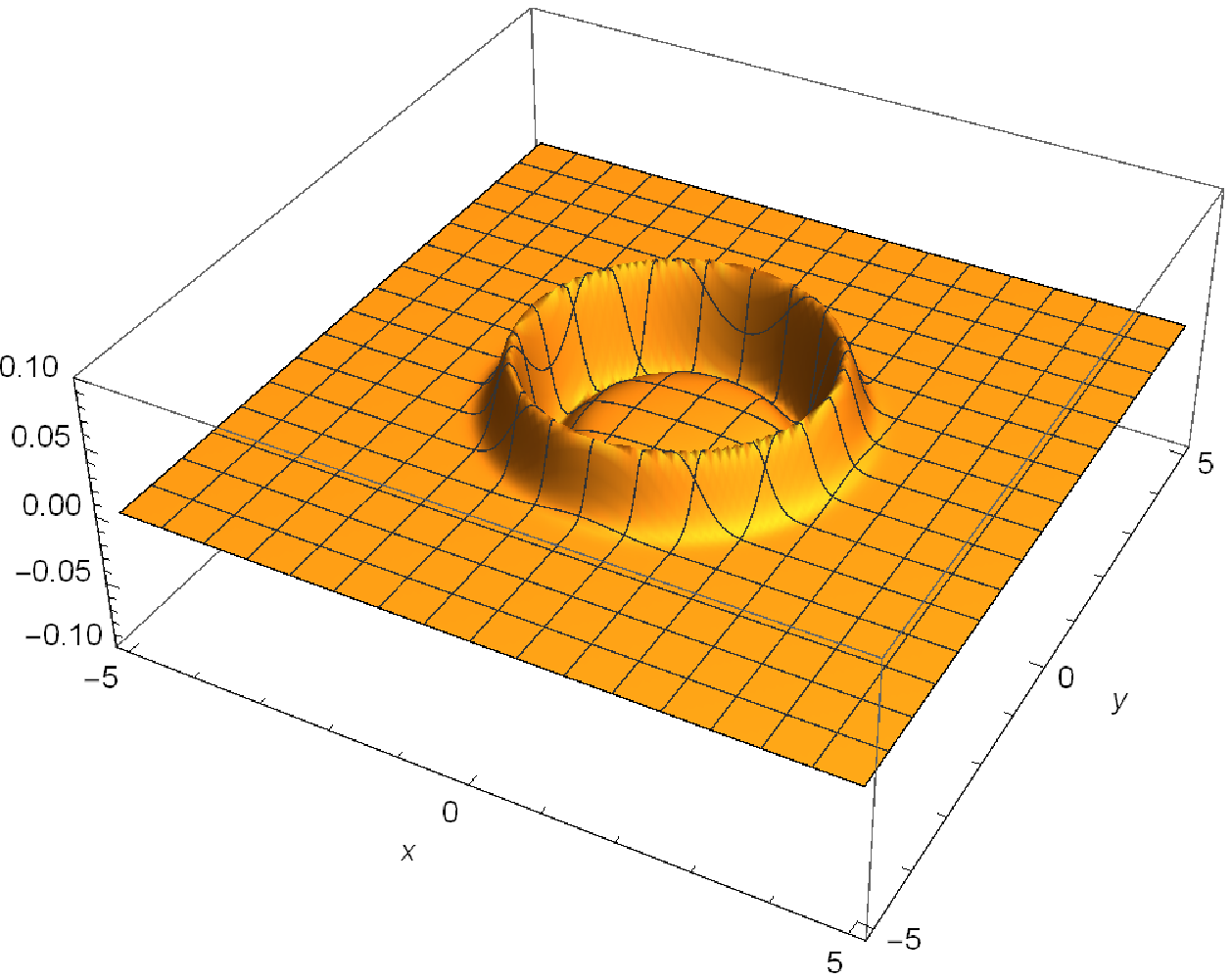}
    \caption{$\sigma^{3/2} F^{(1)}(t,x,y)$ at $t=2$.}
    \label{fig_t2f11_2d}
\end{figure}

\begin{figure}[H]
    \centering
\includegraphics[keepaspectratio, scale=0.60]{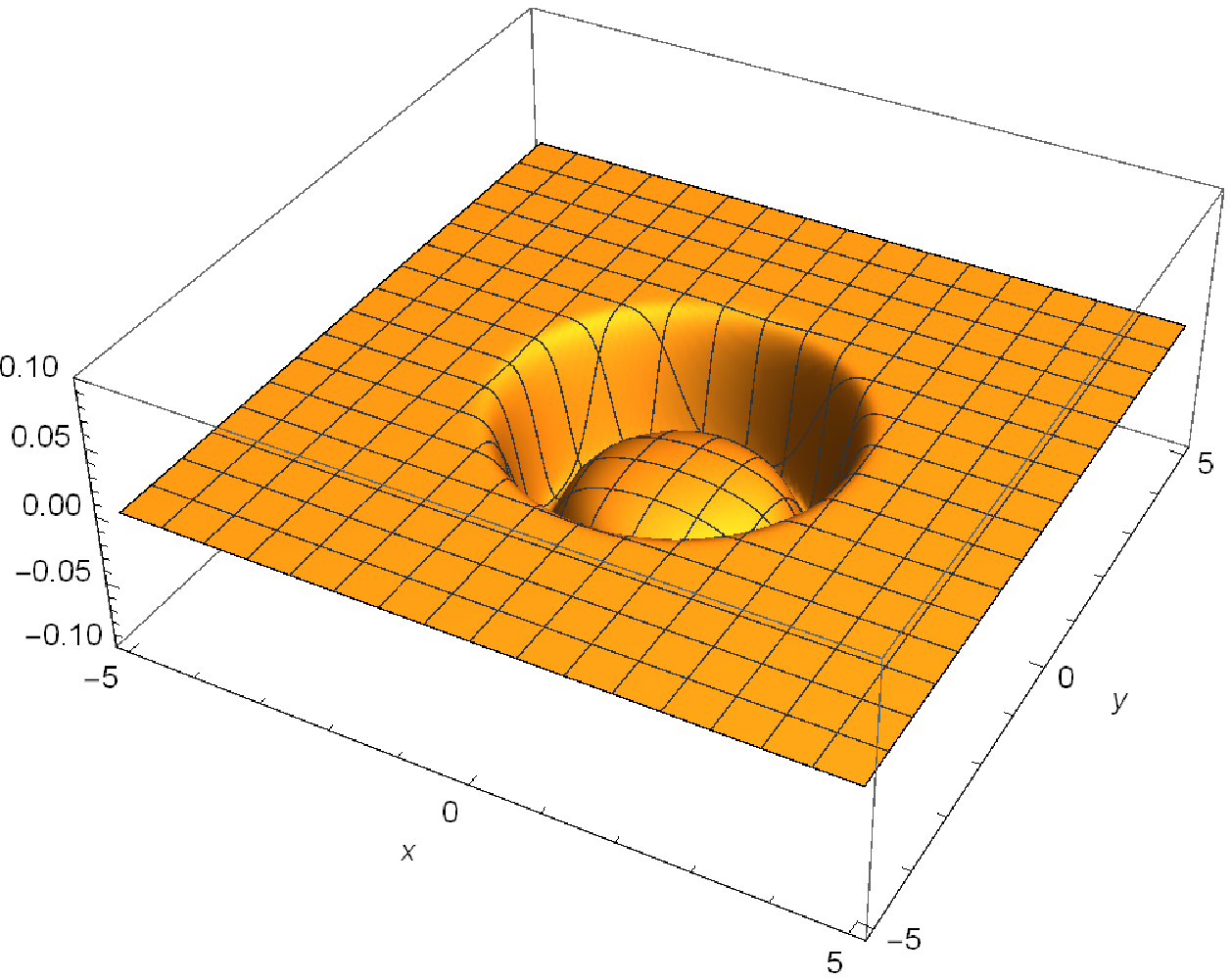}
    \caption{$\sigma^{1/2} F^{(2)}(t,x,y)$ at $t=2$.}
    \label{fig_t2f12_2d}
\end{figure}

\begin{figure}[H]
    \centering
\includegraphics[keepaspectratio, scale=0.60]{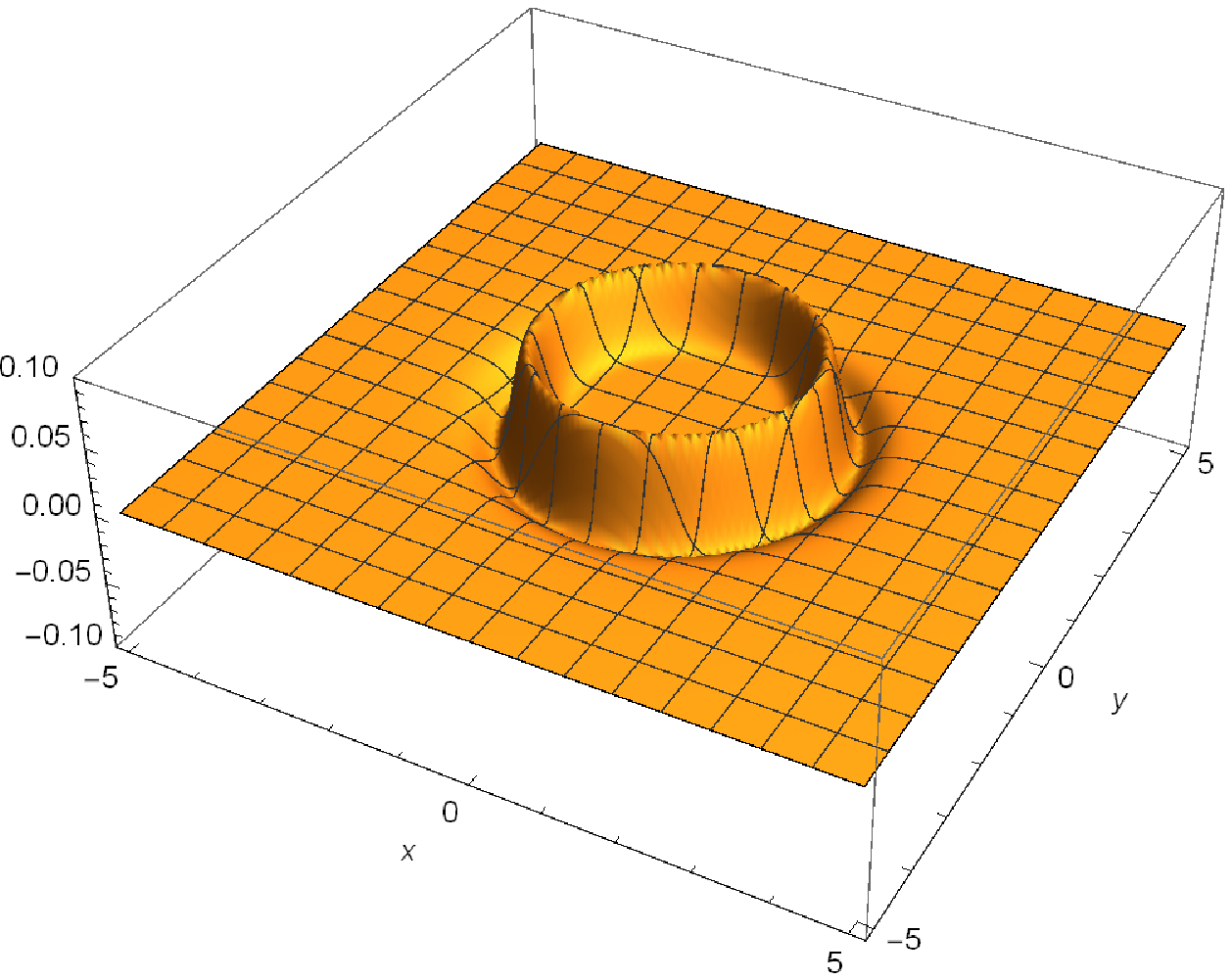}
    \caption{$\sigma^{3/2} G^{(1)}(t,x,y)$ at $t=2$.}
    \label{fig_t2g11_2d}      
\end{figure}

\begin{figure}[H]
    \centering
\includegraphics[keepaspectratio, scale=0.60]{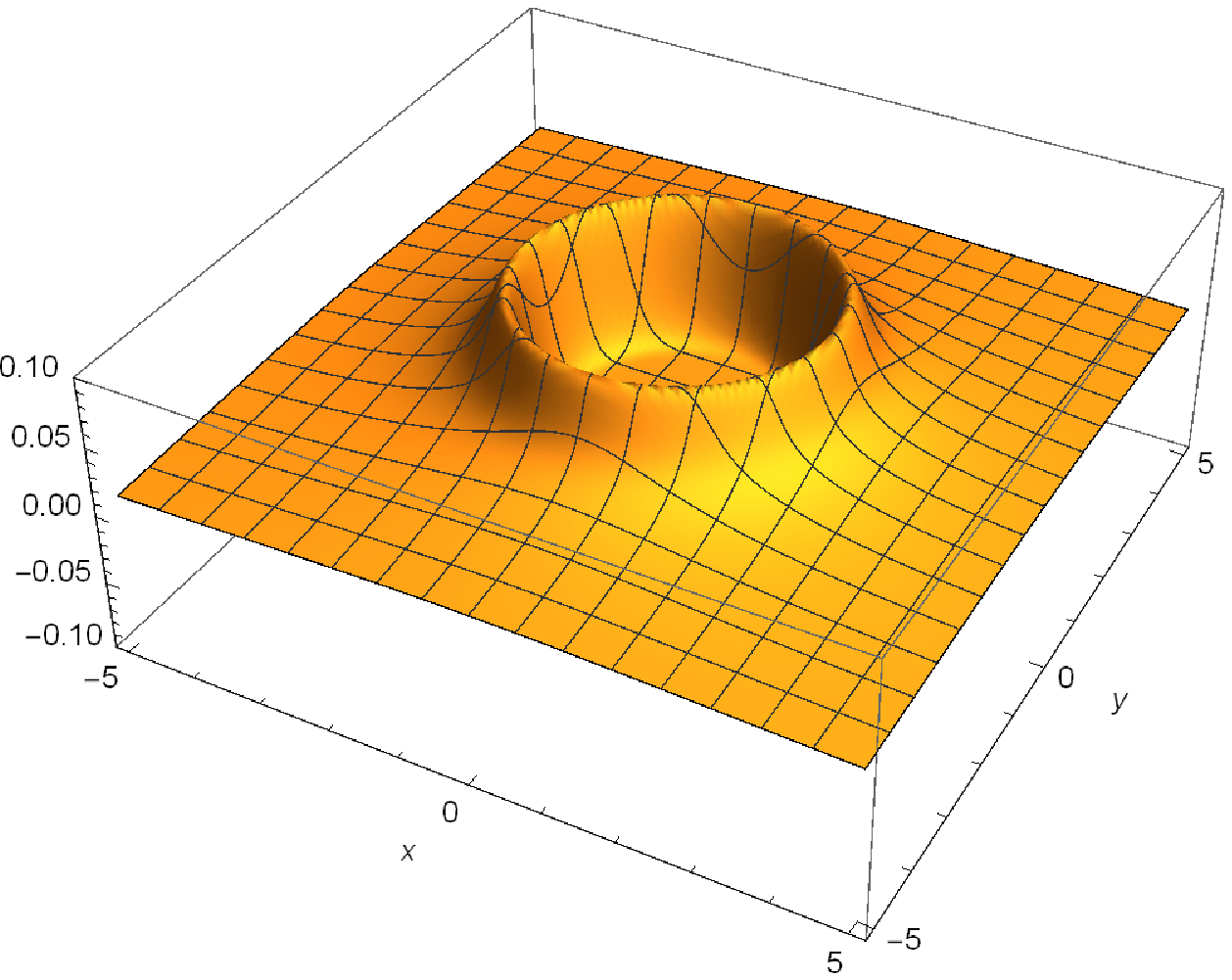}
    \caption{$\sigma^{1/2} G^{(2)}(t,x,y)$ at $t=2$.}
    \label{fig_t2g12_2d}
\end{figure}

\begin{figure}[H]
    \centering
    \includegraphics[keepaspectratio, scale=0.60]{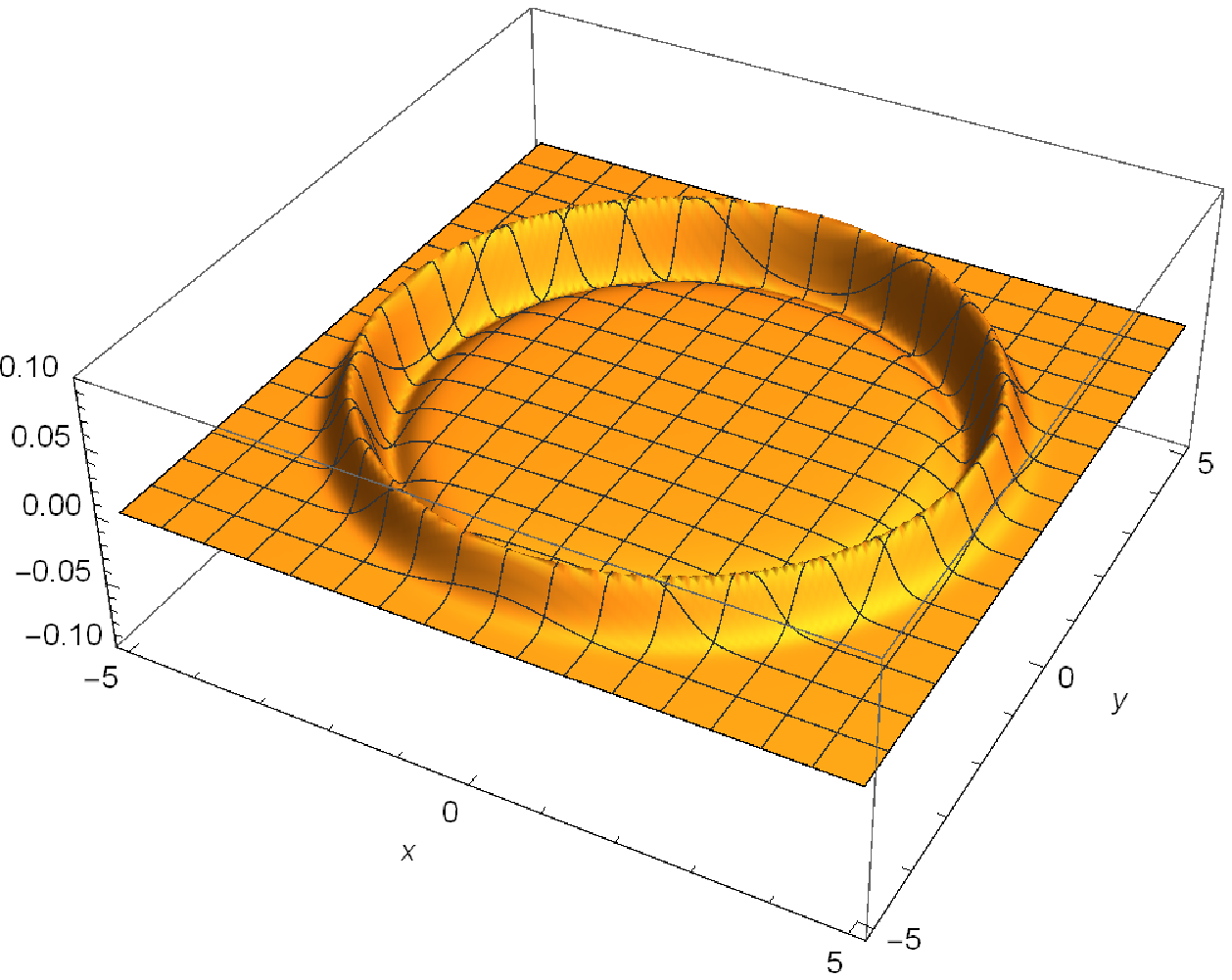}
    \caption{$\sigma^{3/2} F^{(1)}(t,x,y)$ at $t=4$.}
    \label{fig_t4f11_2d}
\end{figure}

\begin{figure}[H]
    \centering
\includegraphics[keepaspectratio, scale=0.60]{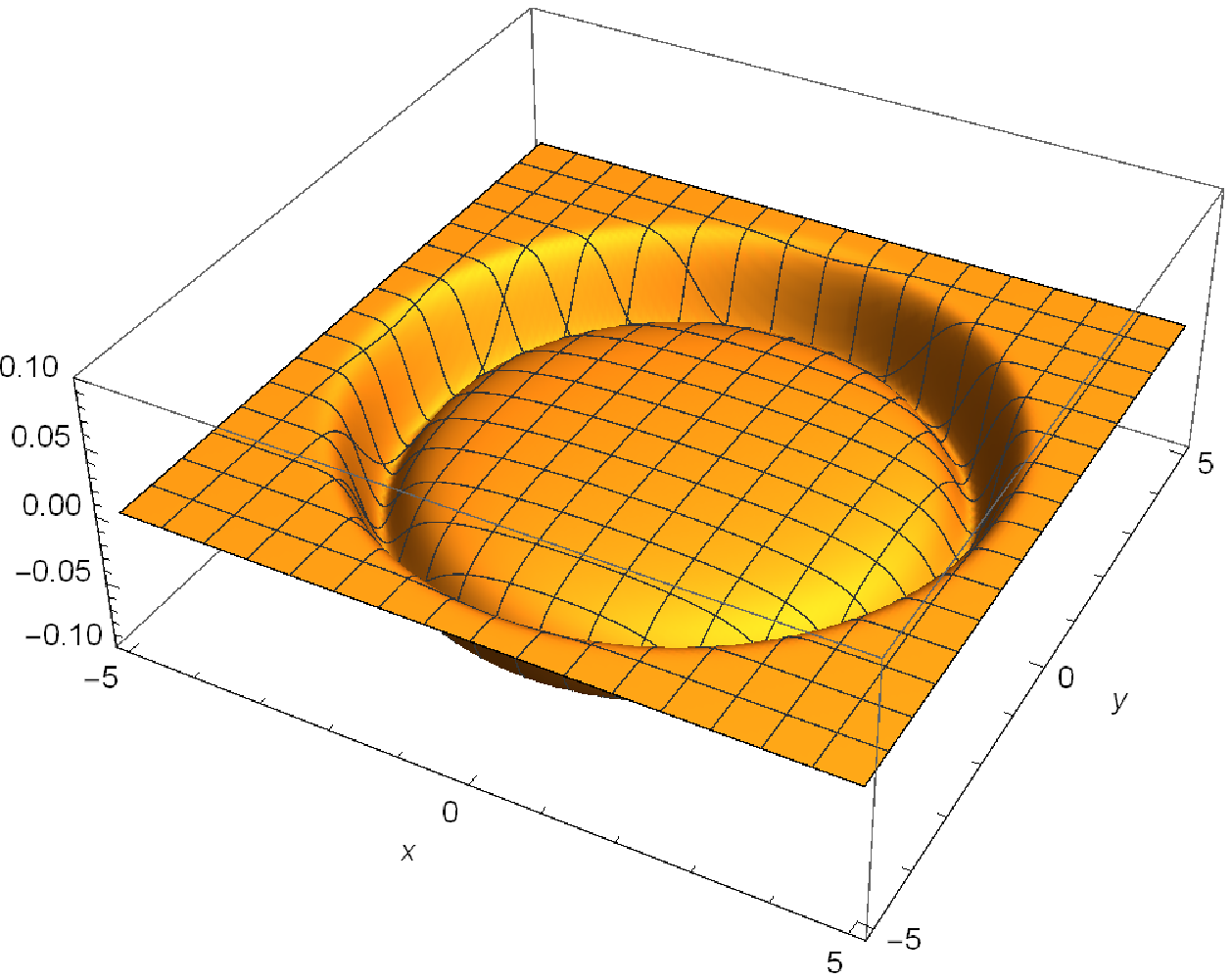}
    \caption{$\sigma^{1/2} F^{(2)}(t,x,y)$ at $t=4$.}
    \label{fig_t4f12_2d}
\end{figure}

\begin{figure}[H]
    \centering
\includegraphics[keepaspectratio, scale=0.60]{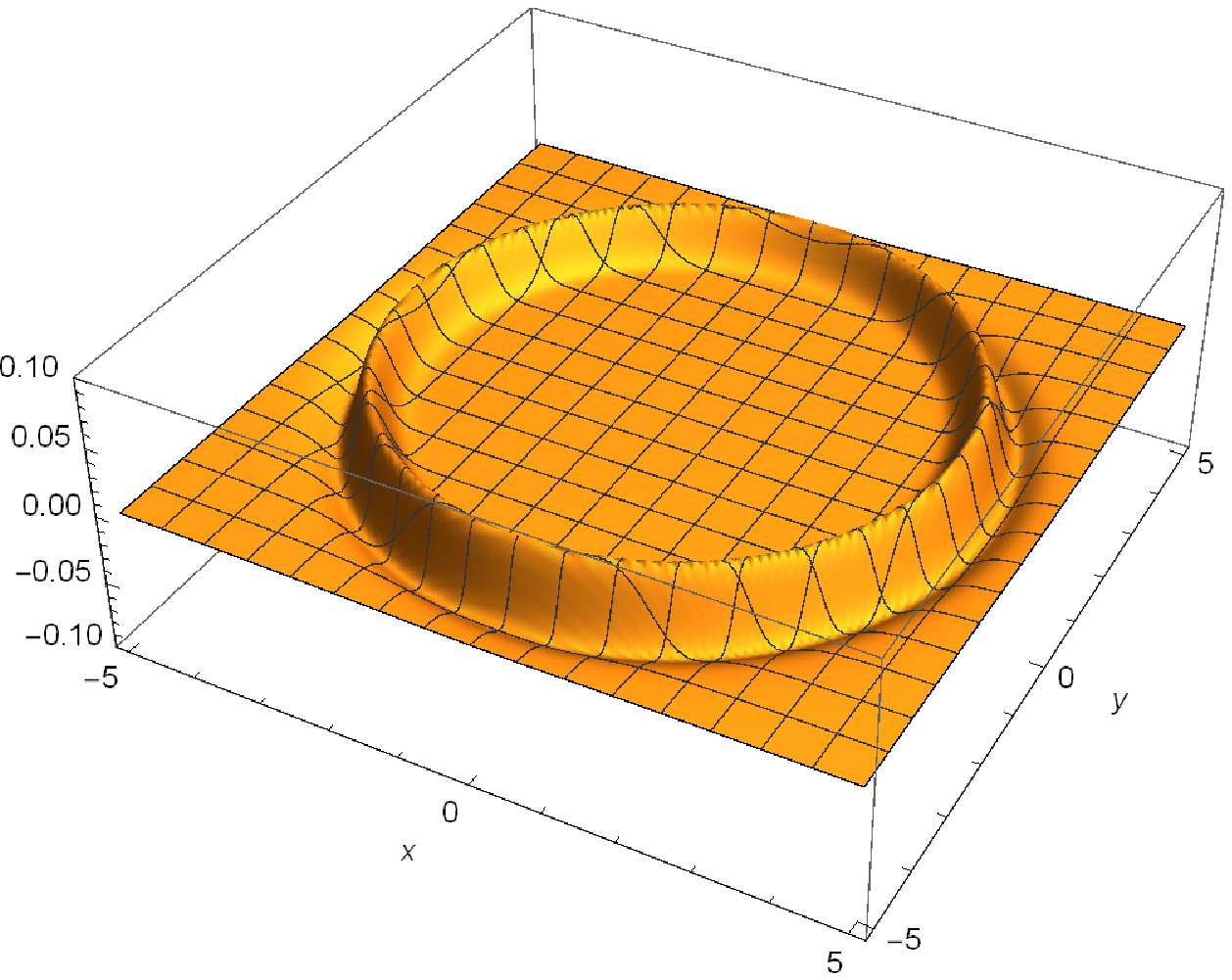}
    \caption{$\sigma^{3/2} G^{(1)}(t,x,y)$ at $t=4$.}
    \label{fig_t4g11_2d}      
\end{figure}

\begin{figure}[H]
    \centering
\includegraphics[keepaspectratio, scale=0.60]{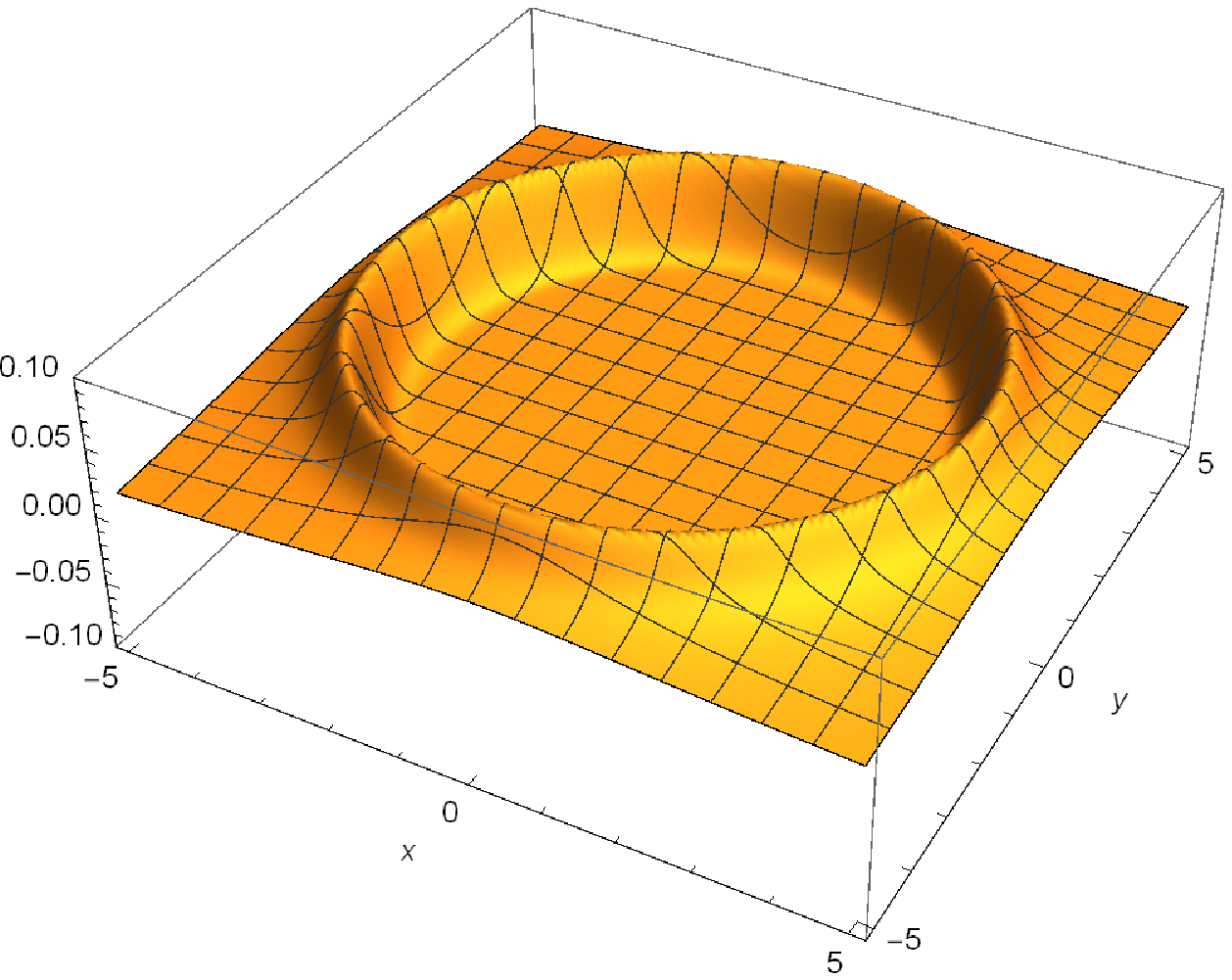}
    \caption{$\sigma^{1/2} G^{(2)}(t,x,y)$ at $t=4$.}
    \label{fig_t4g12_2d}
\end{figure}

\begin{figure}[H]
         \centering
          \includegraphics[keepaspectratio,scale=0.70]{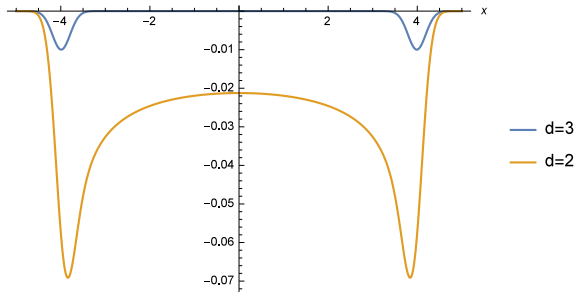}
                          \caption{Comparison of Figs.  \ref{fig_t4f12_3d} and \ref{fig_t4f12_2d} at $y=0$. }
                          \label{fig_comp_f12}
\end{figure}

Figures \ref{fig_t0f11_2d}-\ref{fig_t4g12_2d} show the time evolution of the QIC mode. In these figures, the weighting function is made to be dimensionless by using $\sigma$. Notice that $F^{(2)}(0,x,y) =G^{(1)}(0,x,y) =0$. The parameters characterizing the detector are fixed at $\sigma=0.2$ and $\bm{x}_0=0$. The behavior in this case seems to be qualitatively same as in $(3+1)$-dimensional Minkowski spacetime. For example, the weighting functions are peaked in the region corresponding to the light cone. In the $(2+1)$-dimensional case, however, the weighting functions have a broader tail inside the light cone than those in $(3+1)$-dimensional case, since the strong Huygens principle is violated in the former case. In Fig. \ref{fig_comp_f12}, Figs.\ref{fig_t4f12_3d} and \ref{fig_t4f12_2d} are compared at $y=0$. For $d=3$, the function is strongly localized around the light cone $x=\pm 4$, while it has a tail inside the light cone for $d=2$.

\section{Noise reduction by measurements outside the causal future}\label{sec_ec}
From Eq. \eqref{eq_lin_map}, it can be seen that even when $\hat{O}$ is strictly localized in a spatial region, the operator $f_\Psi(\hat{O})$ has a broader support, meaning that the QIC is a delocalized mode. As mentioned in the Introduction, this is because information is stored in non-local correlations due to the spatial entanglement of the field in its ground state. To explore the physical implications of this tail in $f_{\Psi}(\hat{O})$, let us consider the following information transmission protocol: 
\begin{enumerate}
 \item Encoding:\\
Alice does nothing to the field when she wants to encode $0$.
She turns on the ``switch'' of her UDW detector (i.e. she couples to the field) if she wants to encode $1$. 
We assume that at the initial time the qubit of Alice's detector is in the ground state $\ket{g}_{A}$ and the field is in the vacuum state $\ket{0}$.  
For a delta function switching function, the encoding process is implemented by the unitary operator
\begin{align}
 \hat{U}_A\equiv e^{-\ii\lambda_A \hat\mu_A(t_{\mathrm{enc.}}) \otimes \hat{O}_A}, 
\end{align}
where $t_{\mathrm{enc.}}$ is the time when Alice encodes the information and $\hat{\mu}_A$ is a monopole operator of Alice's detector expressed by
\begin{align}
 \mu_A(t)=e^{-\ii\Omega_A t}\ket{g}\bra{e}+e^{\ii\Omega_A t}\ket{e}\bra{g}
\end{align}
with the ground state $\ket{g}$ and excited state $\ket{e}$. The parameter $\Omega_A>0$ denotes the energy gap of Alice's qubit. The operator $\hat{O}_A$ is given by
\begin{align}
    &\hat{O}_A=\int \text{d}^d\bm{x}\left( v_{A}^{(1)}(\bm{x})\hat{\phi}(t_{\mathrm{enc.}},\bm{x})\nonumber\right.\\
    &\left.\quad\quad\quad\quad\quad\quad\quad\quad+v_{A}^{(2)}(\bm{x})\hat{\Pi}(t_{\mathrm{enc.}},\bm{x})\right)
\end{align}
for real functions $v_A^{(1)}(\bm{x})$ and $v_A^{(2)}(\bm{x})$ which have finite support. 
For example, $\lambda_A=0$ and $\lambda_A=1$ correspond to the cases where she encodes $0$ and $1$, respectively. 
 \item Decoding:\\
Bob tries to decode information from the field by using UDW detectors. To investigate the enhancement of decoding due to correlations, let us assume that he prepares three detectors $B_1, B_2$ and $B_3$. We assume that the detectors are located inside, on and outside the smeared light cone of Alice's encoding operation, respectively. For simplicity, we assume that the detectors are initially in their ground states $\ket{g}_{B_i}$ and pretimed to interact instantaneously with the field at $t=t_{\mathrm{dec. }}>t_{\mathrm{enc.}}$. The decoding unitary operation is expressed as
\begin{align}
  &\hat{U}_B=e^{-\ii\lambda_{B_1}\hat{\mu}_{B_1}(t_{\mathrm{dec.}})\otimes \hat{O}_{B_1}}\nonumber\\
  &\times e^{-\ii\lambda_{B_2}\hat{\mu}_{B_2}(t_{\mathrm{dec.}})\otimes \hat{O}_{B_2}}e^{-\ii\lambda_{B_3}\hat{\mu}_{B_3}(t_{\mathrm{dec.}})\otimes \hat{O}_{B_3}},
\end{align}
where $\hat{\mu}_{B_i}$ is the monopole operator of the detector $B_i$. 
Since the detectors are spatially separated, the $\hat{O}_{B_i}$ commute with each other. After the interaction, projective measurements are performed for the detectors and Bob gathers the measurement results to decode the information. The probability distribution of the measurement results is given by
\begin{align}
 &p_{B_1B_2B_3}(b_1,b_2,b_3|\lambda_A)\nonumber\\
 &\equiv \Braket{\Phi|\hat{U}_A^\dag\hat{U_B}^\dag \hat{E}_{(z_1,z_2,z_3)}\hat{U}_B\hat{U}_A|\Phi},
\end{align}
where $\ket{\Phi}\equiv \ket{g}_{B_1}\ket{g}_{B_2}\ket{g}_{B_3}\ket{g}_{A}\ket{0}$ and $\hat{E}_{(b_1,b_2,b_3)}$ is a projection-valued measure defined by
\begin{align}
 &\hat{E}_{(b_1,b_2,b_3)}\nonumber\\
 &\equiv \ket{b_1}_{B_1}\bra{b_1}_{B_1}\otimes \ket{b_2}_{B_2}\bra{b_2}_{B_2}\otimes \ket{b_3}_{B_3}\bra{b_3}_{B_3}
\end{align}
for $b_i=e,g$. Bob tries to recover the bit Alice sent by using (some of) the detectors' results. 
When Bob uses some of his detectors, the probability distribution of the bits he receives is calculated as the marginal distribution. For example, if Bob uses the detector $B_2$, it is given by
\begin{align}
 p_{B_2}(b_2|\lambda_A)\equiv \sum_{b_1,b_3=e,g}p_{B_1B_2B_3}(b_1,b_2,b_3).
\end{align}
\end{enumerate}

When Alice encodes $0$ with probability $q$, the joint probability distribution is given by
\begin{align}
 p_{AB}(a,b)=
\begin{cases}
q\,p_B(b|\lambda_A=0)\quad &(\text{if } a=0) \\
(1-q)p_B(b|\lambda_A=1)\quad &(\text{if } a=1)
\end{cases} ,
\end{align}
where $B$ denotes one of $\{B_1,B_2,B_3,B_1B_2,B_2B_3,B_1B_3,B_1B_2B_3\}$ depending on the detectors that Bob uses. Let us adopt the classical channel capacity as a quantifier of the efficiency of information transmission, which is given by
\begin{align}
 C_B\equiv \sup_{q} I(A;B), 
\end{align}
where $I(A;B)$ is the mutual information defined by
\begin{align}
 I(A;B)=\sum_a\sum_b p_{AB}(a,b)\log{\left(\frac{p_{AB}(a,b)}{p_A(a)p_B(b)}\right)},
\end{align}
where the marginal distributions are given by
\begin{align}
 p_A(a)=\sum_b p_{AB}(a,b)=
\begin{cases}
 q\quad &(\text{if } a=0)\\
 (1-q)\quad &(\text{if } a=1)
\end{cases}
\end{align}
and
\begin{align}
 p_B(b)\equiv \sum_a p_{AB}(a,b)=q p_B(b|0)+(1-q)p_B(b|1).
\end{align}

\begin{figure}[H]
         \centering
          \includegraphics[keepaspectratio, scale=0.50]{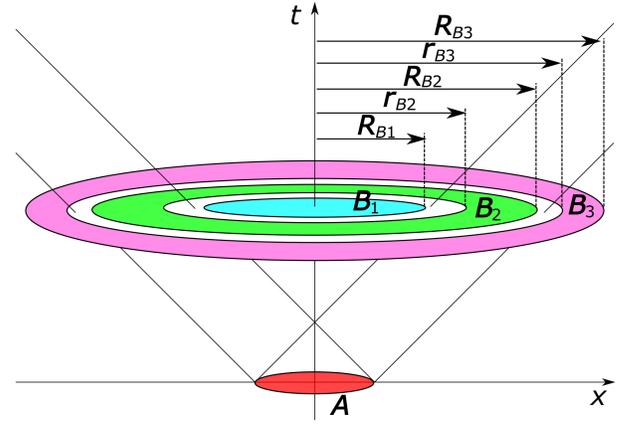}
                          \caption{Schematic figure of the spatial distribution of detectors. For simplicity, we have set $r_{B_1}=0$. The detectors $B_1$, $B_2$ and $B_3$ are located inside, on and outside the smeared light cone of the region where the detector $A$ is located, respectively. }
                          \label{fig_distr}
\end{figure}

As a simple case where the smearing functions of Alice's detector have finite support, let us adopt hard sphere smearing functions:
\begin{align}
 v_A^{(1)}(\bm{x})&=
\begin{cases}
 1\quad & (\text{if } |\bm{x}|<R_A)\\
 0\quad & (\text{otherwise})
\end{cases},\quad
 v_A^{(2)}(\bm{x})=0 .
\end{align}
For Bob's detectors, we also adopt compact smearing functions similar to Alice's:
\begin{align}
 v_{B_i}^{(1)}(\bm{x})&=
\begin{cases}
 1\quad &(\text{if } r_{B_i}<|\bm{x}|<R_{B_i})\\
 0\quad & (\text{otherwise})
\end{cases},\quad
 v_{B_i}^{(2)}(\bm{x})=0 .
\end{align}
To make sure that detectors $B_1,B_2$ and $B_3$ are located inside, on and outside the smeared light cone, the radii have to satisfy
\begin{align}
\begin{split}
 &r_{B_1}<R_{B_1}<\Delta t -R_A,\\
 &\Delta t -R_A< r_{B_2}<R_{B_2}<R_A+\Delta t ,\\
 &R_A+\Delta t< r_{B_3}<R_{B_3},
\end{split}
\label{eq_det_position}
\end{align}
where we have defined $\Delta t \equiv t_{\mathrm{dec.}}-t_{\mathrm{enc.}}$. 
The spatial distribution of the detectors is summarized in Fig. \ref{fig_distr}. 

The probability distribution can be straightforwardly calculated, and is given by
\begin{align}
&  p_{\lambda_A}(z_1,z_2,z_3)\nonumber\\
 &=\frac{1}{2}\sum_{s_A=\pm}\sum_{s_1,s_2,s_3,s_1',s_2',s_3'=\pm}\nonumber\\
 &\quad\times\Braket{g|s_i}\Braket{s_i|U_i(t)|z_i}\Braket{z_i|U_i(t)^\dag|s_i'} \Braket{s_i'|g})\nonumber\\
&\quad \times \exp\left(-\frac{1}{2}\sum_{i=1}^3 \lambda_{B_i}(s_i-s_i')\sum_{j=1}^3 \lambda_{B_j}(s_j-s_j')\right.\nonumber\\
&\left.\quad\quad\times \int\frac{\text{d}^d\bm{k}}{(2\pi)^d 2|\bm{k}|}\tilde{v}_{B_i}^{(1)}(\bm{k})\tilde{v}_{B_j}^{(1)}(\bm{k})^*\right)\nonumber\\
&\quad \times \exp\left(2\lambda_A s_A \sum_{i=1}^3 \lambda_{B_i}\left(s_{i}-s_{i}'\right)\right.\nonumber\\
&\quad\left.\quad\times\mathrm{Im}\left(\int \frac{\text{d}^d\bm{k}}{(2\pi)^d 2|\bm{k}|}e^{-\ii|\bm{k}|\Delta t}\tilde{v}_{B_i}^{(1)}(\bm{k})\tilde{v}_{A}^{(1)}(\bm{k})^*\right)\right).
\end{align}
Here we have introduced the eigenvectors of the Pauli $x$ operator $\ket{\pm}\equiv \frac{1}{\sqrt{2}}(\ket{e}\pm \ket{g})$. The detailed derivation can be found in Appendix \ref{appendix_jpd}. It should be noted that the result is independent of the energy gap of the detectors since the detectors remains in their ground state before the instantaneous interaction with the field.

By using this formula, the classical channel capacity is numerically evaluated in $(3+1)$- and $(2+1)$-dimensional Minkowski spacetimes. The results are summarized in Table \ref{table_channel_capacities}.
\begin{table*}[tb]
\centering
 \begin{tabular}{|c||c|c|c|c|c|c|c|}\hline 
&$C_{B_1}$ & $C_{B_2}$ & $C_{B_3}$ & $C_{B_1B_2}$ & $C_{B_2 B_3}$ & $C_{B_1 B_3}$ & $C_{B_1B_2B_3}$\\ \hline\hline
 $d=3$ &0 &0.0000339083 &0 &0.0000345126 &0.0000373605 & 0&0.0000379689 \\ \hline
 $d=2$ & 0.00167331 &0.00872886 & 0 &0.0102214 &0.0140338 & 0.00167926 &0.0154962\\ \hline
 \end{tabular}
\vspace*{5mm}
\caption{Classical channel capacities. The radii of detectors and the time difference are fixed $R_A=1$, $r_{B_1}=0$, $R_{B_1}=0.9$, $r_{B_2}=1.1$, $R_{B_2}=2.9$, $r_{B_3}=3.1$, $R_{B_3}=4$, and $\Delta t=2$ so that Eq.\eqref{eq_det_position} is satisfied. The coupling constants are fixed as $\lambda_{B_1}=\lambda_{B_2}=\lambda_{B_3}=0.2$. The subscripts represent the detectors which the receiver (Bob) adopts. The detectors $B_1$, $B_2$ and $B_3$ are respectively located inside, on, and outside the smeared light cone of the region where the encoding operation has been performed.}\label{table_channel_capacities} 
\end{table*}

First, $C_{B_3}$ vanishes in both cases, reflecting the fact that there is no superluminal signaling. However, this does not mean that the detector $B_3$ is useless in decoding the information. For example, $ C_{B_2}<C_{B_2B_3}$ holds in both cases. It means that the measurement result of detector $B_3$ enhances the channel capacity once it is processed with the result of $B_2$. This can be interpreted as follows: quantum fields are noisy as media of communication since they have spatial entanglement. Nevertheless, the noises are non-locally correlated. This suggests that by using the measurement result on the detector $B_3$, we can reduce the noise in the measurement result on the detector $B_2$. As a consequence, the channel capacity can be enhanced, as we show to be the case. 

Second, note that $C_{B_1}$ vanishes in $(3+1)$-dimensional Minkowski spacetime, while it does not in the $(2+1)$-dimensional case. This is an explicit consequence of the violation of the strong Hyugens principle in $(2+1)$-dimensional Minkowski spacetime. In both spacetimes, however, the measurement result on detector $B_1$ is also useful if it is combined with the one on detector $B_2$, which can be seen from the fact that in both cases $C_{B_2}<C_{B_1B_2}$ holds. Therefore, even when the strong Huygens principle is valid, the detector inside the light cone is also useful in communication. 

Finally, it should be noted that the QIC identifies the noises which may be used to enhance the channel capacity. Suppose that Bob adopts another UDW detector $B_4$ whose measurement operation commutes with both $\hat{O}_A$ and $f_{\Psi}(\hat{O}_A)$. Since the QIC mode is not correlated with the modes orthogonal to it, no information is gained from $B_4$ even when it is combined with another detector e.g., $B_2$. 

\section{Quantum shockwave communication and multi-mode QIC}\label{sec_mqic}
So far, we have seen that the notion of a QIC can be used to identify the information carrier if the encoding operation is generated by a single Hermitian operator. For example, this analysis can be used in the case where Alice uses an UDW detector which instantaneously couples to the field. 
However, from the viewpoint of information transmission, this restriction makes the problem too simple. For example, it is known that the quantum channel capacity always vanishes when Alice uses a simple-generated encoding unitary \cite{simidzija2019transmission}.
Furthermore, quantum shockwave communication protocols \cite{ahmadzadegan2018quantum} cannot be analyzed by using the single-mode QIC. 

In this section, we first present a general protocol to identify multiple modes in a pure state which carry information. For an encoding operation generated by $k$ generators, (at most) $k$ modes are the information carrier. We call this a $k$-mode QIC, as it is a natural extension of the single-mode QIC. 

\subsection{Multimode quantum information capsule}
Assume that the encoding process is expressed by quantum operations generated by a finite number of operators $\{\hat{O}_i\}_{i=1}^N$, each of which is given by
\begin{align}
 \hat{O}_i = \int \text{d}^d \bm{x} \left(v_i^{(1)}(\bm{x})\hat{\phi}(t_i,\bm{x})+v_i^{(2)}(\bm{x})\hat{\Pi}(t_i,\bm{x})\right),
\end{align} 
where $v_i^{(1)}(\bm{x})$ and $v_i^{(2)}(\bm{x})$ are real functions. For example, this condition is satisfied when Alice adopts $k$ UDW inertial detectors with interaction Hamiltonians
\begin{align}
 \hat{H}_i&=\lambda_i\chi_i(t)\hat{\mu}_i(t)\otimes \hat{O}_i(t),\\
 \chi_i(t)&=\delta(t-t_i) ,\\
 \hat{O}(t)&= \int \text{d}^d \bm{x} \left(v_i^{(1)}(\bm{x})\hat{\phi}(t,\bm{x})+v_i^{(2)}(\bm{x})\hat{\Pi}(t,\bm{x})\right)
\end{align}
for $i=1,\cdots,k$. 
Here, $\lambda_i$ denotes the coupling constant, $\hat{\mu}_i(t)$ is an observable of the $i$th detector and $v_i^{(1)}(\bm{x}),v_i^{(2)}(\bm{x})$ are the smearing functions. 

When $k=1$, the single-mode QIC formula uniquely identifies the information carrier mode which is characterized by
\begin{align}
 \hat{Q}_1\equiv  \frac{1}{\alpha_1}\hat{O}_1,\quad \hat{P}_1\equiv \frac{1}{\alpha_1}f_\Psi\left(\hat{O}_1\right),
\end{align}
where $ \alpha_1\equiv \sqrt{2\Braket{\Psi|\hat{O}_1^2|\Psi}}$ is a normalization factor.
Since the QIC mode is in a pure state, the Gaussian state $\ket{\Psi}$ is expressed as
\begin{align}
 \ket{\Psi}=\ket{0}_1\otimes \ket{\Psi'}_{\bar{1}}\label{eq_tensordecomposition},
\end{align}
where $\ket{\Psi'}_{\bar{1}}$ denotes the state for the subsystem $\bar{1}$ complement to the subsystem characterized by $(\hat{Q}_1,\hat{P}_1)$.
For our purpose, we do not need to calculate $\ket{\Psi'}_{\bar{1}}$ itself explicitly. It should be noted that $\ket{\Psi}_{\bar{1}}$ is also a Gaussian state.

The key idea to extend this analysis to $k=2$ is to decompose the operator $\hat{O}_2$ into the contributions for the subsystems $1$ and $\bar{1}$. Defining
\begin{align}
 \hat{O}_2'\equiv \hat{O}_2-\left(\beta_{2,1}\hat{Q}_1+\gamma_{2,1}\hat{P}_1\right),
\end{align}
where
\begin{align}
 \beta_{2,1}&\equiv\frac{1}{\ii}\Braket{\Psi|\left[\hat{O}_2,\hat{P}_1\right]|\Psi},\\
 \gamma_{2,1}&\equiv-\frac{1}{\ii}\Braket{\Psi|\left[\hat{O}_2,\hat{Q}_1\right]|\Psi} ,
\end{align}
the operator $\hat{O}_2'$ commutes with $\hat{Q}_1$ and $\hat{P}_1$. Therefore, it is an operator on the subsystem $\bar{1}$. 
Since the subsystems $1$ and $\bar{1}$ share no correlations in $\ket{\Psi}$, the operator $f_\Psi(\hat{O}_2')$ must commute with both $\hat{Q}_1$ and $\hat{P}_1$. See Appendix \ref{app_com} for a more formal proof.
Therefore, the mode defined by
\begin{align}
 \hat{Q}_2&\equiv\frac{1}{\alpha_2}\left(\hat{O}_2-\left(\beta_{2,1}\hat{Q}_1+\gamma_{2,1}\hat{P}_1\right)\right)\\
 \hat{P}_2&\equiv f_{\Psi}\left(\hat{Q}_2\right)= \frac{1}{\alpha_2}\left(f_\Psi\left(\hat{O}_2\right)-\left(\beta_{2,1}\hat{P}_1-\gamma_{2,1}\hat{Q}_1\right)\right)
\end{align}
is orthogonal to the mode $(\hat{Q}_1,\hat{P}_1)$ and is initially in a pure state in the standard form. Here, we have used the linearity of $f_\Psi$ and Eq.\eqref{eq_f^2}. 
The factor $\alpha_2$ is fixed so that
\begin{align}
 \Braket{\Psi|\hat{Q}_2^2|\Psi}=\frac{1}{2}
\end{align}
is satisfied. Since
\begin{align}
 \Braket{\Psi|\hat{O}_2^2|\Psi}= \alpha_2^2 \Braket{\Psi|\hat{Q}_2^2|\Psi}+\frac{1}{2}\left(\beta_{2,1}^2+\gamma_{2,1}^2\right)
\end{align}
holds, $\alpha_2$ is determined as
\begin{align}
 \alpha_2\equiv \sqrt{2\Braket{\Psi|\hat{O}_2^2|\Psi}- \left(\beta_{2,1}^2+\gamma_{2,1}^2\right)}.
\end{align}

By repeating this procedure, we obtain the general protocol to identify the modes in which information would be encoded. Recursively, we obtain
\begin{align}
\begin{split}
  \hat{Q}_i&\equiv \frac{1}{\alpha_i}\left(\hat{O}_i -\sum_{j=1}^{i-1}\left(\beta_{i,j}\hat{Q}_j+\gamma_{i,j}\hat{P}_j\right)\right)\\
 \hat{P}_i&\equiv  \frac{1}{\alpha_i}\left(f_\Psi\left(\hat{O}_i\right)-\sum_{j=1}^{i-1}\left(\beta_{i,j}\hat{P}_j-\gamma_{i,j}\hat{Q}_j\right)\right),
\end{split}\label{eq_mqic}
\end{align}
where
\begin{align}
 \beta_{i,j}&\equiv \frac{1}{\ii}\Braket{\Psi|\left[\hat{O}_i,\hat{P}_j\right]|\Psi}\nonumber\\
 &=\frac{1}{\alpha_j}\left(\frac{1}{\ii}\Braket{\Psi|\left[\hat{O}_i,f_{\Psi}\left(\hat{O}_j\right)\right]|\Psi}\right.\nonumber\\
 &\left.\quad\quad\quad\quad\quad\quad-\sum_{k=1}^{j-1}\left(\beta_{j,k}\beta_{i,k}+\gamma_{j,k}\gamma_{i,k}\right)\right)\\
 \gamma_{i,j}&\equiv -\frac{1}{\ii}\Braket{\Psi|\left[\hat{O}_i,\hat{Q}_j\right]|\Psi}\nonumber\\
 &=\frac{1}{\alpha_j}\left(-\frac{1}{\ii}\Braket{\Psi|\left[\hat{O}_i,\hat{O}_j\right]|\Psi}\right.\nonumber\\
& \left.\quad\quad\quad\quad\quad\quad-\sum_{k=1}^{j-1}\left(\beta_{j,k}\gamma_{i,k}-\gamma_{j,k}\beta_{i,k}\right)\right)\\
 \alpha_i&\equiv \sqrt{2\Braket{\Psi|\hat{O}_i^2|\Psi}-\sum_{j=1}^{i-1}\left(\beta_{i,j}^2+\gamma_{i,j}^2\right)} .
\end{align}
The modes defined here are initially in a pure state carrying the information encoded by operations generated by $\{\hat{O}_i\}_{i=1}^k$. Hence we call this set of modes a $k$-mode QIC. Technically speaking, we have assumed that $\alpha_i\neq 0$, which usually holds. In the case where $\alpha_i=0$ for some $i$, it implies that $\hat{O}_i$ is written as a linear combination of $\{(\hat{Q}_j,\hat{P}_j)\}_{j=1}^{i-1}$. Therefore, $(i-1)$ modes play the role of information carrier for the $i$th encoding operation and we can simply skip the recursion process for this operation. In this sense, the protocol to identify QIC works without any exception. Hereafter, we assume that $\alpha_i\neq 0 $ for notational simplicity.

It should be noted that a $k$-mode QIC is unique as a subsystem of the information carrier. 
By decomposing the subsystem into $k$ independent modes, it is possible to visualize the propagation of modes by plotting their weighting functions. Although the plots will help to get an intuition about where information propagates, we need to be careful since they may look different if one adopts another decomposition. 
Hereafter, we adopt $k$ modes in Eq. \eqref{eq_mqic} to visualize the QIC. For this decomposition, the following properties are satisfied: (i) each mode is initially in a pure state in the standard form, and (ii) when information of the $j$th detector is encoded in the field, the $i(>j)$th mode is independent of the encoded information. The QIC operators $\{(\hat{Q}_i,\hat{P}_i)\}_{i=1}^k$ at $t$ can be expressed by weighting functions $F_i^{(1)},F_i^{(2)},G_i^{(1)},G_i^{(2)}$ satisfying
\begin{align}
 \hat{Q}_i&=\int \text{d}^d\bm{x}\left( F_i^{(1)}(t,\bm{x})\hat{\phi}(t,\bm{x})+F^{(2)}(t,\bm{x})\hat{\Pi}(t,\bm{x})\right),\\
 \hat{P}_i&=\int \text{d}^d \bm{x} \left( G_i^{(1)}(t,\bm{x})\hat{\phi}(t,\bm{x})+G^{(2)}(t,\bm{x})\hat{\Pi}(t,\bm{x})\right).
\end{align}
From Eq. \eqref{eq_mqic}, we get
\begin{align}
 &F_i^{(l)}(t,\bm{x})\nonumber\\
 &=\frac{1}{\alpha_i}\left(v_i^{(l)}(t,\bm{x})-\left(\sum_{j=1}^{i-1}\beta_{i,j}F_j^{(l)}(t,\bm{x})+\gamma_{i,j}G_j^{(l)}(t,\bm{x})\right)\right),\\
  &G_i^{(l)}(t,\bm{x})\nonumber\\
  &=\frac{1}{\alpha_i}\left(u_i^{(l)}(t,\bm{x})-\left(\sum_{j=1}^{i-1}\beta_{i,j}G_j^{(l)}(t,\bm{x})-\gamma_{i,j}F_j^{(l)}(t,\bm{x})\right)\right),
\end{align}
where $v_i^{(l)}$ and $u_i^{(l)}$ are defined by
\begin{align}
 \hat{O}_i&=\int \text{d}^d\bm{x}\left( v_i^{(1)}(t,\bm{x})\hat{\phi}(t,\bm{x})+v^{(2)}(t,\bm{x})\hat{\Pi}(t,\bm{x})\right),\\
 f_\Psi\left(\hat{O}_i\right)&=\int \text{d}^d\bm{x}\left( u_i^{(1)}(t,\bm{x})\hat{\phi}(t,\bm{x})+u^{(2)}(t,\bm{x})\hat{\Pi}(t,\bm{x})\right) .
\end{align}
These are the formulas for the $k$-mode QIC written in terms of weighting functions.

In the case where $v^{(2)}(\bm{x})=0$ holds, the commutators are simplified and given by
\begin{align}
\begin{split} 
&\frac{1}{\ii}\Braket{\Psi| \left[\hat{O}_i,\hat{O}_j\right]|\Psi}\\
 &=2\mathrm{Im}\left(\int \frac{\text{d}^d\bm{k}}{(2\pi)^d 2|\bm{k}|} e^{-\ii|\bm{k}|(t_i-t_j)}\tilde{v}_i(\bm{k})\tilde{v}_j(\bm{k})^*\right) \\
&\frac{1}{\ii}\Braket{\Psi| \left[\hat{O}_i,f_0\left(\hat{O}_j\right)\right]|\Psi} \\ &=2\mathrm{Re}\left(\int \frac{\text{d}^d\bm{k}}{(2\pi)^d 2|\bm{k}|} e^{-\ii|\bm{k}|(t_i-t_j)}\tilde{v}_i(\bm{k})\tilde{v}_j(\bm{k})^*\right) 
\end{split}\label{eq_commutators}
\end{align}

\subsection{Quantum shockwave in $(3+1)$- and $(2+1)$-dimensional Minkowski spacetimes}
As is done in Ref. \cite{ahmadzadegan2018quantum}, let us investigate the case where Alice uses three UDW detectors which are located in spatially separated regions in the $(3+1)$- and $(2+1)$-dimensional Minkowski spacetimes to create quantum shockwaves. The three-mode QIC visualizes how a shockwave is formed by this encoding process.
We adopt the Gaussian smearing functions
\begin{align}
 v_i^{(1)}(\bm{x})=e^{-\frac{|\bm{x}-\bm{x}_i|^2}{2\sigma^2}},\quad v^{(2)}(\bm{x})=0,
\end{align}
where $\bm{x}_i$ denotes the spatial position of the detector. 
The integral appearing in Eq.\eqref{eq_commutators} can be evaluated in exactly the same way as in Sec. \ref{sec_te}. 
%
%
%
%

\begin{figure}[H]
    \centering
    \includegraphics[keepaspectratio, scale=0.60]{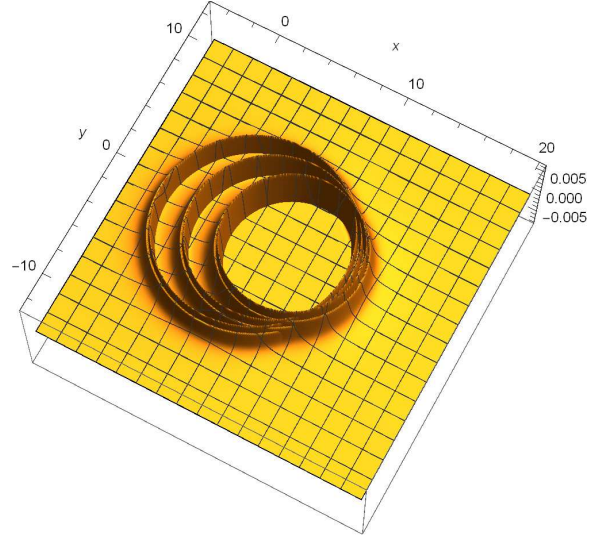}
    \caption{Quantum shockwave forming in $(3+1)$-dimensional Minkowski spacetime at $z=0$. In this figure, four weighting functions for three modes are plotted separately.}
    \label{fig_all_3d}
\end{figure}

\begin{figure}[H]
    \centering
     \includegraphics[keepaspectratio, scale=0.60]{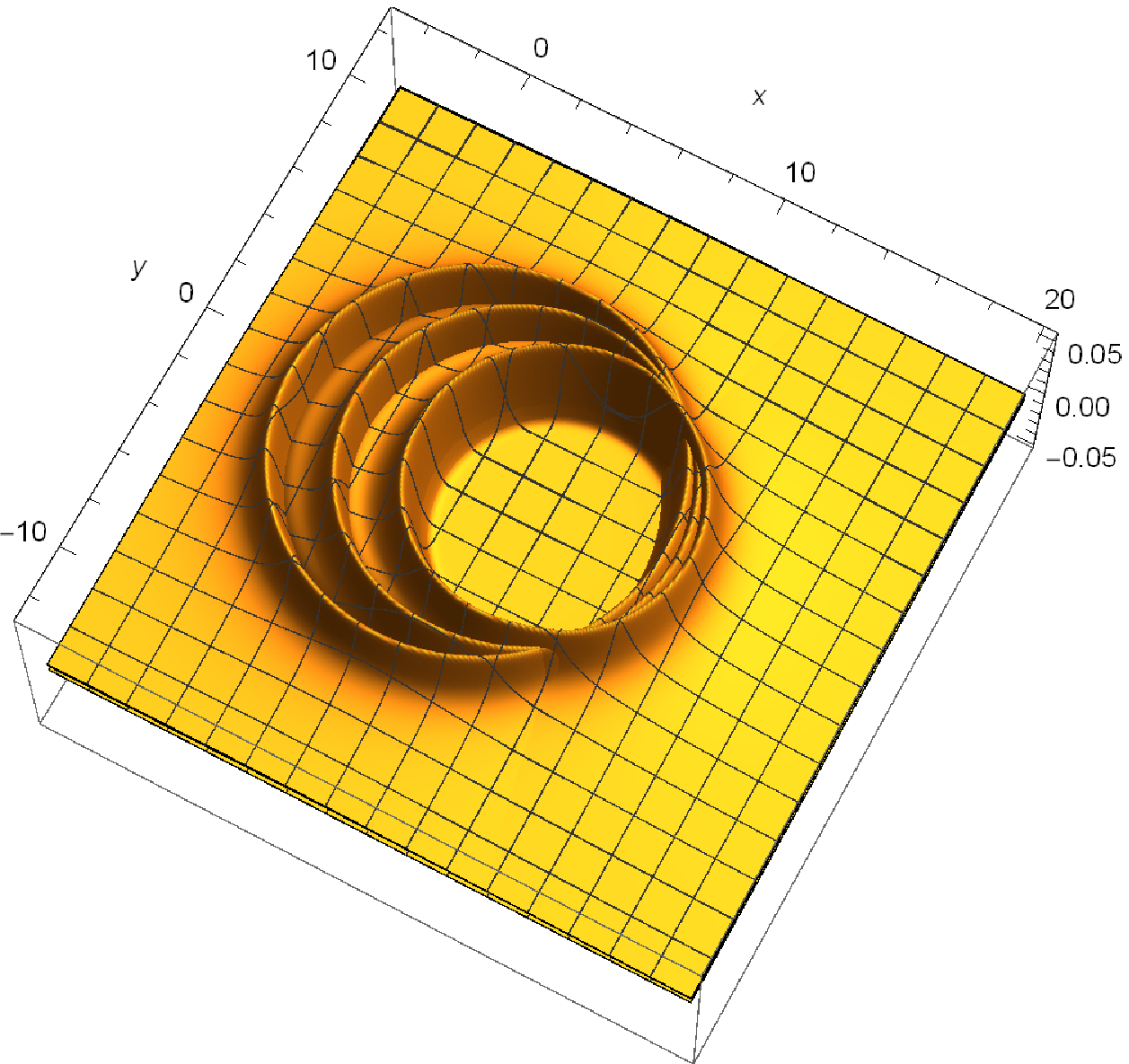}
     \caption{Quantum shockwave forming in $(2+1)$-dimensional Minkowski spacetime. In this figure, four weighting functions for three modes are plotted separately.}\label{fig_all_2d}
\end{figure}

Figures\ref{fig_all_3d} and \ref{fig_all_2d} show the weighting functions of three-mode QIC operators in the $(3+1)$- and $(2+1)$-dimensional cases at $t=8$, where we have fixed $\sigma=0.2$. The spacetime positions of the detectors are set to be $t_i=i$, $x_i=5+1.5 i$ and $y_i=z_i=0$ for $i=1,2,3$. In each figure, $4\times 3=12$ weighting functions are plotted and are made to be dimensionless by using $\sigma$. Notice that some of the weighting functions overlap. We do not specify the correspondence between waves and weighting functions here. Each weighting function is separately plotted in Figs. \ref{fig_f11_3d}-\ref{fig_g32_2d} in Appendix \ref{app_wf}. The wavefront of the shockwave can be easily identified in both cases. 

\begin{figure}[H]
    \centering
    \includegraphics[keepaspectratio,scale=0.8]{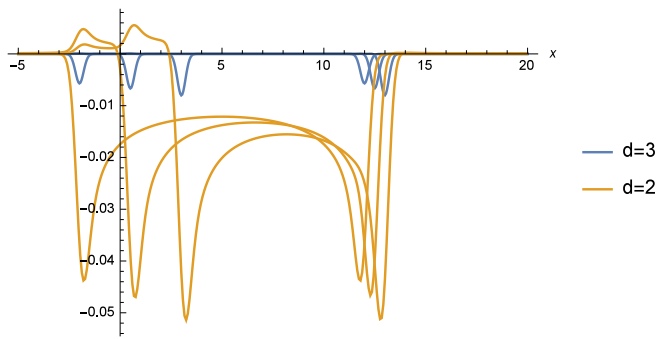}
    \caption{Comparison of $\{\sigma F_i^{(2)}(t,x,0,0)\}_{i=1}^3$ for $d=3$ and $\{\sigma^{1/2}F_i^{(2)}(t,x,0)\}_{i=1}^3$ for $d=2$ at $t=8$. }
    \label{fig_comp_f12_threemodes}
\end{figure}

As we have seen in Sec. \ref{sec_te}, the weighting functions of QIC mode(s) in the $(3+1)$-dimensional case are sharper than those in the $(2+1)$-dimensional case since the strong Huygens principle holds in the former case but not in the latter \cite{AIHPA_1974__20_2_153_0,czapor_hadamards_2007}. To compare the sharpness of the shockwaves in $(3+1)$- and $(2+1)$-dimensional spacetime, $F_i^{(2)}(t,\bm{x})$ is plotted at $y=0$ and $z=0$ in Fig. \ref{fig_comp_f12_threemodes}. It shows that the weighting functions in the $(3+1)$-dimensional case are well localized, while they have a broader tail in the $(2+1)$-dimensional case.


       
\section{Conclusions and outlook}


We applied the method of QICs to study the evolution of the information that is transferred from a qubit particle detector operated by Alice into a quantum field, tracking the  information-carrying disturbances seeded by Alice in the field as they evolve in space and time. 
When allowing Bob to place detectors both inside and outside of the future light cone of Alice's encoding operation one obtains two quantum quantum channels. The first channel is from Alice to those of Bob's detectors which are inside the lightcone and the second channel is from Alice to Bob's detectors outside the lightcone. While the first channel possesses a finite channel capacity, the second channel has, of course, zero capacity due to the spacelike separation. We found that the channel capacity is superadditive in the sense that the capacity of the combined channel is enlarged. This is due to the fact that the vacuum is a spatially entangled state and that, therefore, the quantum noise in the receivers possesses correlations that Bob can use in effect to reduce his signal-to-noise ratio.    

It should be very interesting to investigate to what extent this phenomenon is related to the known phenomenon of the superadditivity of the classical capacity of quantum channels in settings outside quantum field theory. For the literature, see, e.g., Refs. \cite{Hastings2009,Czekaj2009,Sasaki1998,Smith1812}. There, the superadditivity is normally associated with the use of entanglement in the channel inputs. In contrast, in our case here, there is only one input while the superadditivity arises from pre-existing entanglement of quantum noise on the side of the receivers. It will be interesting to further investigate the relationship of these two mechanisms also in light of the known relationship, in the usual settings outside quantum field theory, between the superadditivity of channel capacity and the subadditivity of minimum output entropy; see, e.g., Ref. \cite{Hastings2009}.

Further, we generalized the QIC method to the case of multiple modes. In this generalized setting, Alice and Bob use $N_A$ and $N_B$ emitters and receivers, respectively, to obtain what may be called a quantum MIMO (QMIMO) setup, that generalizes the currently ubiquitously used multiple input, multiple output antenna communication systems (MIMO). Our calculations were simplified by considering the limit of ultrafast couplings of the detectors to the field, described by Dirac delta functions. 
The new multi-mode QIC formula in Eq.\eqref{eq_mqic}, then identifies the multi-mode QIC, i.e., the $(N_A+N_B)$ information-carrying modes of the field  that are in a pure state and that couple to the emitting and receiving UDW detectors. The encoding and decoding processes consists of the interactions among the UDW detectors and the $(N_A+N_B)$-mode oscillators. Each of the QIC-mode oscillators is initially in the ``vaccum'' state, and the generators of interactions are given by
\begin{align}
 \hat{O}_i=\alpha_i\hat{Q}_i+\sum_{j=1}^{i-1}\left(\beta_{i,j}\hat{Q}_j+\gamma_{i,j}\hat{P}_j\right).
\end{align}
The key spatial entanglement of the vacuum state of the field then enters through the calculation of $\alpha_i,\beta_{i,j},\gamma_{i,j}$. 
Calculating channel capacities is hard but one of the advantages of the QIC method is that it enables one to separate the analysis of information communication into two parts: (i) the analysis of the propagation of information-carrying QIC modes in a quantum field  and (ii) the analysis of encoding and decoding process using detectors. 

We demonstrated the new multimode QIC technique for QMIMO by applying it to the case where Alice uses 
suitably lined-up and pretimed emitters to communicate with Bob via quantum shockwaves (see \cite{ahmadzadegan2018quantum}) in the field.
By modulating the entanglement of the emitters, it is possible to modulate the shape of the quantum shockwaves. 

Indeed, it should be very interesting to study the use of the multi-mode QIC technique to investigate the properties of not only the classical but also the quantum channel capacities of QMIMO systems, for example, their superadditivity. 

A technical point in this regard is the fact that in order to be able to perform calculations nonperturbatively, we are working in the limit of short coupling times. It is known that in this limit, single interactions generated by Hamiltonians of the form $A\otimes B$, such as those that arise in quantum field theory, are entanglement breaking \cite{No-go}
and therefore lead to vanishing quantum channel capacities in the case of single modes. It was important, therefore, to generalize to the setting of QMIMO. In QMIMO, if multiple emitters are entangled with an ancilla, then the quantum field can acquire some of that entanglement and transport it to Bob's detectors. The QMIMO channels therefore generally possess a finite quantum channel capacity, i.e., a finite capacity to transmit preexisting entanglement with an ancilla from Alice to Bob.   

Apart from enabling the study of classical and quantum channel capacities through quantum fields, such as their superadditivity, the new methods should also be useful in other contexts of relativistic quantum information theory, such as the harvesting of entanglement from the quantum vacuum, \cite{Reznik2003,Reznik1,Retzker2005,Olson2011,Olson2012,Salton:2014jaa,Pozas-Kerstjens:2015,Pozas2016,Sabinprl,Farming,BeiLok1,PetarHarv}. 

Finally, let us clarify the relationship of the present work to the notion of purification partner modes. For a given mode, a mode which purifies the mode is called its partner. A formula to identify the partner mode is proven for the vacuum state \cite{PhysRevD.91.124060}, for general Gaussian states of a scalar field \cite{Trevison_2019}, and it is generalized for fermionic fields in Ref. \cite{Hackl2019minimalenergycostof}. The partner formulae have been used in the contexts of black hole information loss \cite{tomitsuka2019partner} and entanglement harvesting \cite{Trevison_harvesting,Hackl2019minimalenergycostof}. From the viewpoint of QICs, the partner modes correspond to a class of two-mode QICs. Since our multi-mode QIC formula can identify a $k$-mode QIC with arbitrary $k$, the present results offer wider opportunities for exploring the entanglement structure in quantum fields.



\begin{acknowledgments}
The authors are grateful to M. Hotta, T. Tomitsuka and A. Chatwin-Davies for valuable discussions. 
K.Y. acknowledges the support of JSPS KAKENHI Grant Number 18J20057, and the Graduate Program on Physics for the Universe of Tohoku University. P.S. would like to acknowledge the support of the NSERC CGS-M and CGS-D scholarships.  E.M.M. acknowledges support through an Ontario Early Researcher Award (ERA). A.K. acknowledges support through a Google Faculty Research Award. A.K. and E.M.M. acknowledge support from the National Science and Engineering Research Council of Canada (NSERC).
\end{acknowledgments}

\appendix
\section{The calculation of the joint probability distribution}\label{appendix_jpd}
Here we use the following notation:
\begin{align}
 z_i=e,g,\quad s_i=\pm,\quad \ket{\pm}=\frac{1}{\sqrt{2}}\left(\ket{e}\pm \ket{g}\right).
\end{align}
Since
\begin{align}
 &e^{-\ii\lambda\hat{\sigma}(t)\otimes \hat{O}(t)}\nonumber\\
 &=\left(e^{\ii\Omega \ket{e}\bra{e}}\otimes \mathbb{I}\right)\left(\sum_{s=\pm}\ket{s}\bra{s}\otimes e^{-\ii\lambda s \hat{O}(t)}\right)\left(e^{-\ii\Omega \ket{e}\bra{e}}\otimes \mathbb{I}\right)
\end{align}
holds for any operator $\hat{O}$, we get
\begin{align}
& p(z_1,z_2,z_3)\nonumber\\
 &=\sum_{s_1,s_1',s_2,s_2',s_3,s_3'=\pm}\nonumber\\
 &\quad\prod_{i=1}^3\left(\Braket{g|s_i}\Braket{s_i|U_i(t)|z_i}\Braket{z_i|U_i(t)|s_i'} \Braket{s_i'|g}\right)\nonumber\\
&\quad \times
\left\langle g_A,\Psi \middle| e^{\ii\lambda_A\hat{\sigma}_x^{(A)}\hat{O}_A}e^{\ii \hat{O}_B(s_1,s_2,s_3)}\right.\nonumber\\
&\left.\quad\quad\quad\times e^{-\ii \hat{O}_B(s_1',s_2',s_3')}e^{-\ii\lambda_A\hat{\sigma}_x^{(A)}\hat{O}_A}\middle| g_A,\Psi\right\rangle,
\end{align}
where we have defined
\begin{align}
\hat{O}_B(s_1,s_2,s_3)\equiv \sum_{i=1}^3\lambda_{B_i}s_i\hat{O}_{B_i}.
\end{align}
A straightforward calculation shows that
\begin{align}
&\left\langle g_A,\Psi \middle| e^{\ii\lambda_A\hat{\sigma}_x^{(A)}\hat{O}_A}e^{\ii \hat{O}_B(s_1,s_2,s_3)}\right.\nonumber\\
&\left.\quad\quad\quad\times e^{-\ii \hat{O}_B(s_1',s_2',s_3')}e^{-\ii\lambda_A\hat{\sigma}_x^{(A)}\hat{O}_A}\middle| g_A,\Psi\right\rangle\\
 &=\sum_{s_A=\pm}\Braket{g|s_A}\Braket{s_A|g}\nonumber\\
 &\quad\times\left\langle \Psi\middle|e^{\ii\lambda_As_A\hat{O}_A}e^{\ii \hat{O}_B(s_1,s_2,s_3)}\right.\nonumber\\
 &\left.\quad\quad\quad\times e^{-\ii \hat{O}_B(s_1',s_2',s_3')}e^{-\ii\lambda_As_A\hat{O}_A}\middle|\Psi \right\rangle\nonumber\\
 &=\frac{1}{2} \sum_{s_A=\pm}\left\langle\Psi\middle|e^{\ii\lambda_As_A\hat{O}_A}e^{\ii \hat{O}_B(s_1,s_2,s_3)}\right.\nonumber\\
 &\left.\quad\quad\quad\quad\quad\quad\quad\times e^{-\ii \hat{O}_B(s_1',s_2',s_3')}e^{-\ii\lambda_As_A\hat{O}_A}\middle|\Psi\right\rangle
\end{align}
holds.

From the Baker–Campbell–Hausdorff (BCH) formula, if $\left[A,B\right]\propto \mathbb{I}$, it holds that
\begin{align}
 e^{A}e^{B}=e^{A+B}e^{\frac{1}{2}\left[A,B\right]},
\end{align}
implying that
\begin{align}
 e^{A}e^{B}=e^{B}e^{A}e^{\left[A,B\right]}.
\end{align}
Thus, it holds that
\begin{align}
& e^{\ii\lambda_As_A\hat{O}_A}e^{\ii \hat{O}_B(s_1,s_2,s_3)}e^{-\ii \hat{O}_B(s_1',s_2',s_3')}e^{-\ii\lambda_As_A\hat{O}_A}\nonumber\\
 &= e^{\ii\hat{O}_B(s_1,s_2,s_3)}e^{-\ii\hat{O}_B(s_1',s_2',s_3')}\nonumber\\
 &\quad\times e^{-\lambda_As_A\left[\hat{O}_A,\hat{O}_B(s_1,s_2,s_3)\right]} e^{-\lambda_As_A\left[\hat{O}_B(s_1',s_2',s_3'),\hat{O}_A\right]}\nonumber\\
 &=e^{\ii\left(\hat{O}_B(s_1,s_2,s_3)-\hat{O}_B(s_1',s_2',s_3)\right)}e^{\frac{1}{2}\left[\hat{O}_{B}(s_1,s_2,s_3),\hat{O}_B(s_1',s_2',s_3')\right]} \nonumber\\
 &\quad\times e^{-\lambda_As_A\left[\hat{O}_A,\hat{O}_B(s_1,s_2,s_3)\right]} e^{-\lambda_As_A\left[\hat{O}_B(s_1',s_2',s_3'),\hat{O}_A\right]} .
\end{align}

Now, from the BCH formula, 
\begin{align}
 &\Braket{\Psi|e^{\ii\hat{O}}|\Psi}\nonumber\\
 &=\Braket{\Psi|\exp{\left(\ii\int \text{d}^d\bm{k}\left(c(\bm{k})\hat{a}_{\bm{k}}^\dag+c(\bm{k})^*\hat{a}_{\bm{k}})\right)\right)}|\Psi}\nonumber\\
 &=\Braket{\Psi|\exp{\left(\ii\int \text{d}^d\bm{x}c(\bm{k})\hat{a}_{\bm{k}}^\dag\right)}\exp{\left(\ii\int \text{d}^d\bm{x}c(\bm{k})^*\hat{a}_{\bm{k}}\right)}|\Psi}\nonumber\\
 &\quad\times e^{-\frac{1}{2}\int \text{d}^d\bm{k}|c(\bm{k})|^2}\nonumber\\
 &= e^{-\frac{1}{2}\int \text{d}^d\bm{k}|c(\bm{k})|^2},
\end{align}
where we have introduced annihilation operators $\hat{a}_{\bm{k}}$ that annihilate the Gaussian state $\ket{\Psi}$, i.e., $\hat{a}_{\bm{k}}\ket{\Psi}=0$. On the other hand,
\begin{align}
 \Braket{\Psi|\hat{O}^2|\Psi}=\int \text{d}^d\bm{k}|c(\bm{k})|^2.
\end{align}
Thus, 
\begin{align}
 \Braket{\Psi|e^{i\hat{O}}|\Psi}=e^{-\frac{1}{2}\Braket{\Psi|\hat{O}^2|\Psi}}.
\end{align}

So far, we have shown
\begin{align}
 &\Braket{\Psi| e^{\ii\lambda_As_A\hat{O}_A}e^{\ii \hat{O}_B(s_1,s_2,s_3)}e^{-i \hat{O}_B(s_1',s_2',s_3')}e^{-i\lambda_As_A\hat{O}_A}|\Psi}\nonumber\\
 &=e^{-\frac{1}{2}\Braket{\Psi|\left(\hat{O}_B(s_1,s_2,s_3)-\hat{O}_B(s_1',s_2',s_3)\right)^2|\Psi}}\nonumber\\
 &\quad \times e^{\frac{1}{2}\Braket{\Psi|\left[\hat{O}_{B}(s_1,s_2,s_3),\hat{O}_B(s_1',s_2',s_3')\right]|\Psi}}\nonumber\\
&\quad \times  e^{-\lambda_As_A\Braket{\Psi|\left[\hat{O}_A,\hat{O}_B(s_1,s_2,s_3)\right]|\Psi}}\nonumber\\
&\quad\times e^{-\lambda_As_A\Braket{\Psi|\left[\hat{O}_B(s_1',s_2',s_3'),\hat{O}_A\right]|\Psi}} .
\end{align}
Each element can be evaluated by the same way we have done in Sec. \ref{sec_te} for $\ket{\Psi}=\ket{0}$. The first factor is calculated as follows:
\begin{align}
 &\Braket{0|\left(\hat{O}_B(s_1,s_2,s_3)-\hat{O}(s_1',s_2',s_3')\right)^2|0}\nonumber\\
 &=\sum_{i=1}^3 \lambda_{B_i}(s_i-s_i')\sum_{j=1}^3 \lambda_{B_j}(s_j-s_j')\nonumber\\
 &\quad\times \int \text{d}^d\bm{x}\text{d}^d\bm{y}\, v_i^{(1)}(\bm{x}) v_j^{(1)}(\bm{y})\int \frac{\text{d}^d\bm{k}}{(2\pi)^d 2|\bm{k}|}e^{\ii\bm{k}\cdot(\bm{x}-\bm{y})}\nonumber\\
 &=\sum_{i=1}^3 \lambda_{B_i}(s_i-s_i')\sum_{j=1}^3 \lambda_{B_j}(s_j-s_j')\nonumber\\
 &\quad\times\int\frac{\text{d}^d\bm{k}}{(2\pi)^d 2|\bm{k}|}\tilde{v}_i^{(1)}(\bm{k})\tilde{v}_j^{(1)}(\bm{k})^*.
\end{align}
Since the operators $\hat{O}_{B_i}$ commute with each other,
\begin{align}
& \Braket{\Psi|\left[\hat{O}_B(s_1,s_2,s_3),\hat{O}_{B}(s_1',s_2',s_3')\right]|\Psi}=0
\end{align}
holds. Introducing $\Delta t\equiv t_{\mathrm{enc.}}-t_{\mathrm{dec.}}$, we get
\begin{align}
 &\lambda_As_A \Braket{\Psi|\left[\hat{O}_B(s_1,s_2,s_3),\hat{O}_A\right]|\Psi}\nonumber\\
 &=\lambda_A s_A \sum_{i=1}^3 \lambda_{B_i}s_{B_i} \nonumber\\
 &\quad\times\int \text{d}^d\bm{x}\text{d}^d\bm{y}\, v_{B_i}(\bm{x})v_{A}(\bm{y})\nonumber\\
 &\quad\times 2\mathrm{Im}\left(\int \frac{\text{d}^d\bm{k}}{(2\pi)^d 2|\bm{k}|}e^{-\ii|\bm{k}|\Delta t} e^{i\bm{k}\cdot (\bm{x}-\bm{y})}\right)\nonumber\\
 & =\lambda_A s_A \sum_{i=1}^3 \lambda_{B_i}s_{B_i}\nonumber\\
 &\quad \times 2\mathrm{Im}\left(\int \frac{\text{d}^d\bm{k}}{(2\pi)^d 2|\bm{k}|}e^{-\ii|\bm{k}|\Delta t}\tilde{v}_{B_i}(\bm{k})\tilde{v}_{A}(\bm{k})^*\right)
\end{align}
and
\begin{align}
&  e^{-\lambda_As_A\Braket{\Psi|\left[\hat{O}_A,\hat{O}_B(s_1,s_2,s_3)\right]|\Psi}}\nonumber\\
&\quad\times e^{-\lambda_As_A\Braket{\Psi|\left[\hat{O}_B(s_1',s_2',s_3'),\hat{O}_A\right]|\Psi}}\nonumber\\
 &=\exp\left(2\lambda_A s_A \sum_{i=1}^3 \lambda_{B_i}\left(s_{B_i}-s_{B_i}'\right)\right.\nonumber\\
 &\left.\quad\times \mathrm{Im}\left(\int \frac{\text{d}^d\bm{k}}{(2\pi)^d 2|\bm{k}|}e^{-\ii|\bm{k}|\Delta t}\tilde{v}_{B_i}(\bm{k})\tilde{v}_{A}(\bm{k})^*\right)\right) .
\end{align}
Thus, we have shown the following formula:
\begin{align}
&  p_{\lambda_A}(z_1,z_2,z_3)\nonumber\\
 &=\frac{1}{2}\sum_{s_A=\pm}\sum_{s_1,s_2,s_3,s_1',s_2',s_3'=\pm}\nonumber\\
 &\quad\times\Braket{g|s_i}\Braket{s_i|U_i(t)|z_i}\Braket{z_i|U_i(t)^\dag|s_i'} \Braket{s_i'|g})\nonumber\\
&\quad \times \exp\left(-\frac{1}{2}\sum_{i=1}^3 \lambda_{B_i}(s_i-s_i')\sum_{j=1}^3 \lambda_{B_j}(s_j-s_j')\right.\nonumber\\
&\left.\quad\quad\times \int\frac{\text{d}^d\bm{k}}{(2\pi)^d 2|\bm{k}|}\tilde{v}_{B_i}^{(1)}(\bm{k})\tilde{v}_{B_j}^{(1)}(\bm{k})^*\right)\nonumber\\
&\quad \times \exp\left(2\lambda_A s_A \sum_{i=1}^3 \lambda_{B_i}\left(s_{i}-s_{i}'\right)\right.\nonumber\\
&\quad\left.\quad\times\mathrm{Im}\left(\int \frac{\text{d}^d\bm{k}}{(2\pi)^d 2|\bm{k}|}e^{-\ii|\bm{k}|\Delta t}\tilde{v}_{B_i}^{(1)}(\bm{k})\tilde{v}_{A}^{(1)}(\bm{k})^*\right)\right).
\end{align}
The first factor of the summand is given by
\begin{align}
 &\Braket{g|s_i}\Braket{s_i|U_i(t)|z_i}\Braket{z_i|U_i(t)|s_i'} \Braket{s_i'|g}\nonumber\\
 &=
\begin{cases}
 \frac{1}{4}s_is'_i\quad& (\text{if } z_i=e)\\
 \frac{1}{4}\quad &(\text{if } z_i=g)
\end{cases}.
\end{align}

\section{The proof of commutativity of $f_\Psi(\hat{O}_2')$ and $(\hat{Q}_1,\hat{P}_1)$}\label{app_com}
Let us first show the following lemma: For any operators $\hat{O}$ and $\hat{O}'$ which are given by linear combinations of canonical variables, it holds that
\begin{align}
    \left[\hat{O},f_\Psi(\hat{O}')\right]=-\left[f_\Psi(\hat{O}),\hat{O}'\right].
\end{align}
Proof: Let $\hat{\Gamma}(\bm{x})\equiv (\hat{\phi}(t,\bm{x}),\hat{\Pi}(t,\bm{x}))^{\mathrm{T}}$ be the set of canonical variables. Let us define
\begin{align}
    \Omega(\bm{x},\bm{y})&\equiv \frac{1}{\ii} \Braket{\Psi|\left[\hat{\Gamma}(\bm{x}),\hat{\Gamma}^{\mathrm{T}}(\bm{x})\right]|\Psi}\nonumber\\
    &=
\begin{pmatrix}
0&\delta^{(d)}(\bm{x}-\bm{y})\\
-\delta^{(d)}(\bm{x}-\bm{y})&0
\end{pmatrix}\\
    M(\bm{x},\bm{y})&\equiv \mathrm{Re}\left(\Braket{\Psi|\hat{\Gamma}(\bm{x}),\hat{\Gamma}^{\mathrm{T}}(\bm{x})|\Psi}\right).
\end{align}
We can interpret these functions as a matrix with continuous indices. For example,
\begin{align}
    \Omega^2(\bm{x},\bm{y})&\equiv \int \text{d}^d\bm{z} \Omega(\bm{x},\bm{z})\Omega(\bm{z},\bm{y})\nonumber\\
    &=-
    \begin{pmatrix}
    \delta^{(d)}(\bm{x}-\bm{y})&0\\
    0&\delta^{(d)}(\bm{x}-\bm{y})
\end{pmatrix}.
\end{align}
In this notation, the operator can be expressed by inner product:
\begin{align}
    \hat{O}\equiv V^{\mathrm{T}}\hat{\Gamma}\equiv \int \text{d}^d\bm{x}\left(v^{(1)}\hat{\phi}(t,\bm{x})+v^{(2)}(\bm{x}))\hat{\Pi}(t,\bm{x})\right),
\end{align}
where $V(\bm{x})\equiv (v^{(1)}(\bm{x}),v^{(2)}(\bm{x}))^{\mathrm{T}}$. Similarly, the other operator is expressed as $\hat{O}'\equiv V^{'\mathrm{ T}}\hat{\Gamma}$. The map $f_\Psi$ in Eq.\eqref{eq_lin_map} can be rewritten as
\begin{align}
    f_\Psi(\hat{O})=(-2\Omega M V)^{\mathrm{T}}\hat{\Gamma}.
\end{align}
Since $\Omega^\mathrm{T}=-\Omega$, it holds that
\begin{align}
    \frac{1}{\ii}\Braket{\Psi|\left[\hat{O},f_\Psi(\hat{O}')\right]|\Psi}    &=V^{\mathrm{T}}\Omega (-2\Omega M V')\nonumber\\
    &=2V^\mathrm{T}M V'\nonumber\\
    &=-\frac{1}{\ii}\Braket{\Psi|\left[f_\Psi(\hat{O}),\hat{O}'\right]|\Psi},
\end{align}
which concludes the proof of lemma.

The commutativity follows immediately from the lemma. Since $\hat{O}_2'$ commutes with both $\hat{Q}_1$ and $\hat{P}_1$, we get
\begin{align}
    \left[f_\Psi(\hat{O}'_2),\hat{Q}_1\right]&=-\left[\hat{O}'_2,f_\Psi(\hat{Q}_1)\right]=-\left[\hat{O}'_2,\hat{P}_1\right]=0.\\
    \left[f_\Psi(\hat{O}'_2),\hat{P}_1\right]&=-\left[\hat{O}'_2,f_\Psi(\hat{P}_1)\right]=\left[\hat{O}'_2,\hat{Q}_1\right]=0.
\end{align}

\section{Plots for weighting functions in Sec. \ref{sec_mqic}}\label{app_wf}
\begin{figure}[H]
    \centering
      \includegraphics[keepaspectratio, scale=0.60]{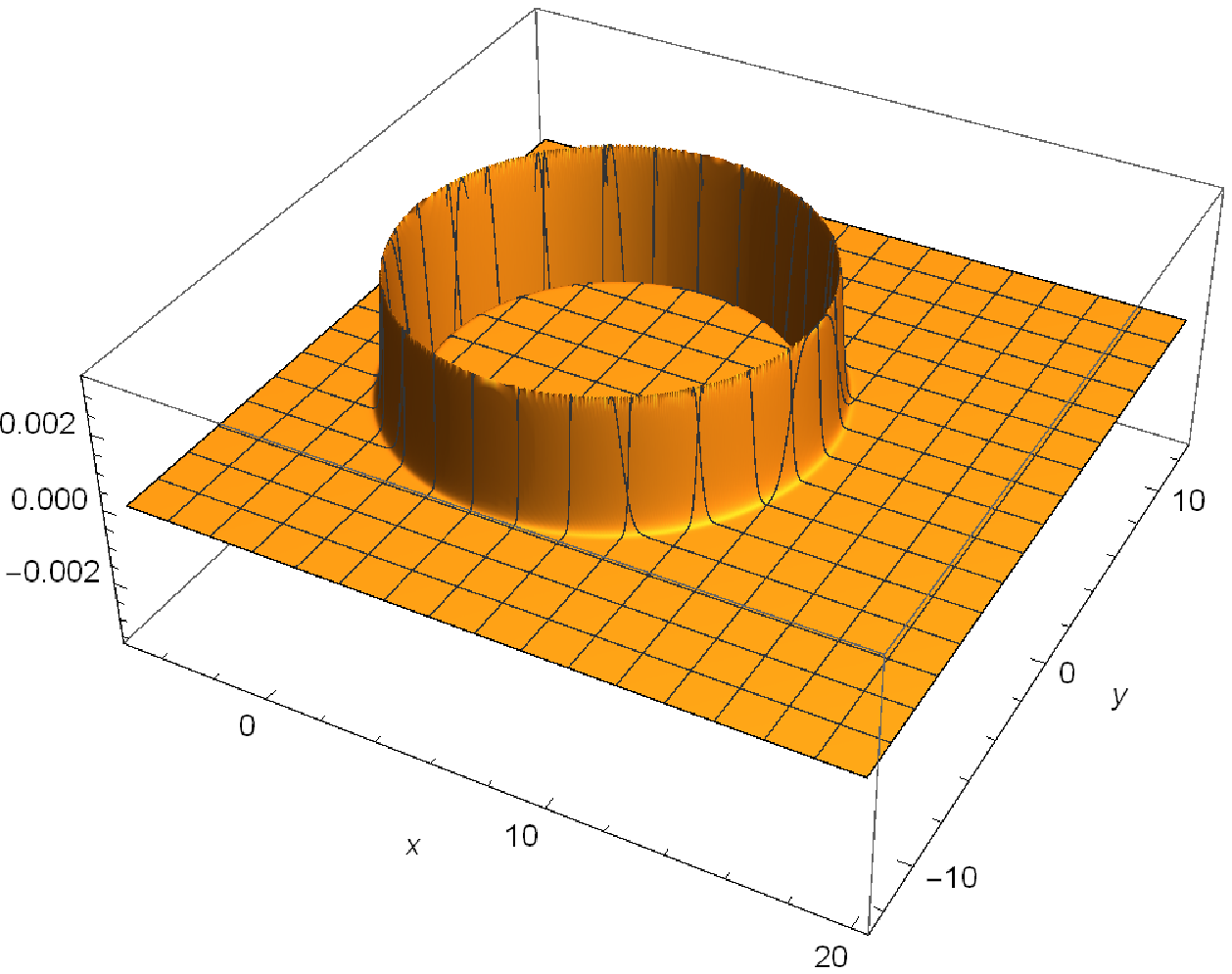}
       \caption{$\sigma^2 F_1^{(1)}(t,x,y,0)$ at $t=8$ for $d=3$.}
       \label{fig_f11_3d}
\end{figure}
\begin{figure}[H]
    \centering
    \includegraphics[keepaspectratio, scale=0.60]{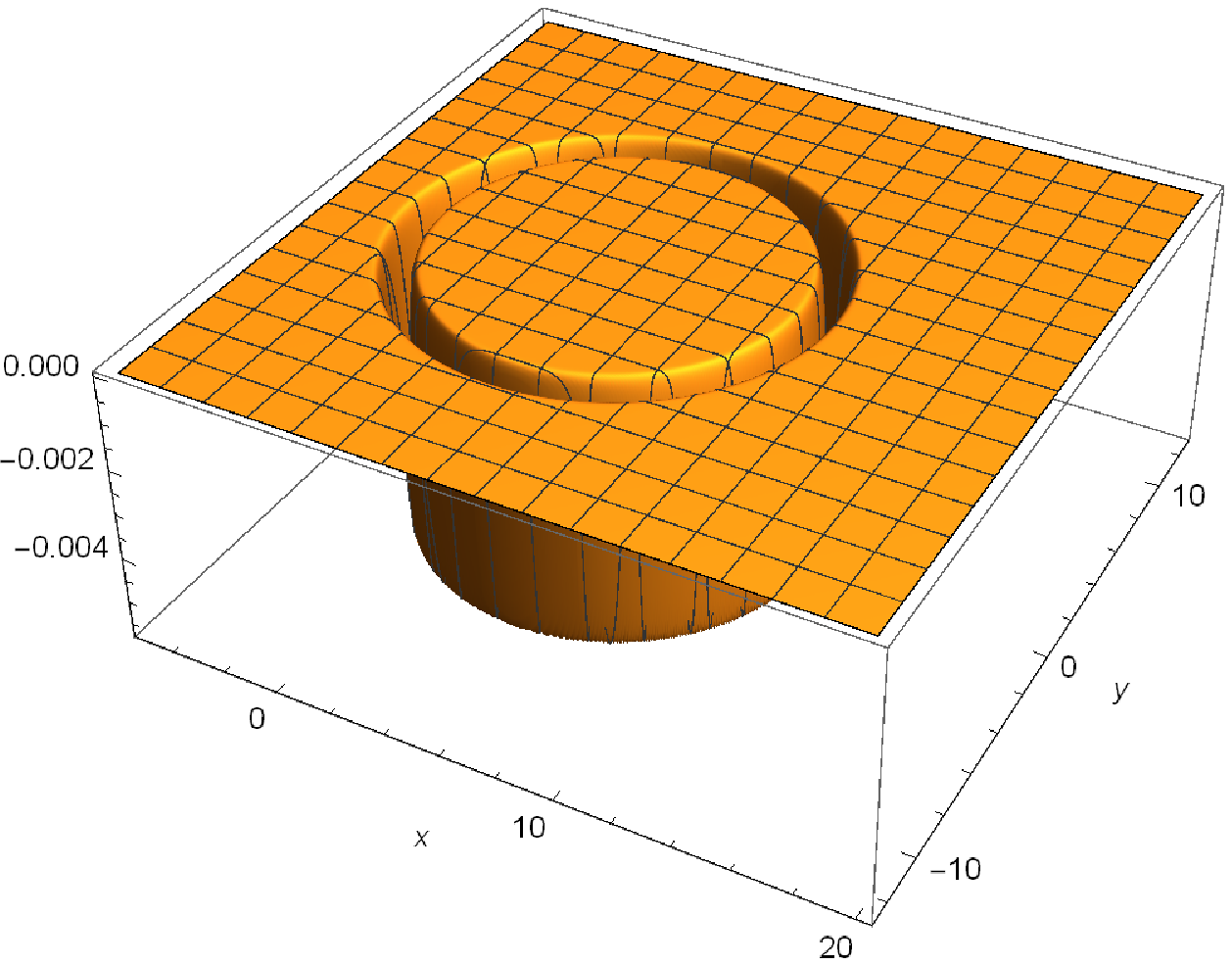}
    \caption{$\sigma F_1^{(2)}(t,x,y,0)$ at $t=8$ for $d=3$.}
    \label{fig_f12_3d}
\end{figure}
\begin{figure}[H]
    \centering
     \includegraphics[keepaspectratio, scale=0.60]{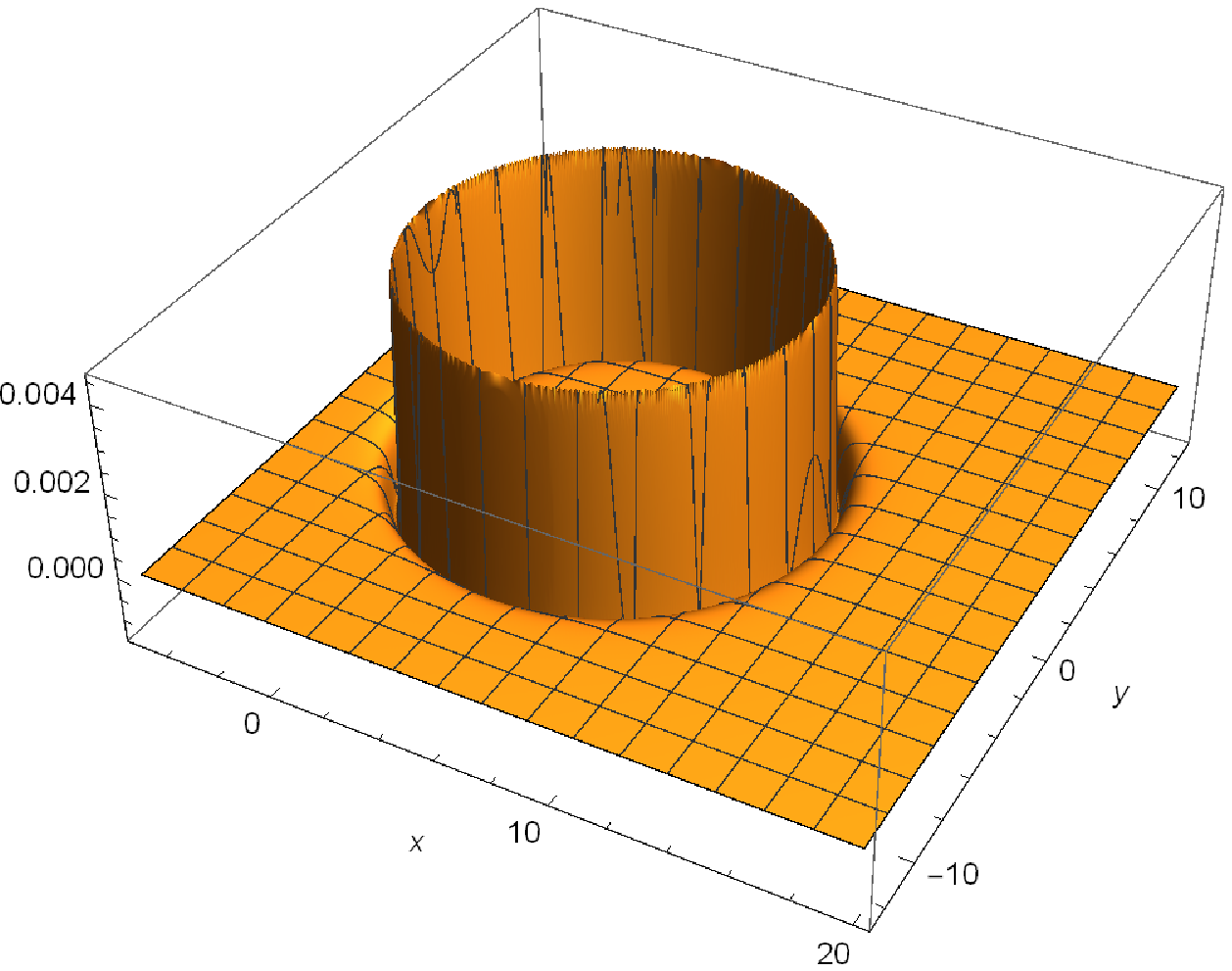}
    \caption{$\sigma^2 G_1^{(1)}(t,x,y,0)$ at $t=8$ for $d=3$.}
    \label{fig_g11_3d}
\end{figure}
\begin{figure}[H]
    \centering
     \includegraphics[keepaspectratio, scale=0.60]{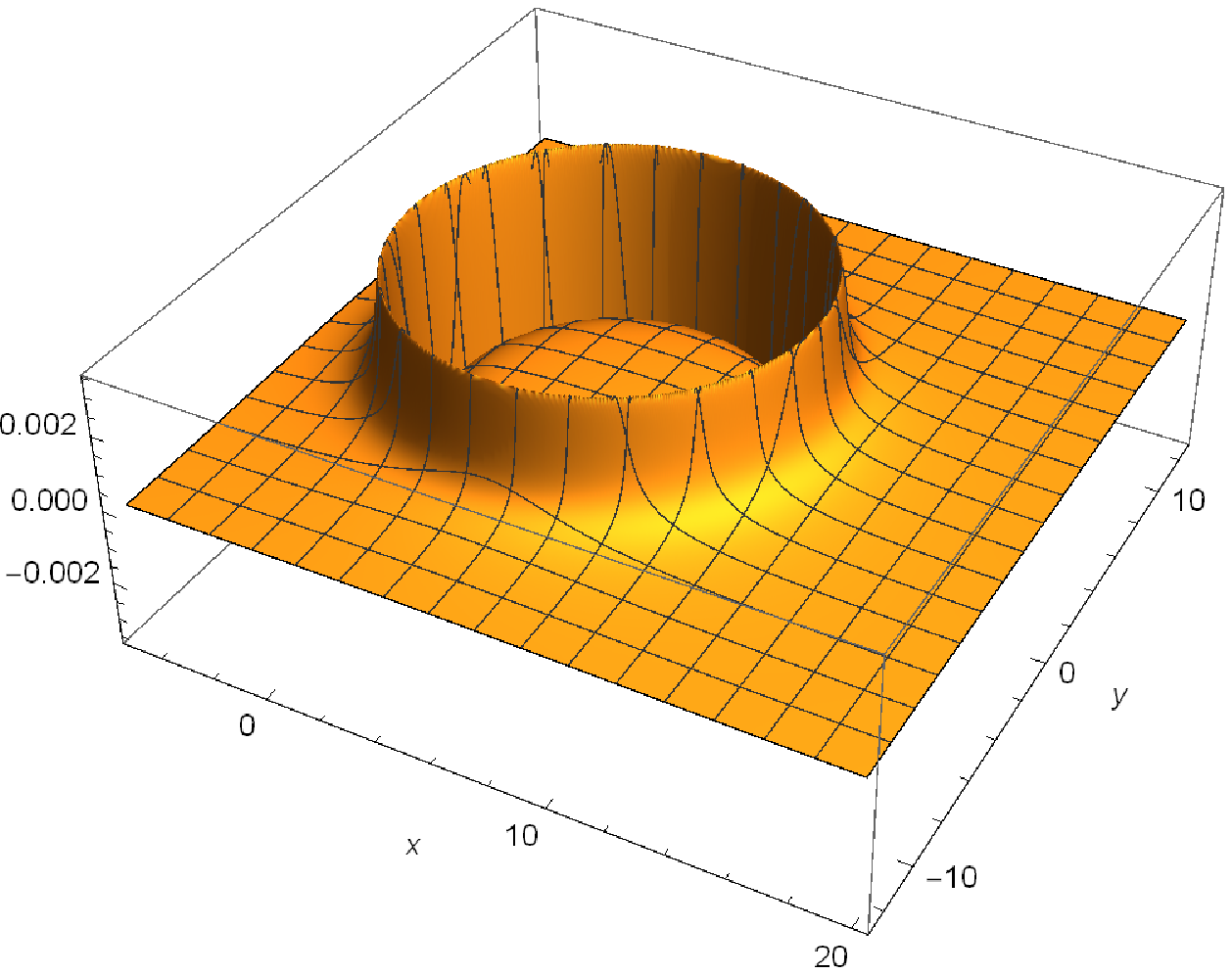}
    \caption{$\sigma G_1^{(2)}(t,x,y,0)$ at $t=8$ for $d=3$.}
    \label{fig_g12_3d}
\end{figure}

\begin{figure}[H]
    \centering
    \includegraphics[keepaspectratio, scale=0.60]{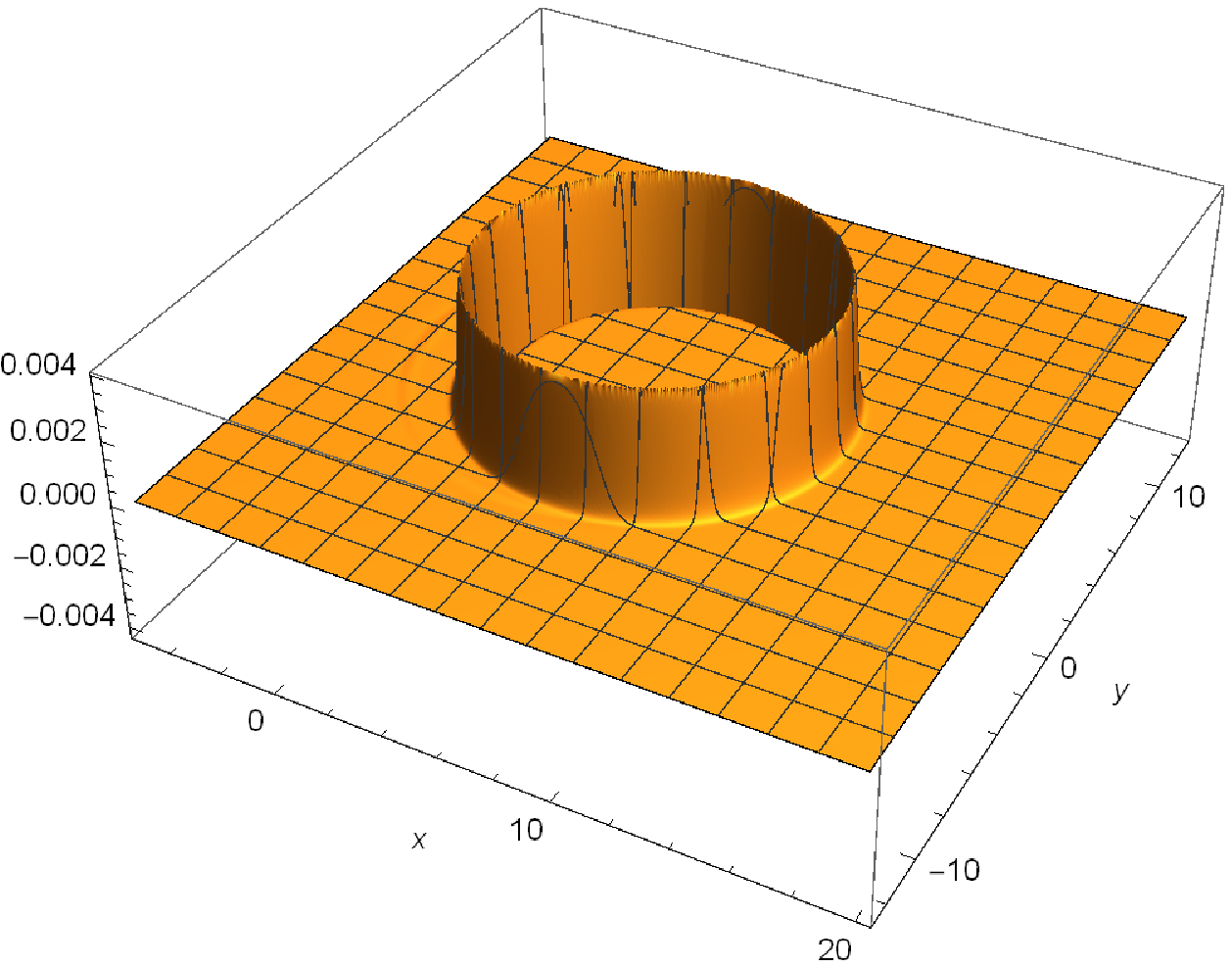}
    \caption{$\sigma^2 F_2^{(1)}(t,x,y,0)$ at $t=8$ for $d=3$.}
    \label{fig_f21_3d}
\end{figure}
\begin{figure}[H]
    \centering
    \includegraphics[keepaspectratio, scale=0.60]{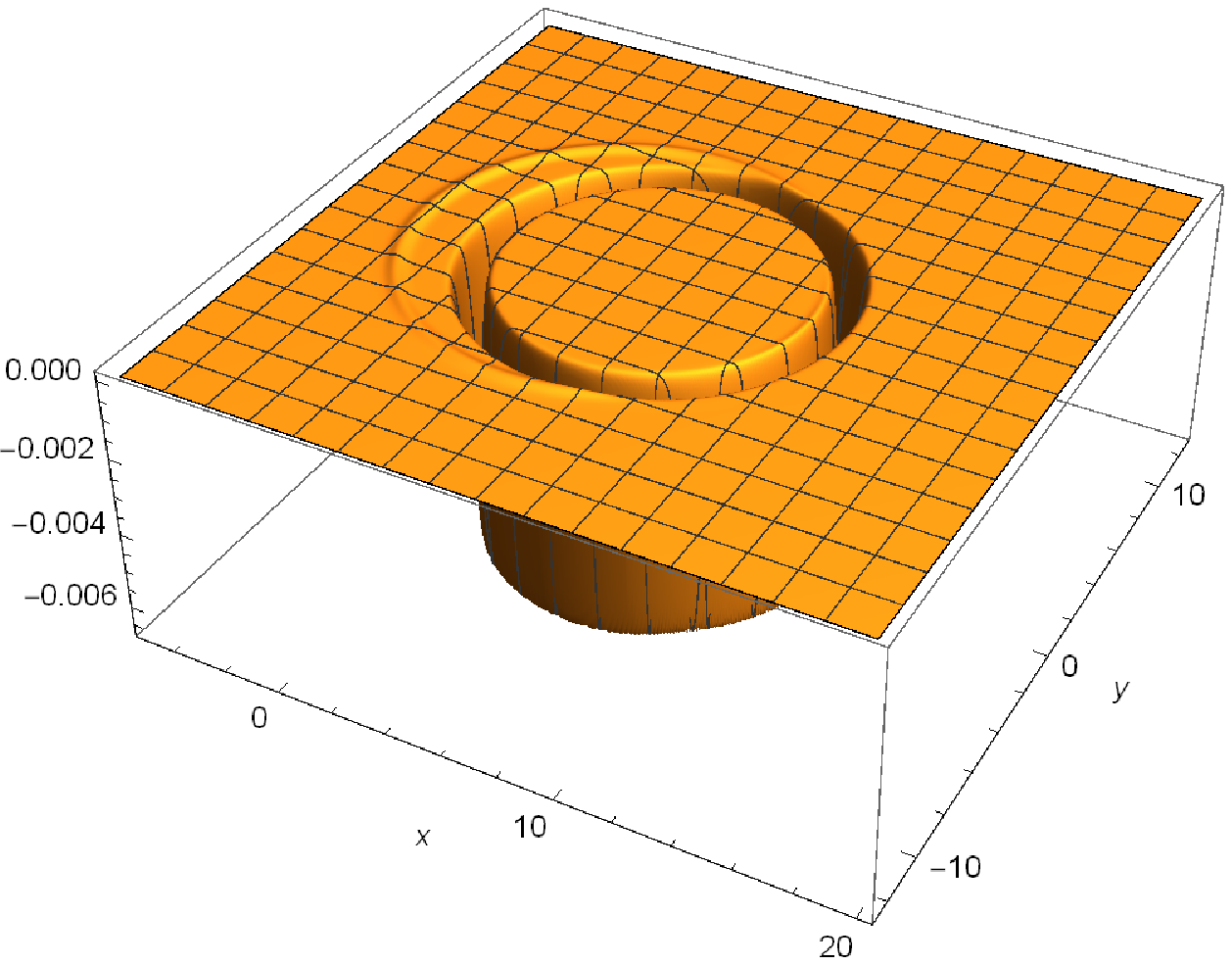}
    \caption{$\sigma F_2^{(2)}(t,x,y,0)$ at $t=8$ for $d=3$.}
    \label{fig_f22_3d}
\end{figure}
\begin{figure}[H]
    \centering
    \includegraphics[keepaspectratio, scale=0.60]{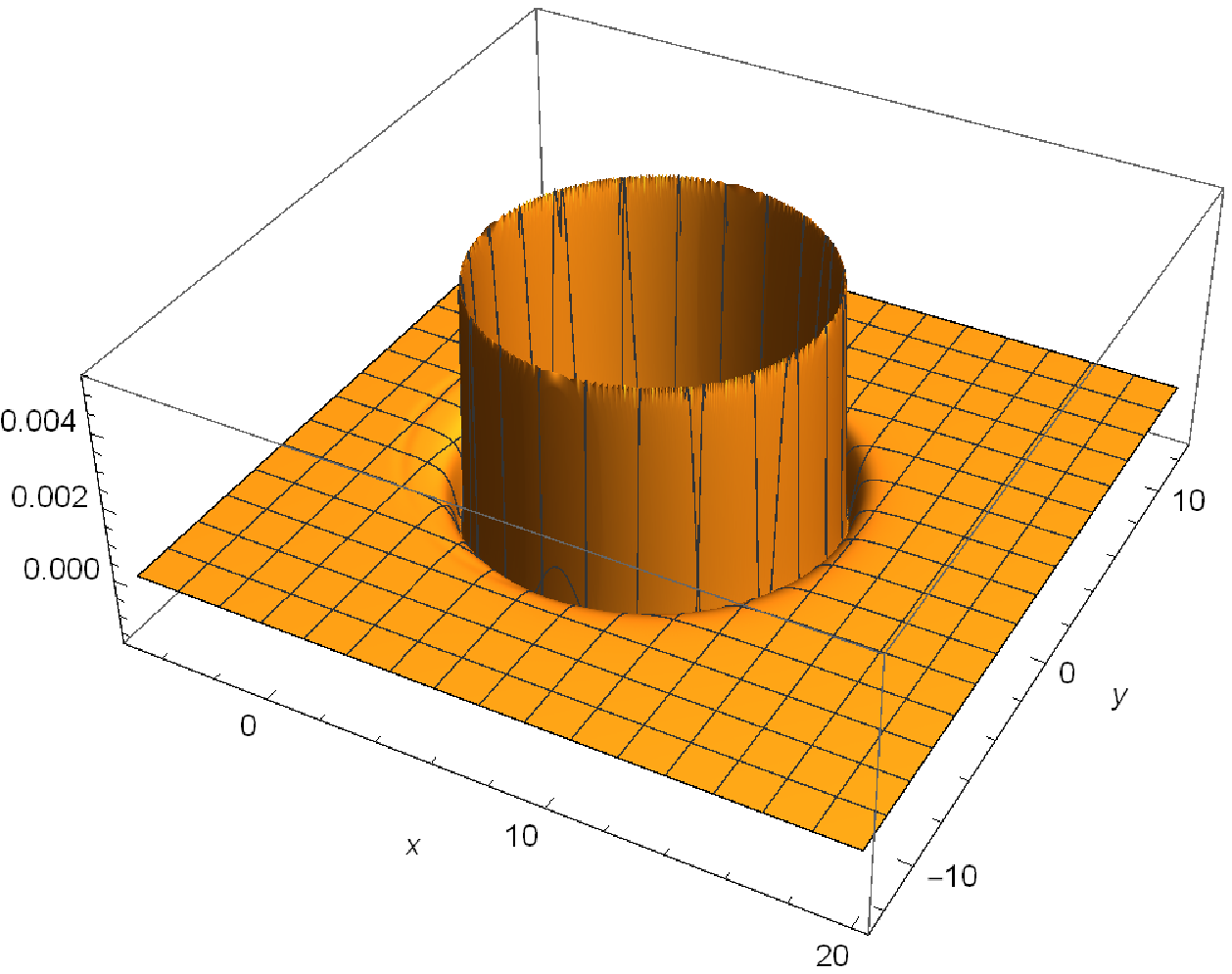}
    \caption{$\sigma^2 G_2^{(1)}(t,x,y,0)$ at $t=8$ for $d=3$.}
    \label{fig_g21_3d}
\end{figure}
\begin{figure}[H]
    \centering
    \includegraphics[keepaspectratio, scale=0.60]{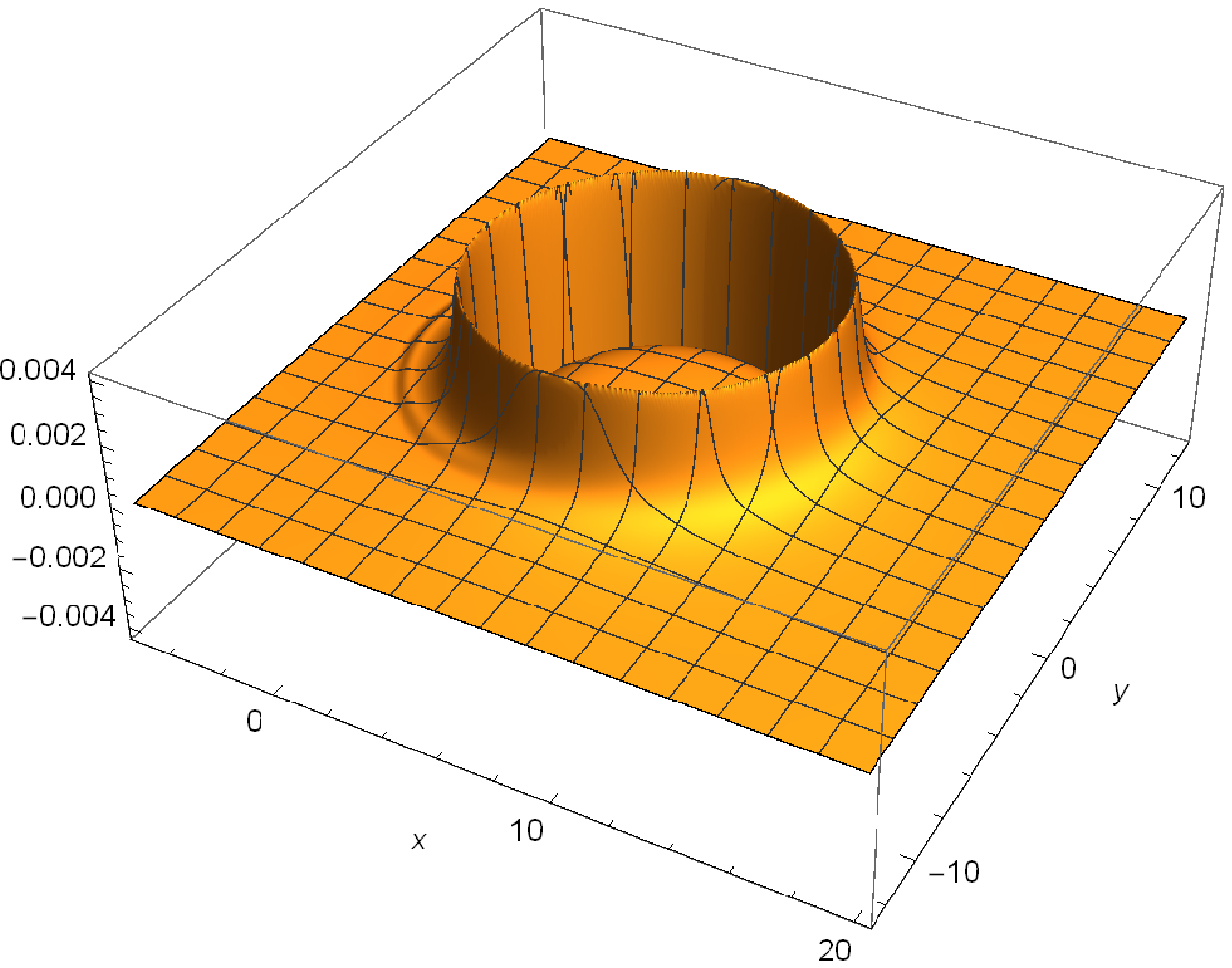}
    \caption{$\sigma G_2^{(2)}(t,x,y,0)$ at $t=8$ for $d=3$.}
    \label{fig_g22_3d}
\end{figure}
      
\begin{figure}[H]
    \centering
    \includegraphics[keepaspectratio, scale=0.60]{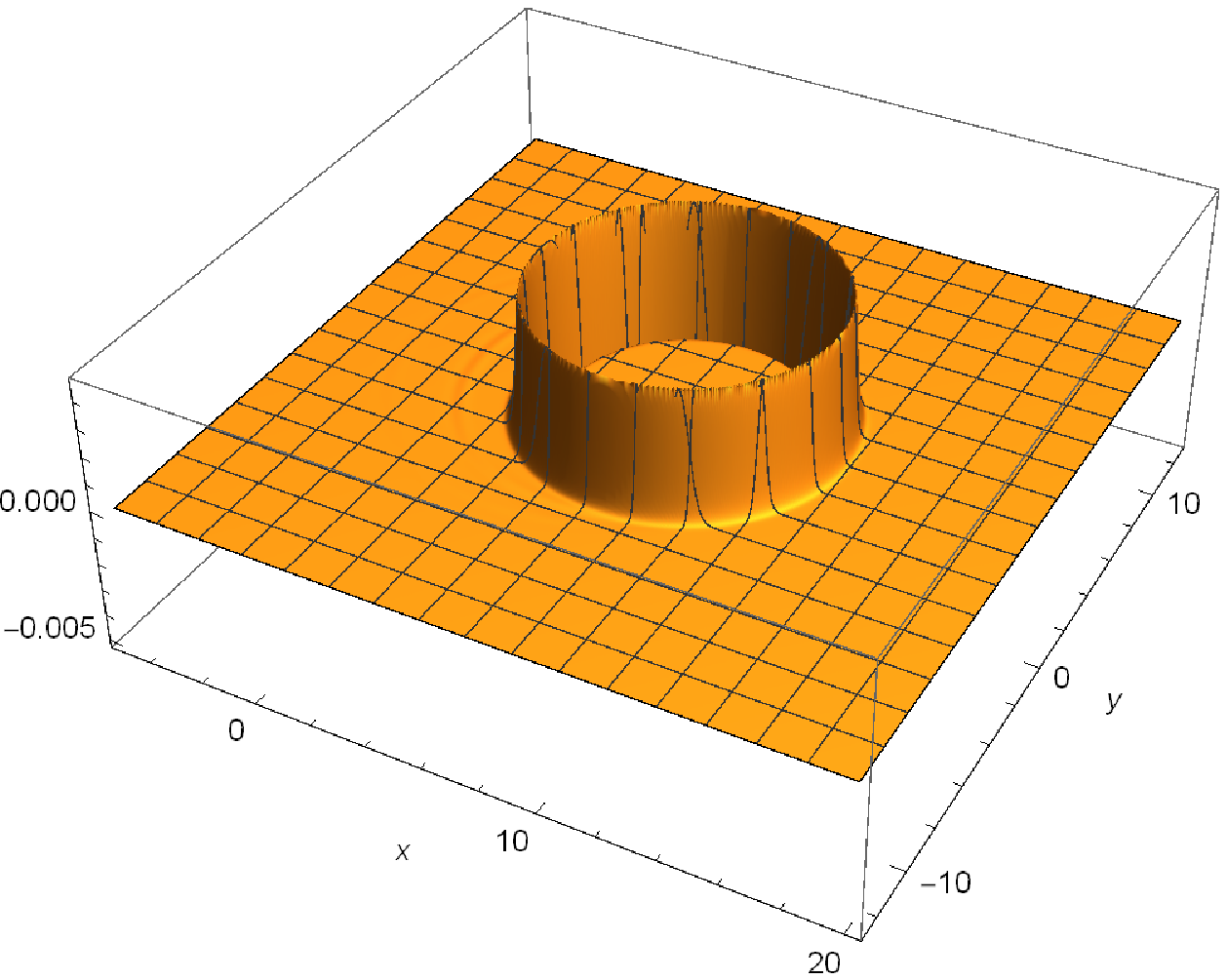}
    \caption{$\sigma^2 F_3^{(1)}(t,x,y,0)$ at $t=8$ for $d=3$.}
    \label{fig_f31_3d}
\end{figure}
\begin{figure}[H]
    \centering
    \includegraphics[keepaspectratio, scale=0.60]{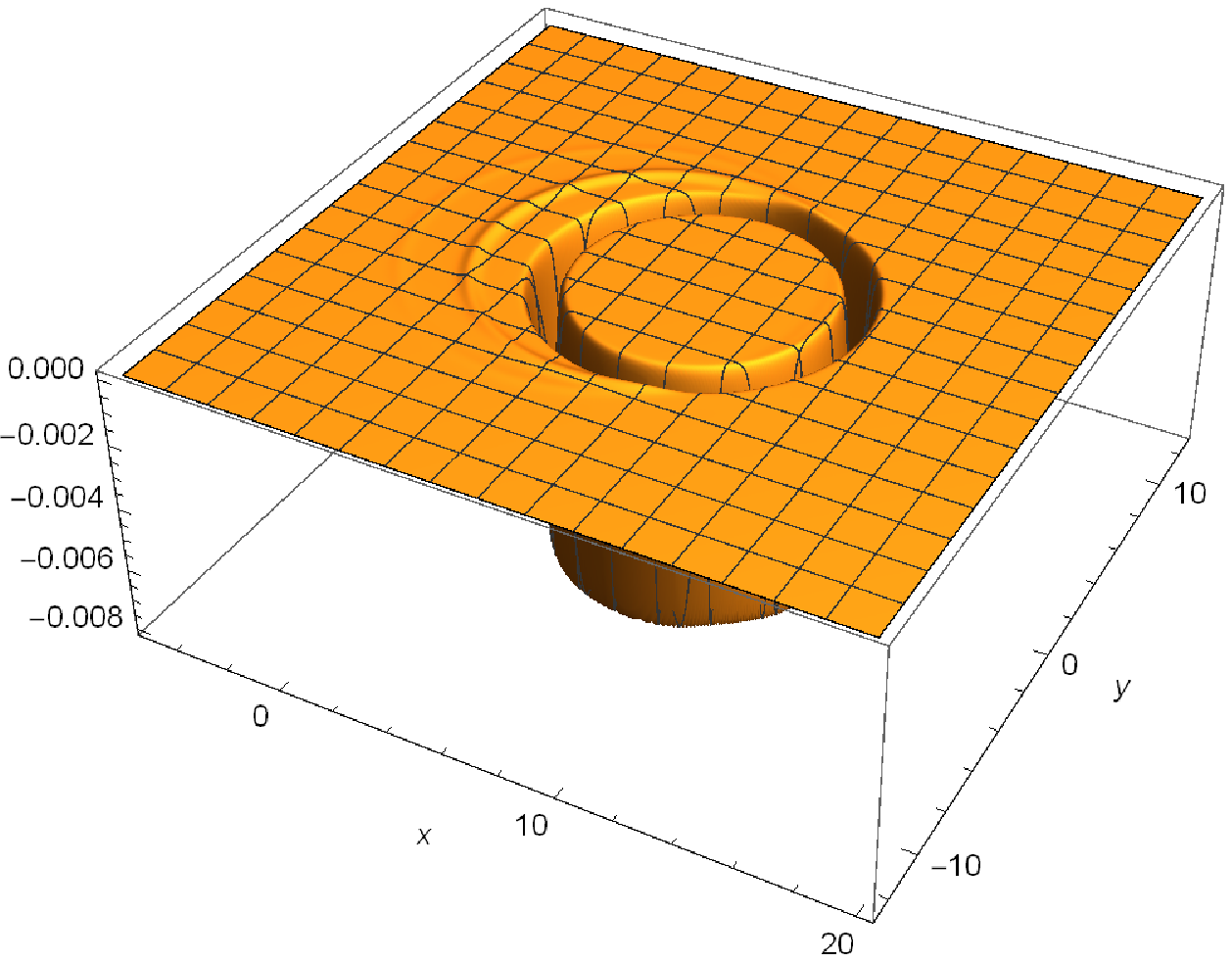}
    \caption{$\sigma F_3^{(2)}(t,x,y,0)$ at $t=8$ for $d=3$.}
    \label{fig_f32_3d}
\end{figure}
\begin{figure}[H]
    \centering
    \includegraphics[keepaspectratio, scale=0.60]{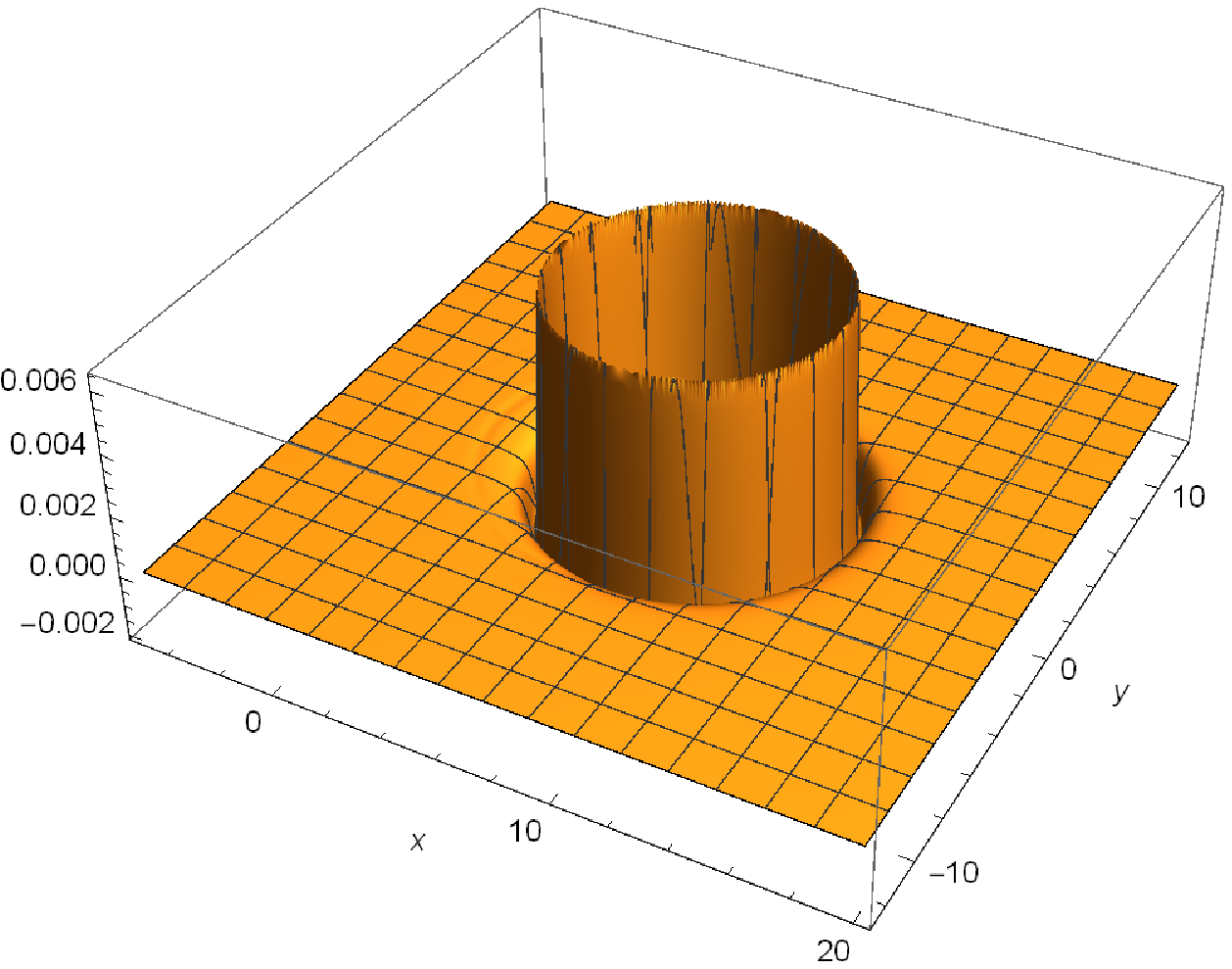}
    \caption{$\sigma^2 G_3^{(1)}(t,x,y,0)$ at $t=8$ for $d=3$.}
    \label{fig_g31_3d}
\end{figure}
\begin{figure}[H]
    \centering
    \includegraphics[keepaspectratio, scale=0.57]{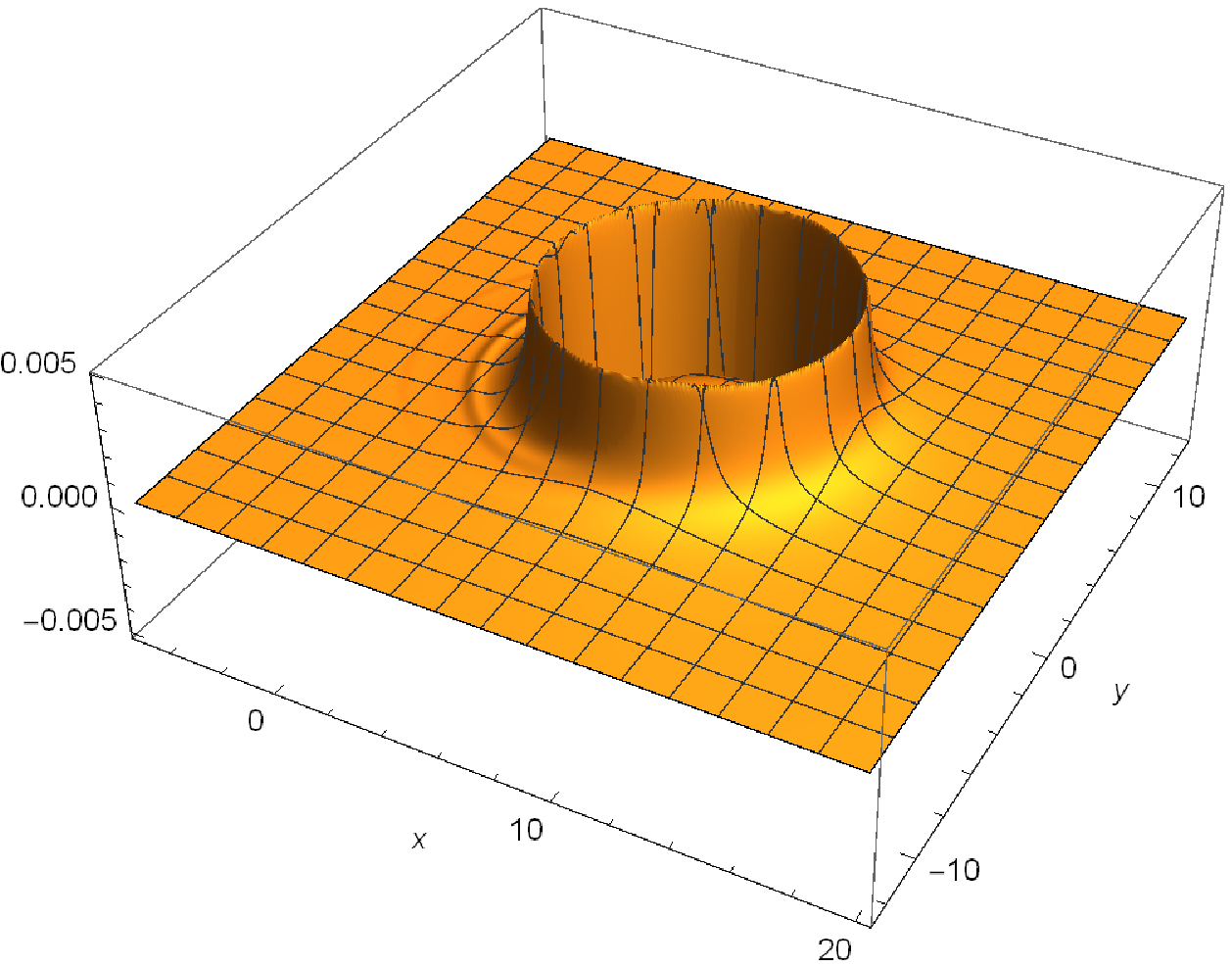}
    \caption{$\sigma G_3^{(2)}(t,x,y,0)$ at $t=8$ for $d=3$.}
    \label{fig_g32_3d}
\end{figure}

\begin{figure}[H]
    \centering
    \includegraphics[keepaspectratio, scale=0.57]{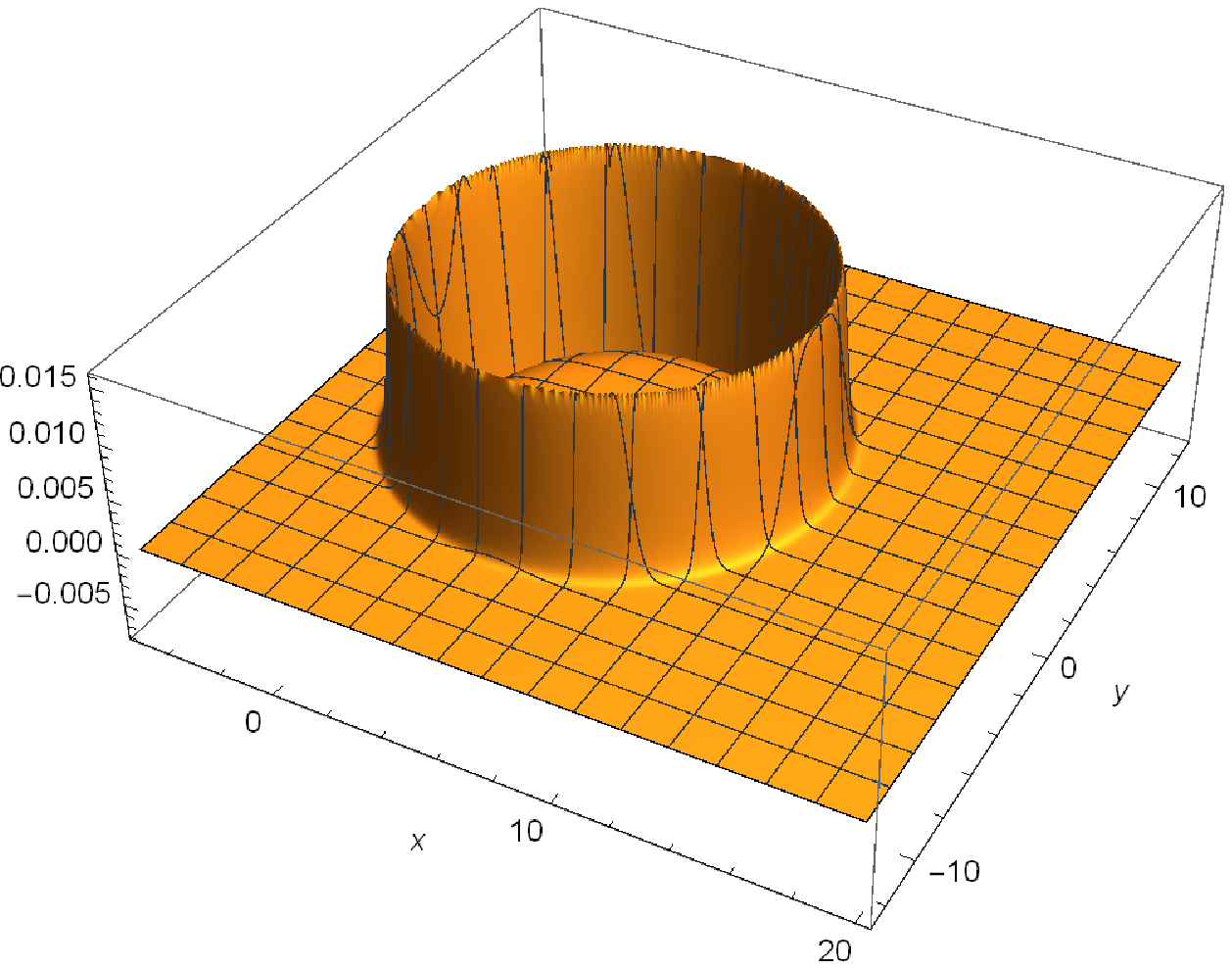}
    \caption{$\sigma^{3/2} F_1^{(1)}(t,x,y)$ at $t=8$ for $d=2$.}
    \label{fig_f11_2d}
\end{figure}
\begin{figure}[H]
    \centering
    \includegraphics[keepaspectratio, scale=0.57]{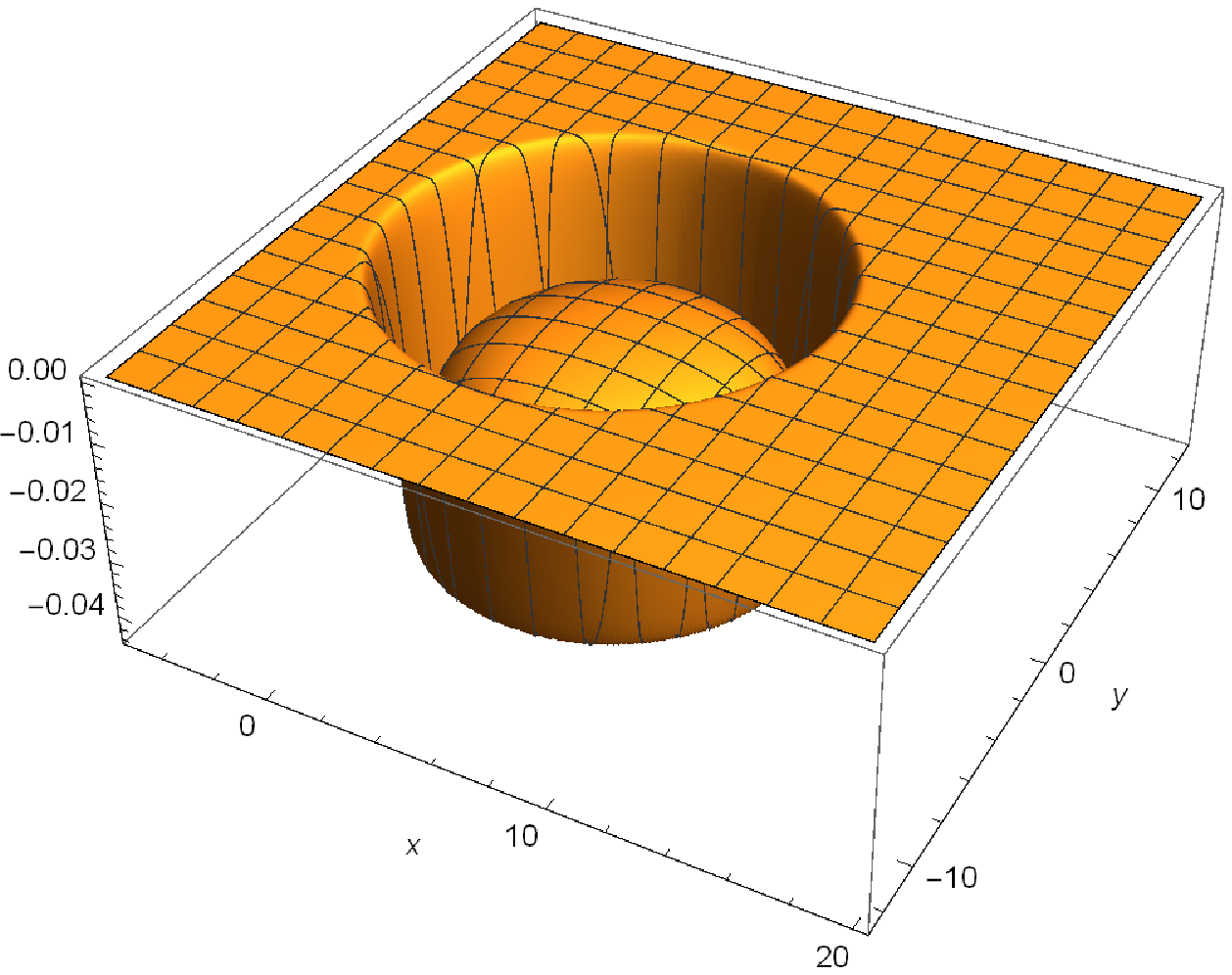}
    \caption{$\sigma^{1/2} F_1^{(2)}(t,x,y)$ at $t=8$ for $d=2$.}
    \label{fig_f12_2d}
\end{figure}
\begin{figure}[H]
    \centering
    \includegraphics[keepaspectratio, scale=0.57]{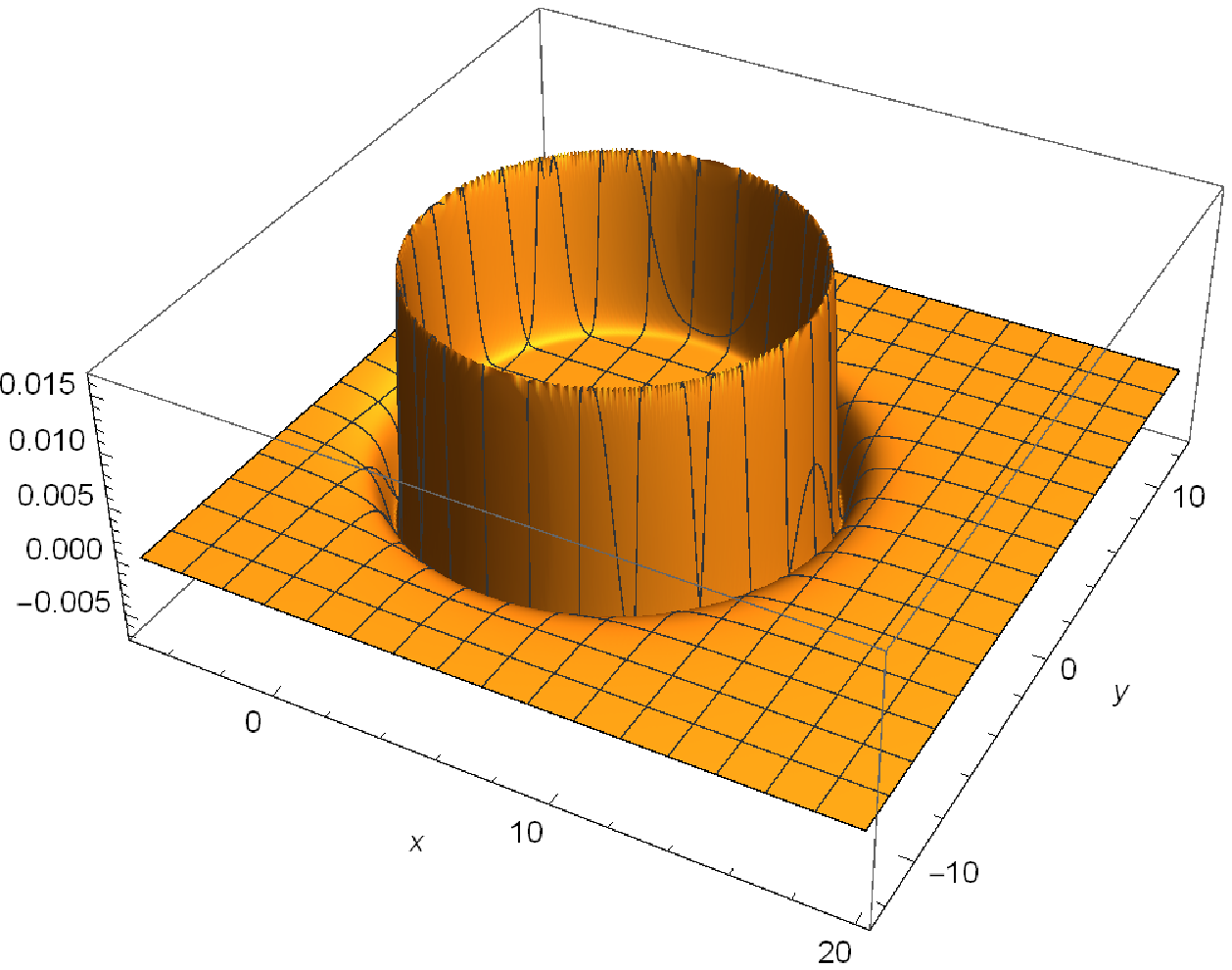}
    \caption{$\sigma^{3/2} G_1^{(1)}(t,x,y)$ at $t=8$ for $d=2$.}
    \label{fig_g11_2d}
\end{figure}
\begin{figure}[H]
    \centering
    \includegraphics[keepaspectratio, scale=0.57]{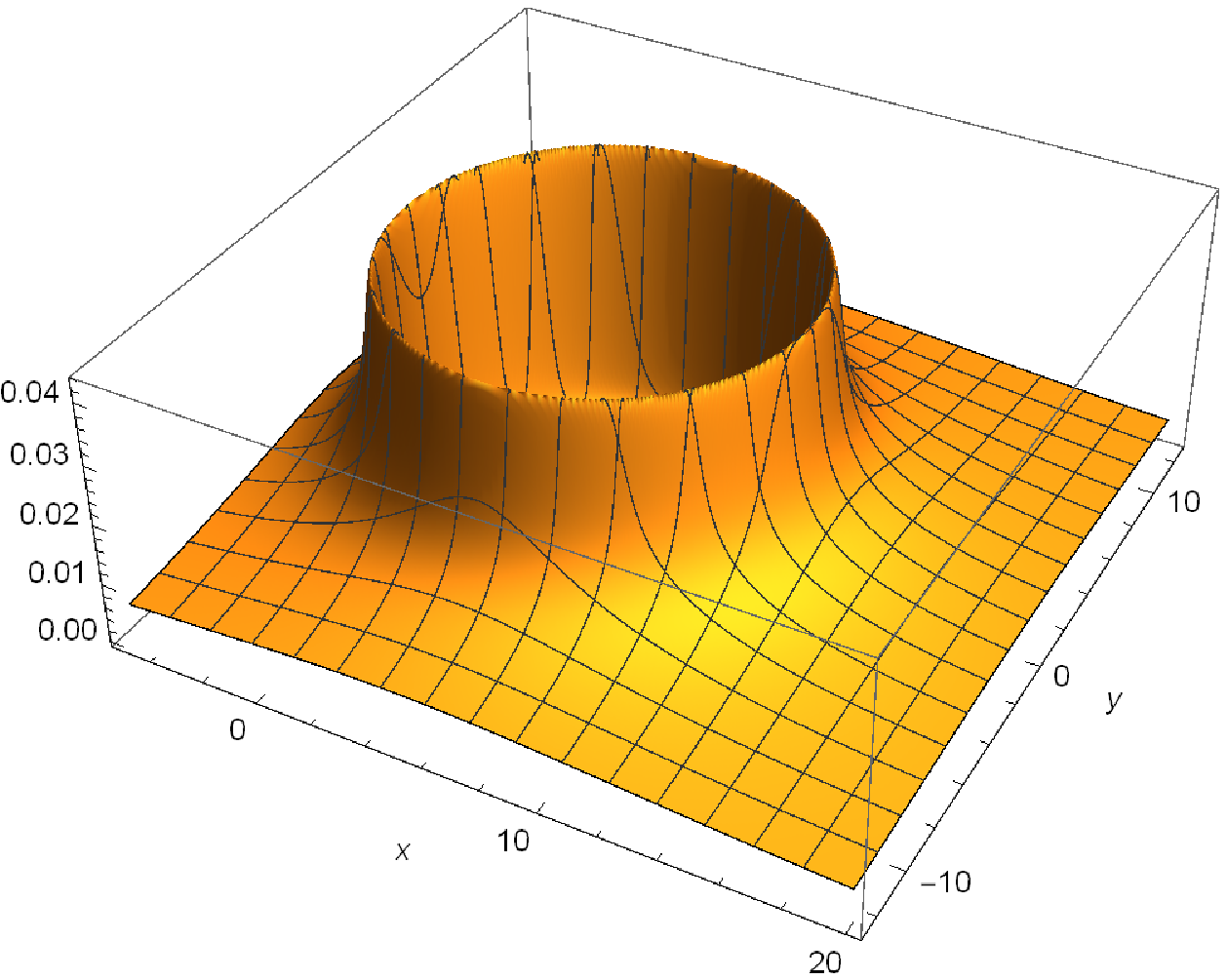}
    \caption{$\sigma^{1/2} G_1^{(2)}(t,x,y)$ at $t=8$ for $d=2$.}
    \label{fig_g12_2d}
\end{figure}
\begin{figure}[H]
    \centering
      \includegraphics[keepaspectratio, scale=0.57]{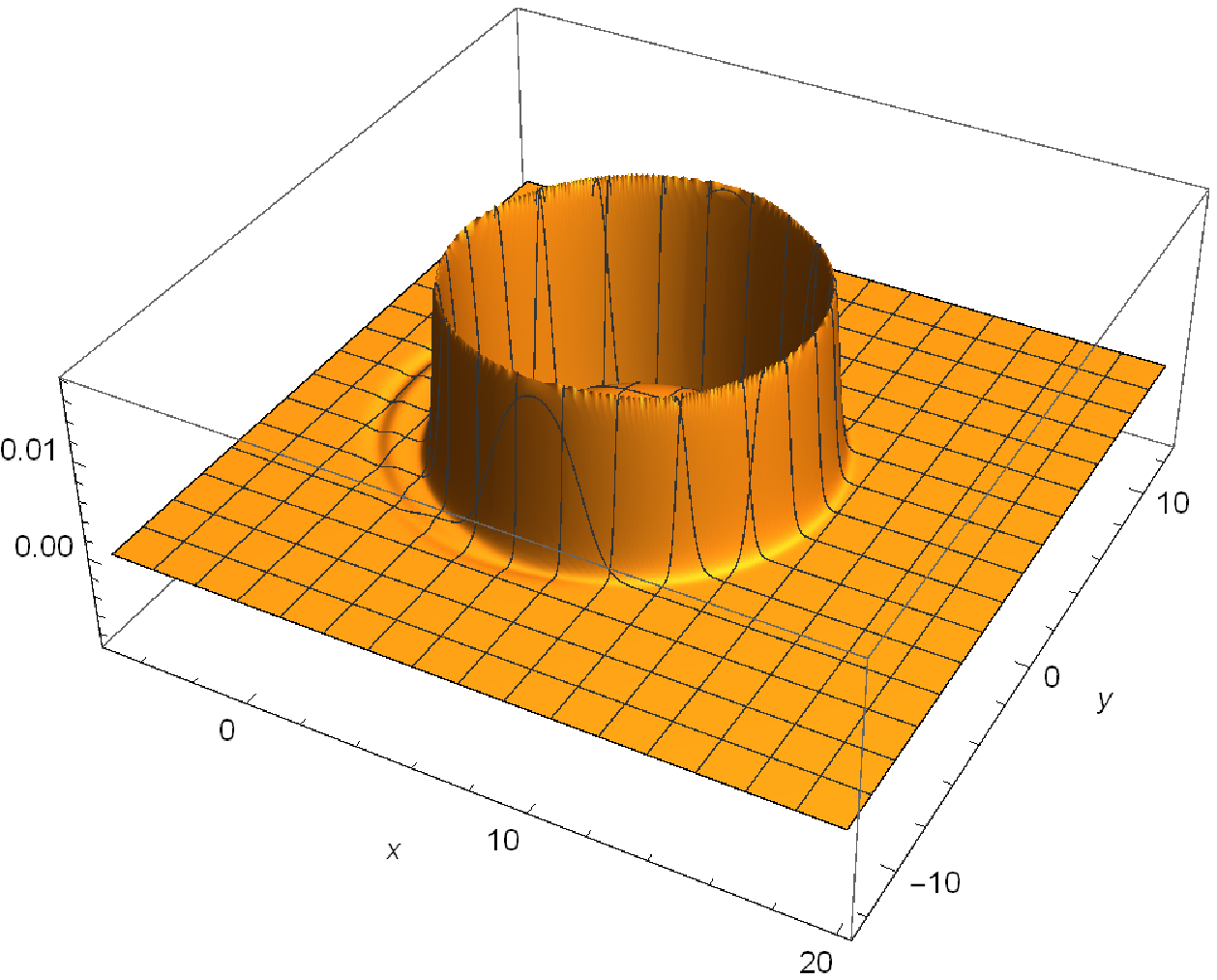}
        \caption{$\sigma^{3/2} F_2^{(1)}(t,x,y)$ at $t=8$ for $d=2$.}
        \label{fig_f21_2d}
\end{figure}
\begin{figure}[H]
    \centering
    \includegraphics[keepaspectratio, scale=0.57]{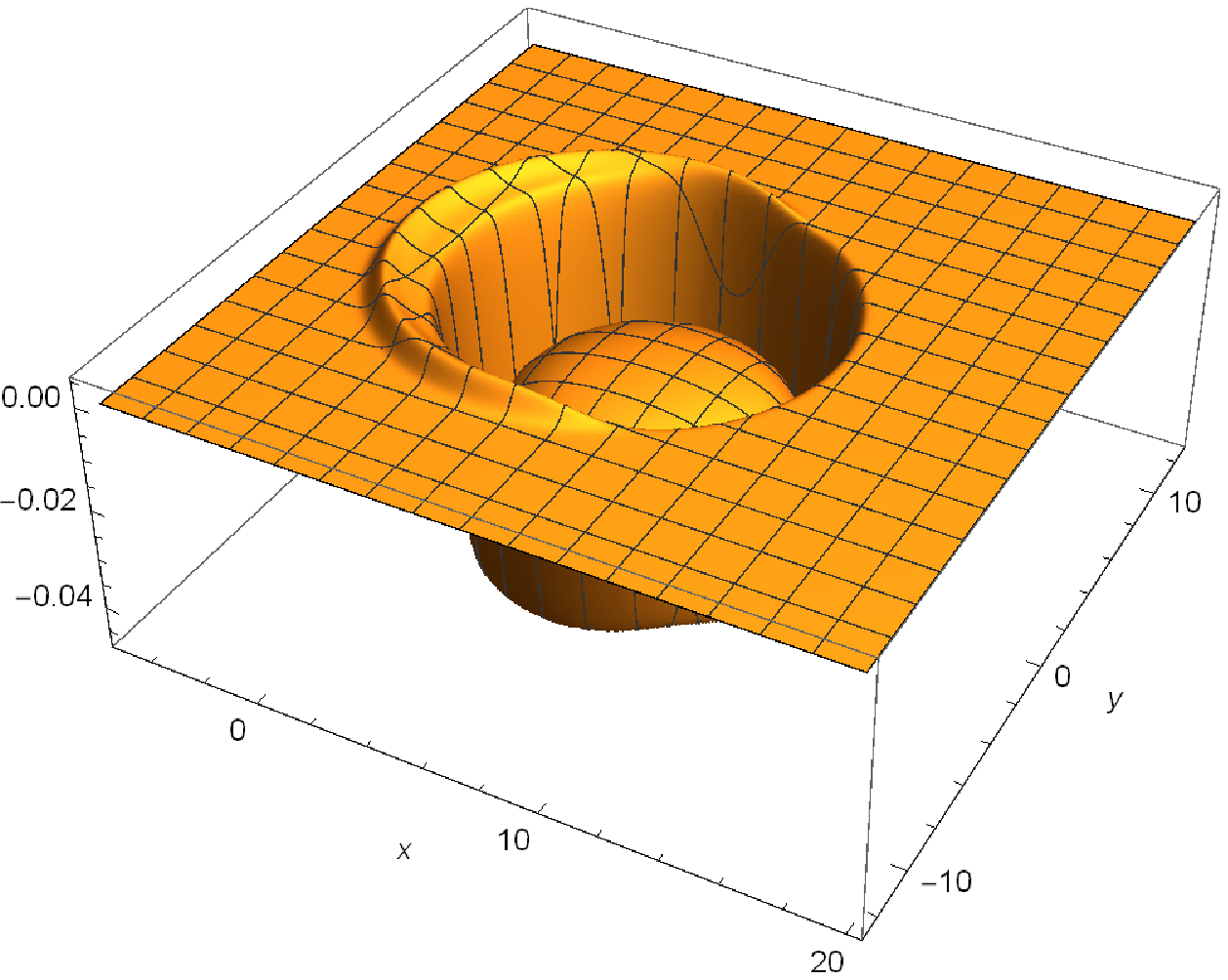}
    \caption{$\sigma^{1/2} F_2^{(2)}(t,x,y)$ at $t=8$ for $d=2$.}
    \label{fig_f22_2d}
\end{figure}
\begin{figure}[H]
    \centering
    \includegraphics[keepaspectratio, scale=0.60]{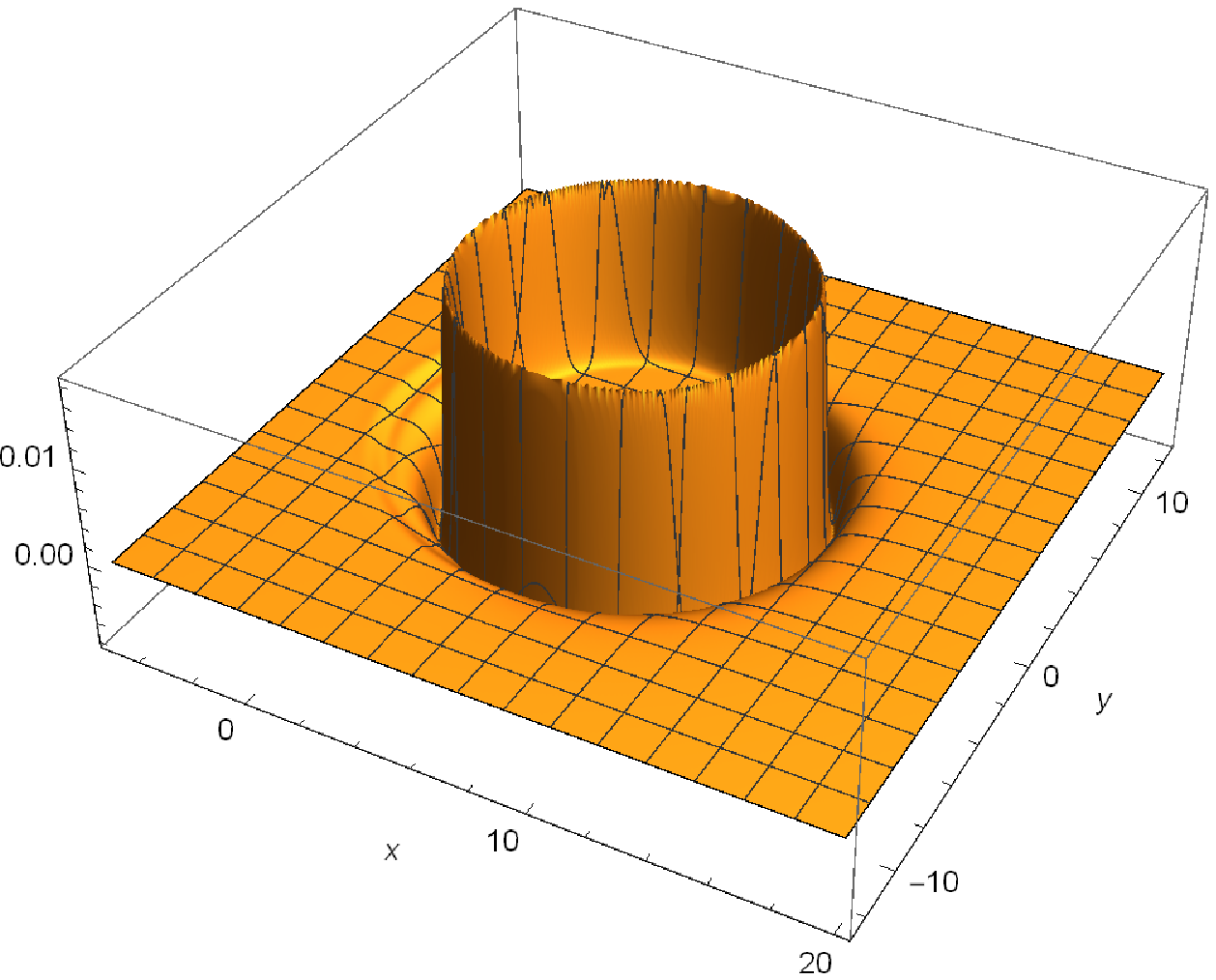}
    \caption{$\sigma^{3/2} G_2^{(1)}(t,x,y)$ at $t=8$ for $d=2$.}
    \label{fig_g21_2d}
\end{figure}
\begin{figure}[H]
    \centering
    \includegraphics[keepaspectratio, scale=0.57]{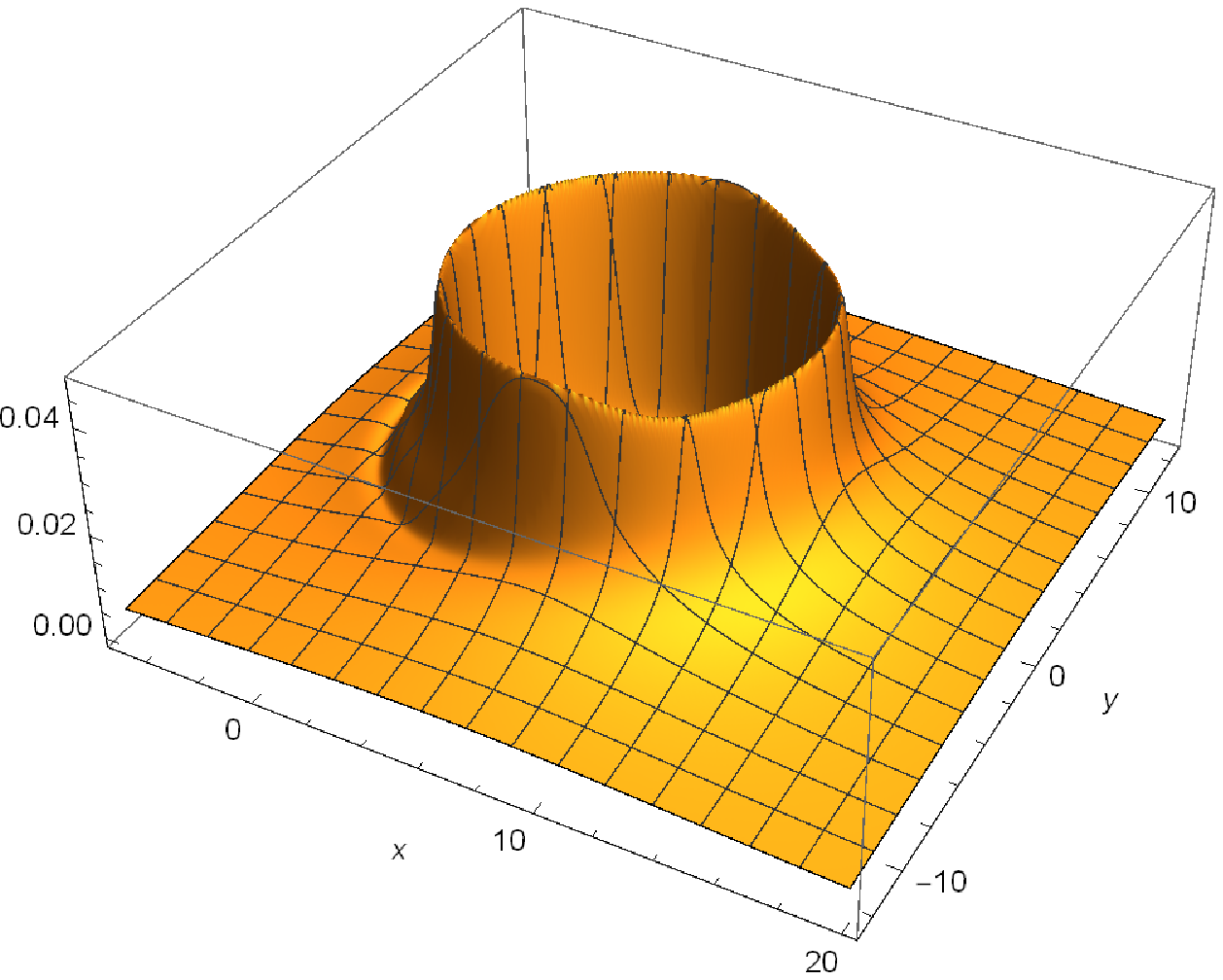}
    \caption{$\sigma^{1/2} G_2^{(2)}(t,x,y)$ at $t=8$ for $d=2$.}
    \label{fig_g22_2d}
\end{figure}
\begin{figure}[H]
    \centering
    \includegraphics[keepaspectratio, scale=0.57]{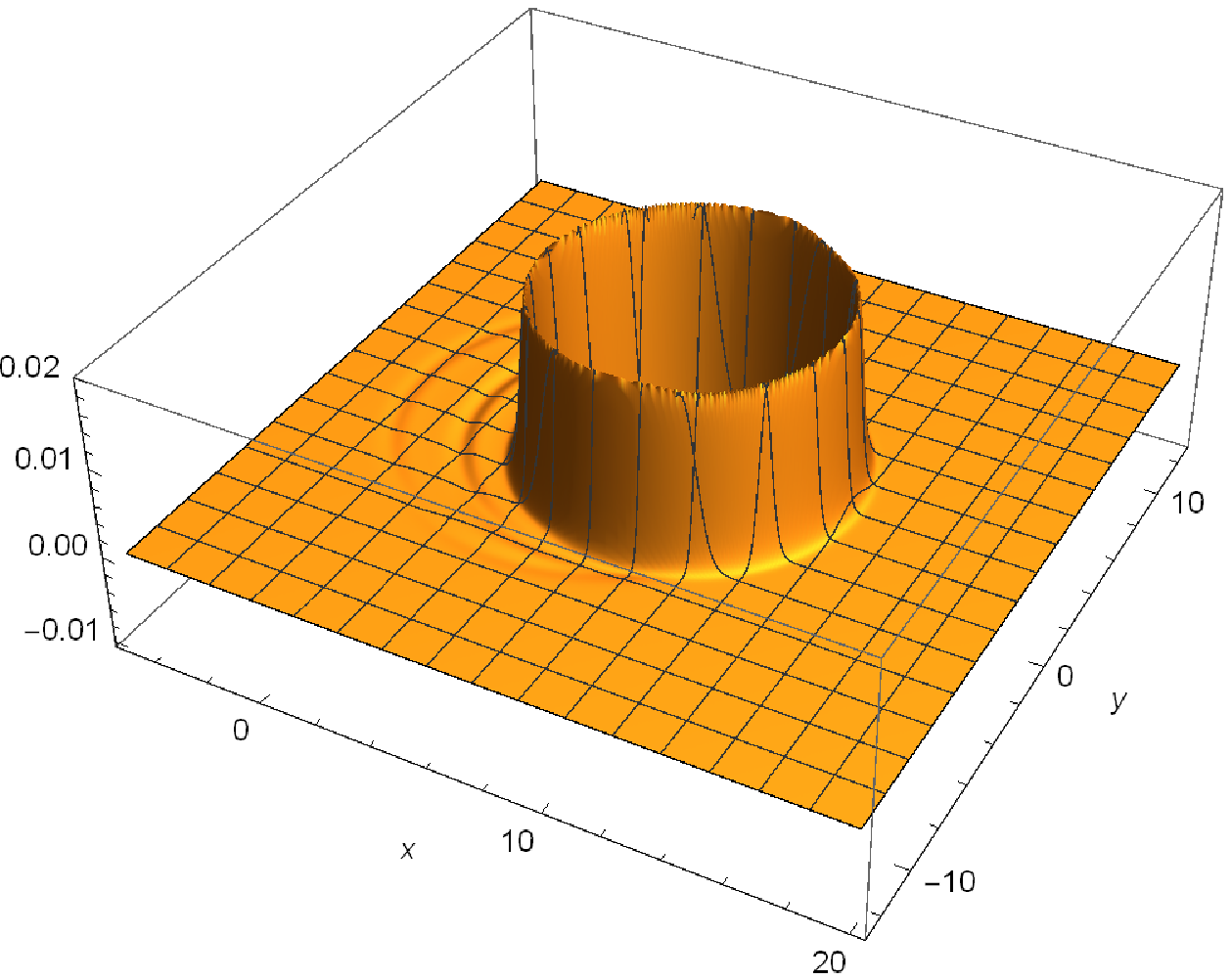}
    \caption{$\sigma^{3/2} F_3^{(1)}(t,x,y)$ at $t=8$ for $d=2$.}
    \label{fig_f31_2d}
\end{figure}
\begin{figure}[H]
    \centering
    \includegraphics[keepaspectratio, scale=0.57]{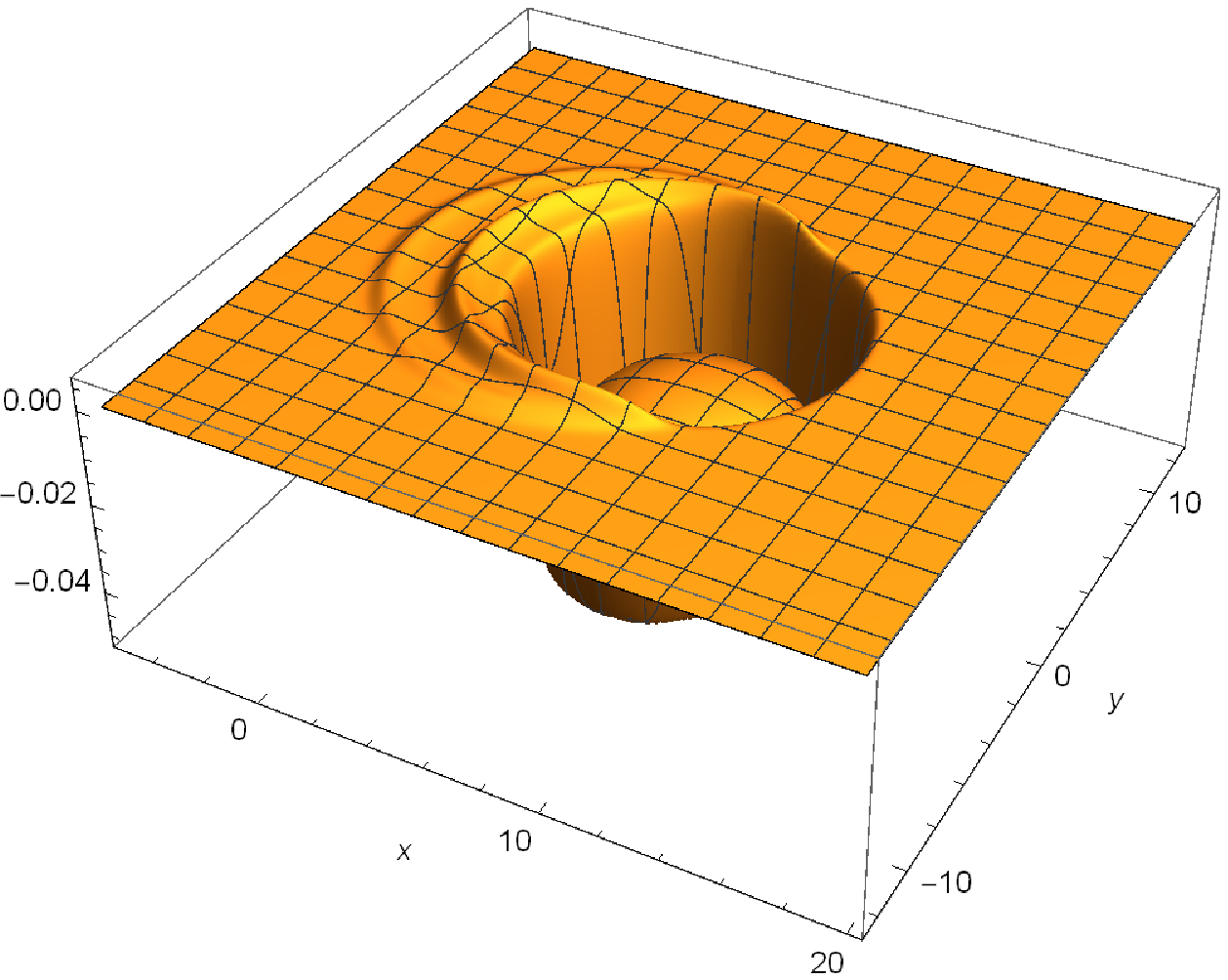}
    \caption{$\sigma^{1/2} F_3^{(2)}(t,x,y)$ at $t=8$ for $d=2$.}
    \label{fig_f32_2d}
\end{figure}
\begin{figure}[H]
    \centering
    \includegraphics[keepaspectratio, scale=0.57]{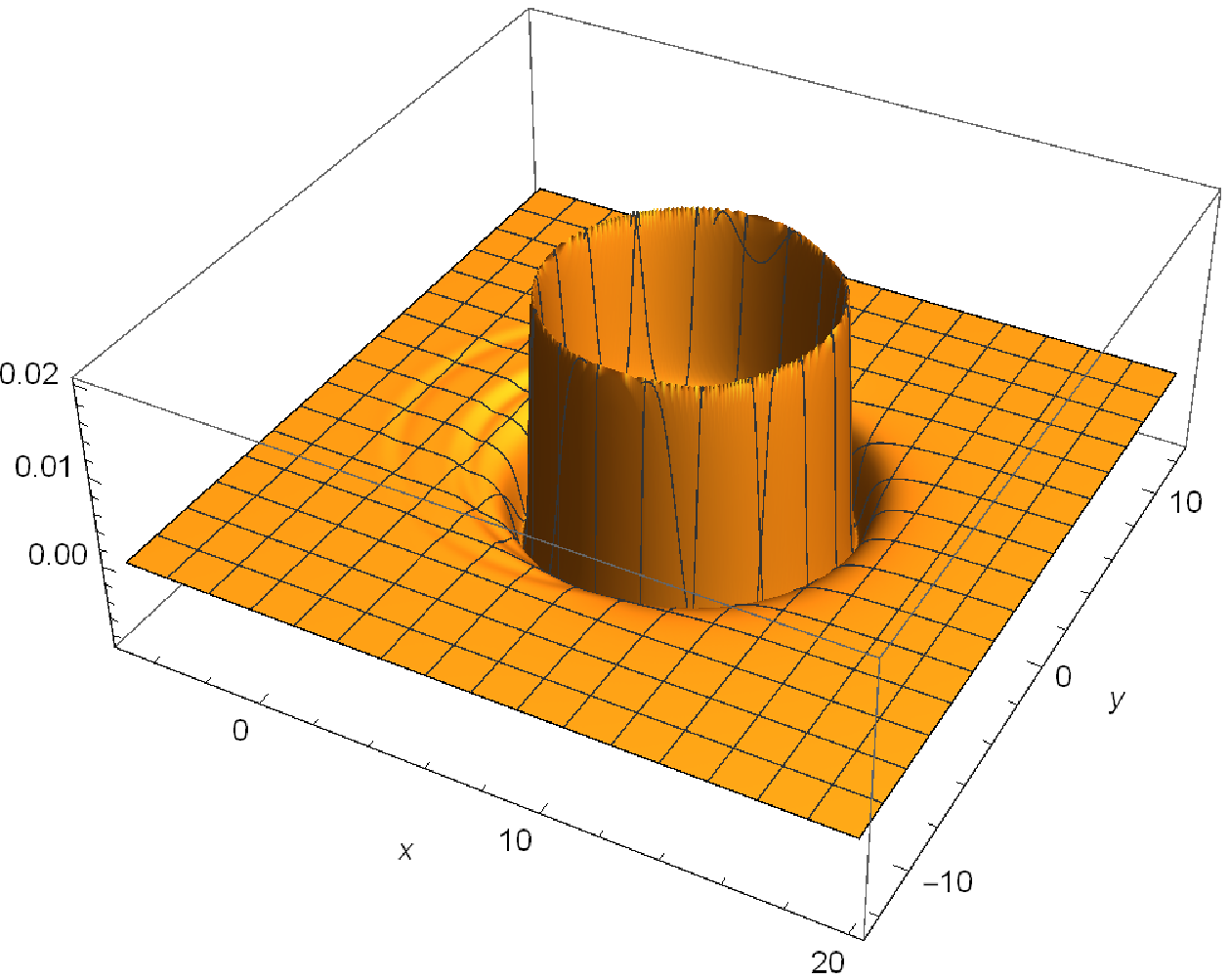}
    \caption{$\sigma^{3/2} G_3^{(1)}(t,x,y)$ at $t=8$ for $d=2$.}
    \label{fig_g31_2d}
\end{figure}
\begin{figure}[H]
    \centering
    \includegraphics[keepaspectratio, scale=0.57]{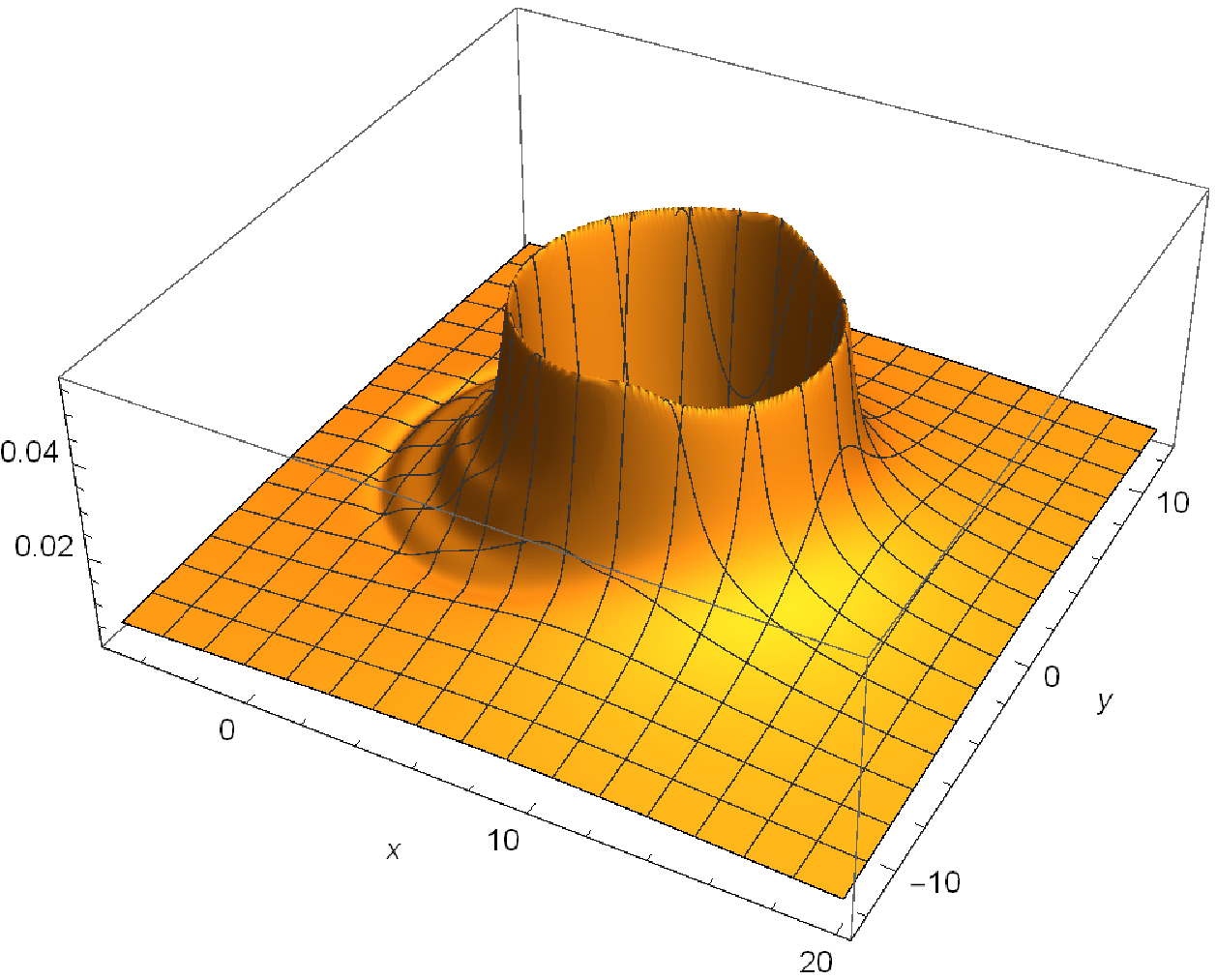}
    \caption{$\sigma^{1/2} G_3^{(2)}(t,x,y)$ at $t=8$ for $d=2$.}
    \label{fig_g32_2d}
\end{figure}

\bibliography{ref}

\end{document}